\documentclass[prd,twocolumn,preprintnumbers,amsmath,amssymb,unsortedaddress]{revtex4}

\usepackage{graphicx}  
\usepackage{dcolumn}   
\usepackage{bm}        

\newcommand{\be}{\begin{equation}}
\newcommand{\ee}{\end{equation}}
\newcommand{\ba}{\begin{eqnarray}}
\newcommand{\ea}{\end{eqnarray}}

\def\nhat{{\hat{\bf n}}}

\def\Lin{{\rm Lin}}

\def\pp1{{\prime}}
\def\pp2{{\prime\prime}}

\def\2D{{\rm 2D}}

\def\Vu{{V_{\mu}}}

\def\pdot{\dot{\Phi}}

\def\Fka{{\mathcal F}(k)}

\def\SN{{\mathcal S}/{\mathcal N}}

\def\h{{\rm h}}

\def\bx{{\bf x}}

\def\bk{{\bf k}}
\def\bq{{\bf q}}

\def\1Loop{{\rm 1Loop}}

\def\rhob{\bar{\rho}}

\def\Msol{h^{-1}M_{\odot}}

\def\Mpc{\, h^{-1}{\rm Mpc}}

\def\Gpc{\, h^{-1}{\rm Gpc}}

\def\kMpc{\, h \, {\rm Mpc}^{-1}}

\def\dx{d^3\!x}
\def\dk{d^3\!k}
\def\dq{d^3\!q}

\def\d{\delta}

\def\del{\nabla}

\def\nbar{\bar{n}}

\def\nbarh{\bar{n}_{\rm h}}

\def\fun#1#2{\lower3.6pt\vbox{\baselineskip0pt\lineskip.9pt
        \ialign{$\mathsurround=0pt#1\hfill##\hfil$\crcr#2\crcr\sim\crcr}}}

\begin{document}


\title{Impact of Scale Dependent Bias and Nonlinear Structure Growth
  on Integrated Sachs-Wolfe Effect: Angular Power Spectra}




\author{Robert E. Smith} 
\affiliation{Institute for Theoretical Physics, University of Zurich,
  Zurich CH 8037} 
\email{res@physik.unizh.ch}

\author{Carlos Hern\'andez-Monteagudo} 
\affiliation{Max-Planck Institute For Astrophysics, P.O. Box 1523,
  85741 Garching, Germany} 
\email{chm@MPA-Garching.MPG.DE}

\author{Uro$\check{\rm s}$ Seljak} \affiliation{Institute for
  Theoretical Physics, University of Zurich, Zurich CH 8037}
\affiliation{Physics Department and Lawrence Berkeley National
  Laboratory, University of California, Berkeley, California, 04720,
  USA. } 
\affiliation{ Ewha University, Seoul 120-750, S. Korea}
\email{seljak@physik.unizh.ch}


\begin{abstract}
We investigate the impact of nonlinear evolution of the gravitational
potentials in the LCDM model on the Integrated Sachs-Wolfe (ISW)
contribution to the CMB temperature power spectrum, and on the
cross-power spectrum of the CMB and a set of biased tracers of the
mass. We use an ensemble of $N$-body simulations to directly follow
the potentials and compare the results to analytic perturbation theory
(PT) methods.  The predictions from the PT match the results to high
precision for $k<0.2 \kMpc$. We compute the nonlinear corrections to the
angular power spectrum and find them to be $<10\%$ of linear theory
for $l<100$. These corrections are swamped by the cosmic variance. On
scales $l>100$ the departures are more significant, however the CMB
signal is more than a factor $10^3$ larger at this scale. Nonlinear
ISW effects therefore play no role in shaping the CMB power spectrum
for $l<1500$.  We analyze the CMB--density tracer cross-spectrum using
simulations and renormalized bias PT, and find good agreement.  The
usual assumption is that nonlinear evolution enhances the growth of
structure and counteracts the linear ISW on small scales, leading to a
change in sign of the CMB-LSS cross-spectrum at small scales. However,
PT analysis suggests that this trend reverses at late times when the
logarithmic growth rate $f=d\ln D/d\ln a<0.5$ or $\Omega_m (z)<0.3$.
Numerical results confirm these expectations and we find no sign
change in ISW-LSS cross-power for low redshifts.  Corrections
due to nonlinearity and scale dependence of the bias are found to be
$<10\%$ for $l<100$, therefore below the signal-to-noise of the
current and future measurements.  Finally, we estimate the
cross-correlation coefficient between the CMB and halos and show that
it can be made to match that for the dark matter and CMB to within
$5\%$ for thin redshift shells, thus mitigating the need to model bias
evolution.
\end{abstract}




\maketitle


\section{Introduction}

Measurements of the temperature fluctuations in the Cosmic Microwave
Background (CMB), provide a unique window onto the primordial Universe
and a means to learn about the physical processes that generated the
initial conditions. This discriminatory power is exemplified by recent
results from the WMAP experiment \citep{Komatsuetal2008}: the
primordial power spectral index is $n_s=0.960\pm0.013$, ruling out the
Harrison-Zel'Dovich spectrum at 3$\sigma$ level.
However, the temperature power spectrum does not
provide a pristine window, but it must be cleaned for the imprint of
foreground signals. One cosmological foreground, is the change in
energy that a CMB photon experiences as it propagates through an
inhomogeneous Universe with time evolving gravitational potentials,
$\pdot$. There are three main effects that may give rise to such
secondary fluctuations:
\begin{itemize} 
\item Linear Integrated Sachs--Wolfe Effect\citep[][hereafter
  ISW]{SachsWolfe1967}: unless the growth of density perturbations
  matches the expansion rate, $\pdot$ will evolve from zero. This will
  lead to a net change in photon temperatures. In LCDM $| \pdot | <0$
  as the potential decays, giving rise to a net positive
  correlation between density and temperature in Fourier space.
\item Rees--Sciama Effect\citep[][hereafter RS]{ReesSciama1968}:
  nonlinear collapse of perturbations to filaments and clusters leads
  to $\pdot \ne 0$ even in the absence of linear ISW, and CMB photons
  change energy as they transit across nonlinear structures.  It is
  usually assumed that nonlinear evolution accelerates the growth of
  structure and counteracts the linear decay of gravitational
  potential in LCDM.  In this paper we show that this is not always
  justified.
\item Birkinshaw--Gull Effect\citep[][hereafter
  BG]{BirkinshawGull1983}: if a mass concentration moves transversely
  to the line of sight, it will create a time variation in the
  potential even if the potential itself is not evolving in time, and
  this will have a dipolar pattern. Consequently, photons which enter
  the potential in the wake, will receive a net energy boost on exit,
  and those which enter ahead will loose energy on exit. However,
  unlike the previous two effects, this contributes only to the CMB
  auto-correlation and not to the cross-correlation of CMB with a
  density tracer.
\end{itemize}
All three effects combine into the nonlinear ISW.  It is well known
that the linear ISW effect leads to fluctuations of the order $\Delta
T\approx 1{\mu\rm K}$ on the largest scales $l<10$ for LCDM \citep[see
  for example][]{Dodelson2003} and has been used to rule out the
self-accelerating branch of DGP model \citep{Fangetal2008}.

The impact of the nonlinear evolution of $\pdot$ on the CMB has been
the subject of a number of studies. However, most of these works
attempt to quantify the effect through the use of simplified analytic
models
\citep{ReesSciama1968,Kaiser1982,MartinezGonzalezetal1990,MartinezGonzalezetal1992,MartinezGonzalezetal1994,Cooray2002a,Cooray2002b}. A
number of studies have employed numerical simulations to track the
evolution of $\pdot$: in a pioneering study, \citet{TuluieLaguna1995}
and \citet{Tuluieetal1996} used ray tracing methods to compute the
change in temperature for individual photon bundles propagating
through inhomogeneous universes. They found that the combined imprint
on the CMB power spectrum, due to the RS and BG effects, were of the
order $\Delta T\sim 1 {\mu\rm K}$ on angular scales $l\sim200$. Owing
to the limited size of their simulations, they were unable to comment
on the effects on the lower multipoles.  \citet{Seljak1996a} related
$\pdot$ to density and momentum using the Poisson and continuity
equation. These predictions were compared to an $N$-body simulation of
the then favored SCDM model, and good agreement was found between the
two as well as to those of \citet{Tuluieetal1996}. However, these
results were obtained in the context of $\Omega_m=1$ model, where no
linear ISW exists, and they could not address $l<100$ behavior, owing
to the limited dynamic range of the simulations.
\citet{Puchadesetal2006} also recently addressed this problem, but
again attention was focused on the large multipole regime.

In a more recent study, \citet{Caietal2008} used a single $N$-body
simulation, the {\tt L-BASIC} simulation, which has $N=488^3$ and
comoving length of $L=1.34\Gpc$, to compute the nonlinear ISW effect.
They measured the $\pdot$ power spectra at each epoch in the
simulation and developed an empirical fitting formula for the
deviations from linear theory.  Using this model they computed the CMB
angular power spectrum and found, on scales $l>50$, that there was
significant nonlinear amplification of power, qualitatively confirming
the earlier halo model predictions of \citet{Cooray2002a}.  However,
these nonlinear corrections occur on angular scales where the primary
anisotropy spectrum is more than two orders of magnitude larger,
rendering them of negligible importance.  \citet{Caietal2008} also
found that there was no evidence for deviations for multipoles
$l<50$. One of the aims of this paper is to  
place more precise constraints on the expected level of
contamination on these large scales.  

The temperature fluctuations induced through the evolving $\pdot$ can
also be observed by correlating the CMB against density perturbations,
as pointed out by \citet{CrittendenTurok1996}, and the large-scale ISW
effect provides an important test for Dark Energy and the curvature of
the Universe. This information can be extracted through the
cross-correlation of the CMB with tracers of the Large-Scale Structure
(hereafter LSS). This analysis has recently been performed by a number
of groups using the WMAP data and several large-scale structure
measurements (e.g. SDSS, NVSS, 2MASS). This work has resulted in up to
$4\sigma$ level detections of the ISW effect
\citep{Scrantonetal2003,BoughnCrittenden2004,Afshordietal2004,Padmanabhanetal2005c,Cabreetal2006,
  Giannantonioetal2006,Rassatetal2007,Giannantonioetal2008,
  Hoetal2008}. In the near future these detections will be improved
upon with PLANCK and the new wide field LSS surveys, such as BOSS,
DES, Pan-STARRS-1 and EUCLID, etc..  However, in a recent paper
\citet{Granettetal2008b} measured the cross-correlation between
superstructures and super-voids with the CMB. On stacking the signal
they found a $\sim4.5\sigma$ detection, in multiple WMAP bands, and
the sign of which appeared consistent with late time ISW. This appears
in stark contrast to expectations from simple signal-to-noise
calculations within the LCDM model
\citep{CrittendenTurok1996,HernandezMonteagudo2008,Douspisetal2008,Frommertetal2008}.
A follow up `consistency' test was performed by
\citet{Granettetal2008c}, the results of which cast some doubt on the
the signal as arising from ISW, at least within the LCDM
model. \citet{Caietal2008} also investigated the ISW-density
cross-correlations, focusing on the nonlinearities arising from the
mass evolution. They found that there was no evidence for enhancement
of evolution of $\pdot$, in agreement with the earlier work of
\citet{VerdeSpergel2002}.  One of the questions we shall address in
this paper is whether selecting biased tracers of LSS relative to the
mass distribution can influence the detection sensitivity for the ISW.

We pursue a two-pronged attack on all of these problems. Our first
avenue will be to use a large ensemble of $N$-body simulations to
directly follow the evolution of $\pdot$. Our second line is analytic,
and we use the nonlinear gravitational perturbation theory (PT) and
renormalized bias frameworks to compute all measured quantities. This
will help us to provide physical insight into the results along the
way. 

The paper breaks down as follows: In \S\ref{sec:ISWtheory} we
summarize the basic theory of the ISW. In \S\ref{sec:simulations} we
describe the ensemble of simulations that we use, and describe our
estimator for measuring $\pdot$ from the simulations. Here we also
present maps, comparing the time evolution of density, and $\pdot$ in
the simulations. In \S\ref{sec:TwoPoint} we investigate the two-point
statistics of $\pdot$, and besides the usual linear analysis we derive
nonlinear expressions within the context of the gravitational
perturbation theory. We evaluate the theory and compare directly with
measurements from the simulations. Then in \S\ref{sec:results1} we
compute the impact on the CMB temperature power spectrum. In
\S\ref{sec:ISWdelta} 
we turn to the cross-correlations with dark matter, followed 
by the correlations with halos in \S\ref{sec:bias} including 
the effects of scale
dependent bias. Again, the theory is compared directly with
measurements from the simulations.  In \S\ref{sec:results2} we perform
the line-of-sight integrals and compute angular cross-power
spectra. Finally, in \S\ref{sec:conc} we summarize our findings and
conclude.


\section{Theoretical background}\label{sec:ISWtheory}

\subsection{The ISW effect}

On arrival at the observer, the CMB photons, which are sourced at the
surface of last scattering, $z\approx1100$, are imprinted with two
sets of fluctuations: the primary anisotropies, which are induced by
the primordial fluctuations, perhaps seeded through the inflationary
mechanism; and the secondary anisotropies, which are induced as the
photons propagate through the clumpy Universe. The primary
anisotropies have been studied in great detail for several
decades\citep[and for a review of the important processes
  see][]{Dodelson2003,Weinberg2008}.  There are a number of physical
mechanisms that give rise to the generation of secondary anisotropies
\citep[for a review see][]{PlanckBlueBook} and one of these is the
redshifting of the photons as they pass through evolving gravitational
potentials.

The temperature fluctuation induced by the gravitational redshift may
be written as \citep{SachsWolfe1967}:
\be \frac{\Delta T(\nhat)}{T_0} = {2 \over c^2}\int_{t_{\rm
ls}}^{t_{0}}dt \dot{\Phi}(\nhat,\chi;t) \ \label{eq:ISW} ,\ee
where $\nhat$ is a unit direction vector on the sphere, $\Phi$ is the
dimensionless metric perturbation in the Newtonian gauge, which
reduces to the usual gravitational potential on small scales, the
`over dot' denotes a partial derivative with respect to the coordinate
time $t$ from the FLRW metric, $\chi$ is the comoving radial geodesic
distance $\chi=\int cdt/a(t)$, and so may equivalently parameterize
time. $t_0$ and $t_{\rm ls}$ denote the time at which the photons are
received and emitted (i.e. last scattering), respectively, $c$ is the
speed of light and $a(t)$ is the dimensionless scale factor.

On scales smaller than the horizon, the perturbed Poisson equation
enables us to relate potential and matter fluctuations
\citep{Peebles1980}:
\be \nabla^2\Phi(\bx;t)=4\pi G\rhob(t)\delta(\bx;t) a^2\!(t) \ , \ee
where $\rhob(t)$ is the mean matter density in the Universe and the
density fluctuation is $\delta(\bx;t)\equiv
[\rho(\bx,t)-\rhob(t)]/\rhob(t)$. Poisson's equation may most easily be
solved in Fourier space, upon which we have,
\be \Phi(\bk;t) = -4\pi G\rhob(t)a^2(t)\frac{\delta(\bk;t)}{k^2}\ . \ee
However, what we are really interested in is the instantaneous time
rate of change of the potential,
\ba \dot{\Phi}(\bk;t) 
& = &
 -\frac{4\pi G}{k^2} \left[\rhob(t)a^3\right] 
\frac{\partial }{\partial t} \left[ \frac{\delta(\bk;t)}{a(t)}\right]\ , \\ 
& = & 
\frac{3}{2}\Omega_{m0} H_0^2 k^{-2} 
\left[ 
\frac{H(t)}{a(t)}\delta(\bk;t)-\frac{\dot{\delta}(\bk;t)}{a(t)}
\right] \ ,
\ea
where $[a^3(t)\rhob(t)]$ is a time independent quantity in the matter
dominated epoch. In the above, we also defined $H(t)\equiv
\dot{a}(t)/a(t)$ and $\Omega_m(t)\equiv \rhob(t)/\rho_{\rm crit}(t)$,
with $\rho_{\rm crit}(t)=3H^2(t)/8\pi G$. All quantities with a
subscript $0$ are to be evaluated at the present epoch. Estimating the
change in the photon temperature due to the evolving potentials
requires knowledge of the evolution of the density perturbation and
its time rate of change. In the linear regime we may solve the
equation of motion for $\delta$ exactly and obtain both of these
quantities.  However, in the nonlinear regime the situation is more
complex and requires numerical simulations or nonlinear models to
proceed. In simulations, measuring $\delta(\bk,a)$ is relatively
straightforward, whereas its time derivative is more complicated. As
was shown by \citet{Seljak1996a} one may obtain this from the
perturbed continuity equation \citep{Peebles1980}:
\be {\bf\del}\cdot\left[1+\delta(\bx;t)\right]{\bf v}_p(\bx;t)=-
a(t)\dot{\delta}(\bx;t) \ , \ee
where ${\bf v}_p(\bx;t)$ is the proper peculiar velocity field. On
defining the pseudo-peculiar momentum field to be, 
\be {\bf p}(\bx;t)\equiv\left[1+\delta(\bx;t)\right]{\bf v}_p(\bx;t)\ , \ee
then in Fourier space we may solve the continuity equation directly to
give us:
\be \dot{\delta}(\bk;t)= i\bk\cdot{\bf p}(\bk;t) /a(t)\ .\ee
Hence, our final expression becomes,
\be \dot{\Phi}(\bk;t) = \Fka \left[
  \frac{H(t)}{a(t)}\delta(\bk;t)-\frac{i\bk\cdot{\bf
      p}(\bk;t)}{a^2(t)} \right] \ , \label{eq:pdot}\ee
where to enable us to pass easily from potential to density we
introduced the quantity
\be {\mathcal F}(k) \equiv \frac{3}{2}\Omega_{m0}
\left(\frac{H_0}{k^2}\right)^2  \ . \ee
%


\section{The ISW from $N$-body simulations}\label{sec:simulations}

\subsection{The {\tt zHORIZON} simulations}

In this study we use a subset of the Z\"urich Horizon, ``{\tt
  zHORIZON}'', simulations.  These are a large ensemble of pure cold
dark matter $N$-body simulations ($N_{\rm sim}=30$), performed at the
University of Z\"urich on the {\tt zBOX2} and {\tt zBOX3}
super-computers. The specific aim for these simulations is to provide
high precision measurements of cosmic structures on scales of the
order $\sim100\Mpc$ and to also provide insight into the rarest
fluctuations within the LCDM model that we should expect to find
within the observable universe.  In this paper we shall only employ
the first 8 {\tt zHORIZON} simulations, since these runs have 11
snapshots logarithmically spaced in the expansion factor from $z=1$ to
$0$, thus giving sufficient time sampling of the simulated density
field to capture the late time evolution.  The expansion factors at
which snapshots are recorded are:
$a=\{1.0,\,0.93,\,0.87,\,0.76,\,0.66,\,0.62,\,0.57,\,0.54,\,0.5\}$.

Each numerical simulation was performed using the publicly available
{\tt Gadget-2} code \citep{Springel2005}, and followed the nonlinear
evolution under gravity of $N=750^3$ equal mass particles in a
comoving cube of length $L=1500\Mpc$. All of the simulations were run
within the same cosmological model, and the particular choice for the
parameters was inspired by results from the WMAP experiment
\citep{Spergeletal2003,Spergeletal2007,Komatsuetal2008}. The
parameters are: $\{\Omega_{m0}=0.25,\,\Omega_{\rm DE,0}=0.75,
\Omega_{\rm b,0}=0.04,\,\sigma_8=0.8,\,n_s=1.0,\,w_0=-1,\,h=0.72\}$,
where these are: the density parameters in matter, dark energy and
baryons; the power spectrum normalization and primordial spectral
index; equation of state parameter for dark energy $p/\rho=w_0$;
dimensionless Hubble parameter. The transfer function for the
simulations was generated using the publicly available {\tt cmbfast}
code \citep{SeljakZaldarriaga1996,Seljaketal2003b}, with high sampling
of the spatial frequencies on large scales. Initial conditions were
lain down at redshift $z=50$ using the serial version of the publicly
available {\tt 2LPT} code \citep{Scoccimarro1998,Crocceetal2006}.

Dark matter halo catalogs were generated for all snapshots of each
simulation using the Friends-of-Friends (FoF) algorithm
\citep{Davisetal1985}, with the linking-length parameter set to the
standard $b=0.2$. For this we used the fast parallel {\tt B-FoF} code,
kindly provided by V.~Springel. The minimum number of particles for
which an object was considered to be a bound halo, was set to 30
particles. This gave a minimum host halo mass of $\sim1.5\times10^{13}
M_{\odot}/h$.




\subsection{Estimating the ISW effect in simulations}

In order to estimate $\pdot$, we require estimates of both the density
field and pseudo-peculiar momentum field in Fourier space
(c.f. Eq.~(\ref{eq:pdot}). The dark matter density field can be
written as a sum over Dirac delta functions,
\be \rho(\bx)= \sum_{l=1}^{N} m_l \delta^D(\bx-\bx_l) \ ,\ee
where $m_l$ is the mass of the $l$th particle and we take all
particles to have equal mass. The density field averaged on a cubical
lattice can then be obtained through the convolution,
\ba
\rho_{g}(\bx_{ijk}) 
& = & 
\frac{1}{V_{W}}\int \dx \rho(\bx) W(\bx_{ijk}-\bx) \ ; \nonumber \\
& = &
\frac{m}{\Vu}\sum_{l}^{N} W(\bx_{ijk}-\bx_l) \ ,
\ea
where $W$ represents the dimensionless window function of the mass
assignment scheme, and where the normalization factor is $V_{W} = \int
\dx' W(\bx-\bx')$. The filter function $W$ that we adopt throughout is
the `cloud-in-cell' charge assignment scheme
\citep{HockneyEastwood1988}. Hence, our estimate for the density
fluctuation is
\ba 
1+\widehat{\delta(\bx)}
 & = & 
\frac{1}{N}\frac{\Vu}{V_{W}} \sum_{l}^{N} W(\bx_{ijk}-\bx_l)  \ ,\nonumber \\
& = & 
\frac{N_{\rm cell}}{N} \sum_{l}^{N} W(\bx_{ijk}-\bx_l)  \ ,
\ea
where $N_{\rm cell}=\Vu/V_{W}$ is the total number of grid cells.

The pseudo-momentum field may be estimated in a similar fashion.  For
convenience we write,
\be {\bf p} = \left[1+\delta(\bx)\right]{\bf u}(\bx) a(t) \ , \ee
where ${\bf u}={\bf v}_p/a$ is the comoving peculiar velocity
field. The particle momentum field is then written as
\be \left[(1+\delta){\bf u}\right](\bx) = \frac{\Vu}{N}
\sum_{l}^{N}\delta^{D}(\bx-\bx_l) {\bf u}_l \ . \ee
This may be convolved with the mass assignment scheme to obtain the
mesh averaged quantity
\be  \left[(1+\delta){\bf u}\right](\bx_{ijk}) =
\frac{1}{N}\frac{\Vu}{V_{W}} \sum_l^{N} {\bf u}_l W(\bx_{ijk}-\bx_l) \ .\ \ee
Thus our estimate for the pseudo-momentum field is given by
\be \widehat{\bf p}(\bx_{ijk}) = a(t) \frac{N_{\rm cell}}{N}\sum_l^{N}
    {\bf u}_l W(\bx_{ijk}-\bx_l) \ . \ee

The density Fourier modes were then estimated using the publicly
available {\tt FFTW} routines \citep{FFTW}, and each resulting mode
was corrected for the convolution with the mass-assignment window
function. For the CIC algorithm this corresponds to the following
operation:
\be \delta_{\rm d}(\bk)=\delta_{\rm g}(\bk)/W_{\rm CIC}(\bk) \ ,\ee
where
\be W_{\rm CIC}(\bk)=\prod_{i=1,3}\left\{\left[\frac{\sin 
\left[\pi k_i/2k_{\rm Ny}\right]}{\left[\pi k_i/2k_{\rm Ny}\right]}\right]^2\right\} \ee
and where sub-script d and g denote discrete and grid quantities, and
where $k_{\rm Ny}=\pi N_{\rm g}/L$ is the Nyquist frequency, and
$N_{\rm g}$ is the number of grid cells \citep{HockneyEastwood1988}.

To obtain the real space $\pdot(\bx,t)$, we solved for $\pdot(\bk,t)$
in Fourier space using Eq.~(\ref{eq:pdot}), set the unobservable $k=0$
mode to zero, and inverse transformed back to real space.


\begin{figure*}
\centering{
  \includegraphics[width=8.8cm,clip=]{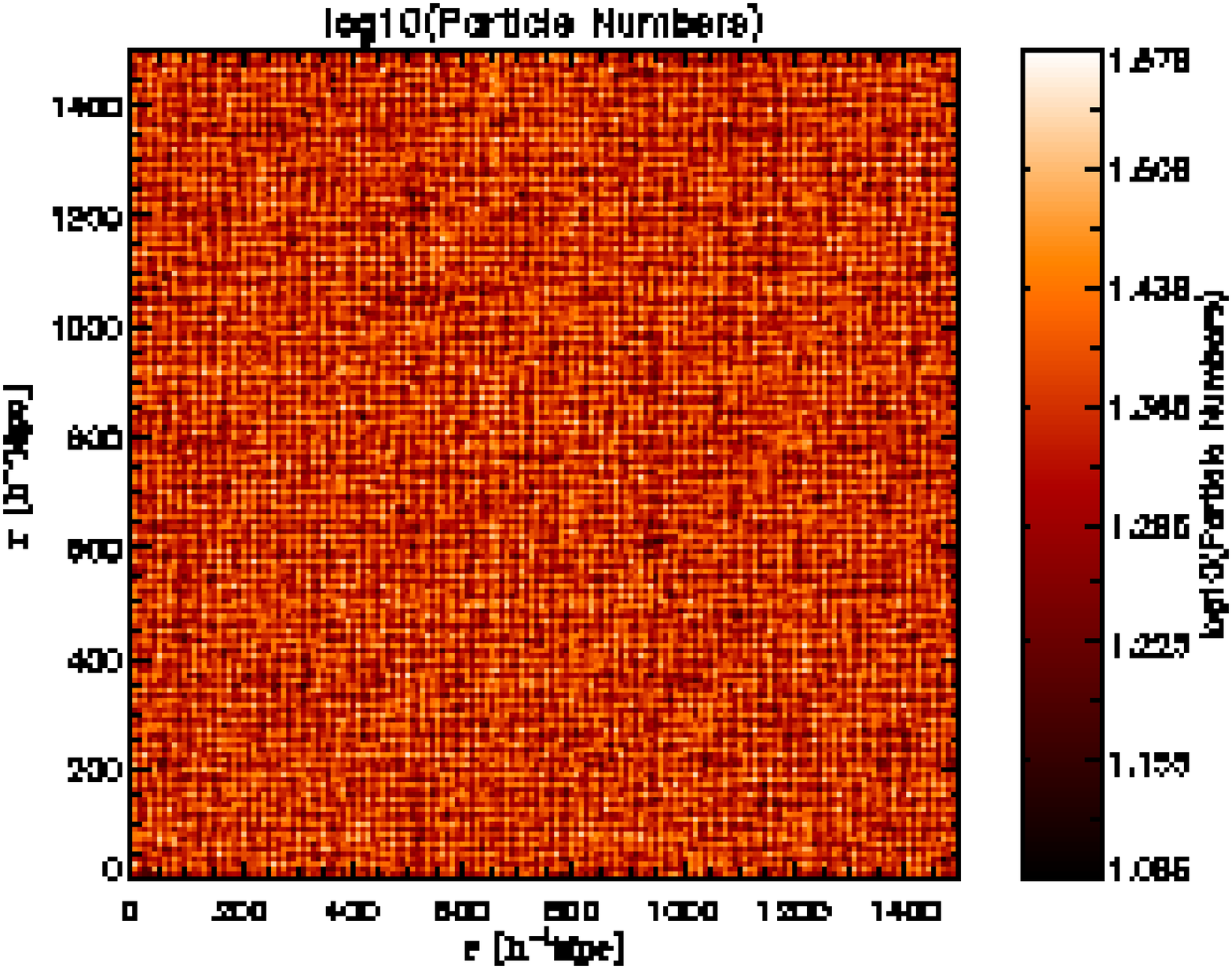}
  \includegraphics[width=8.8cm,clip=]{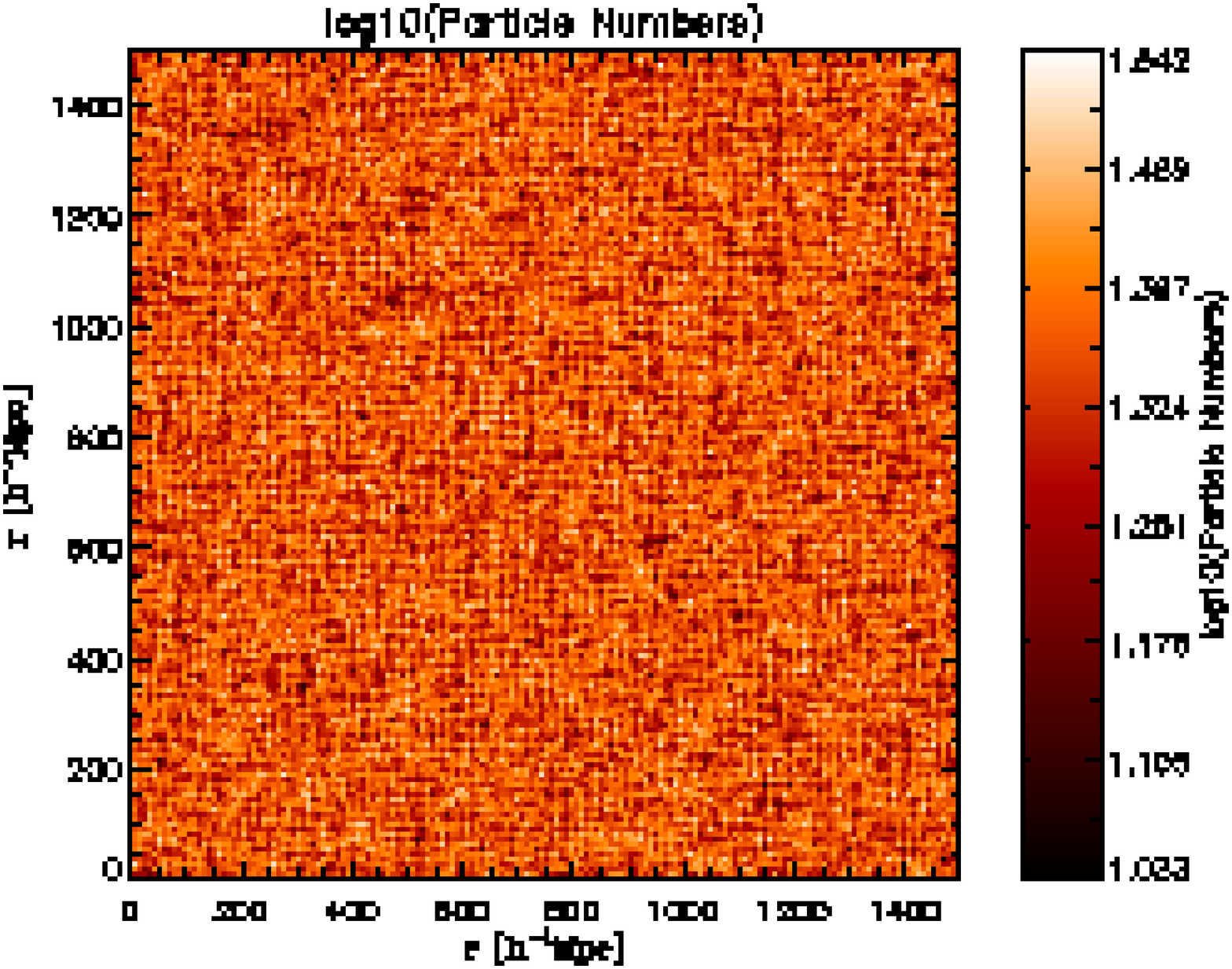}}
  \centering{
  \includegraphics[width=8.8cm,clip=]{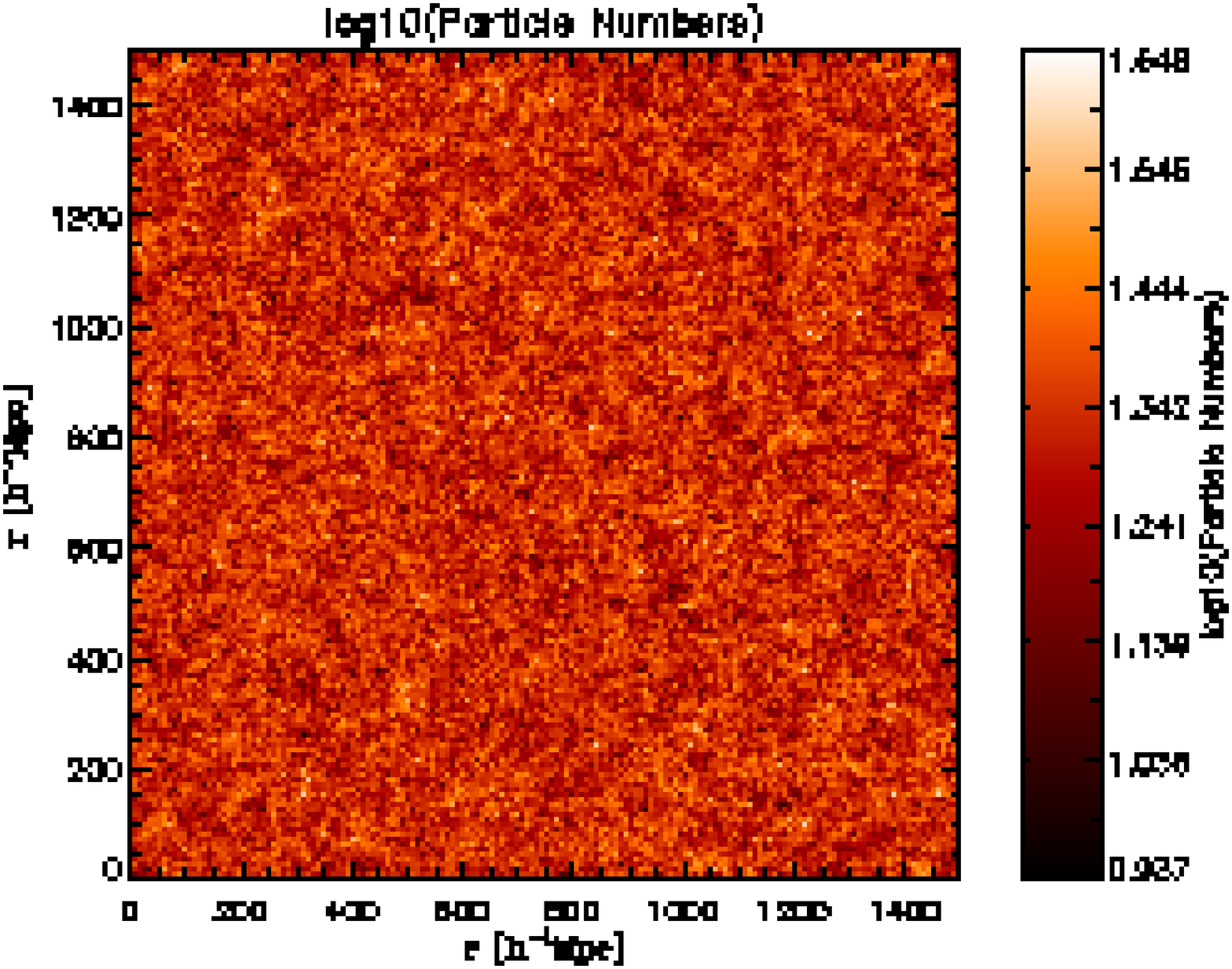}
  \includegraphics[width=8.8cm,clip=]{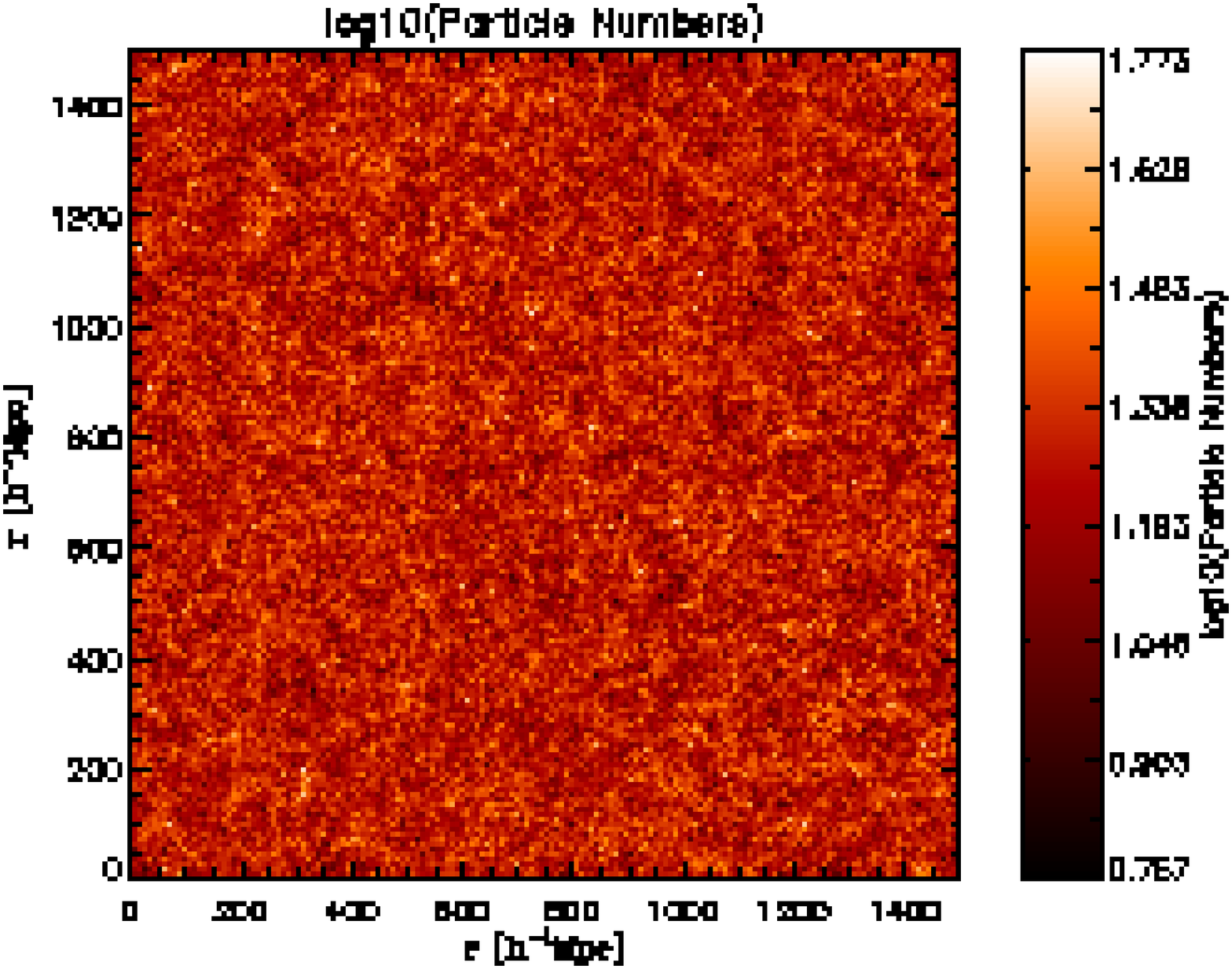}}
  \centering{
  \includegraphics[width=8.8cm,clip=]{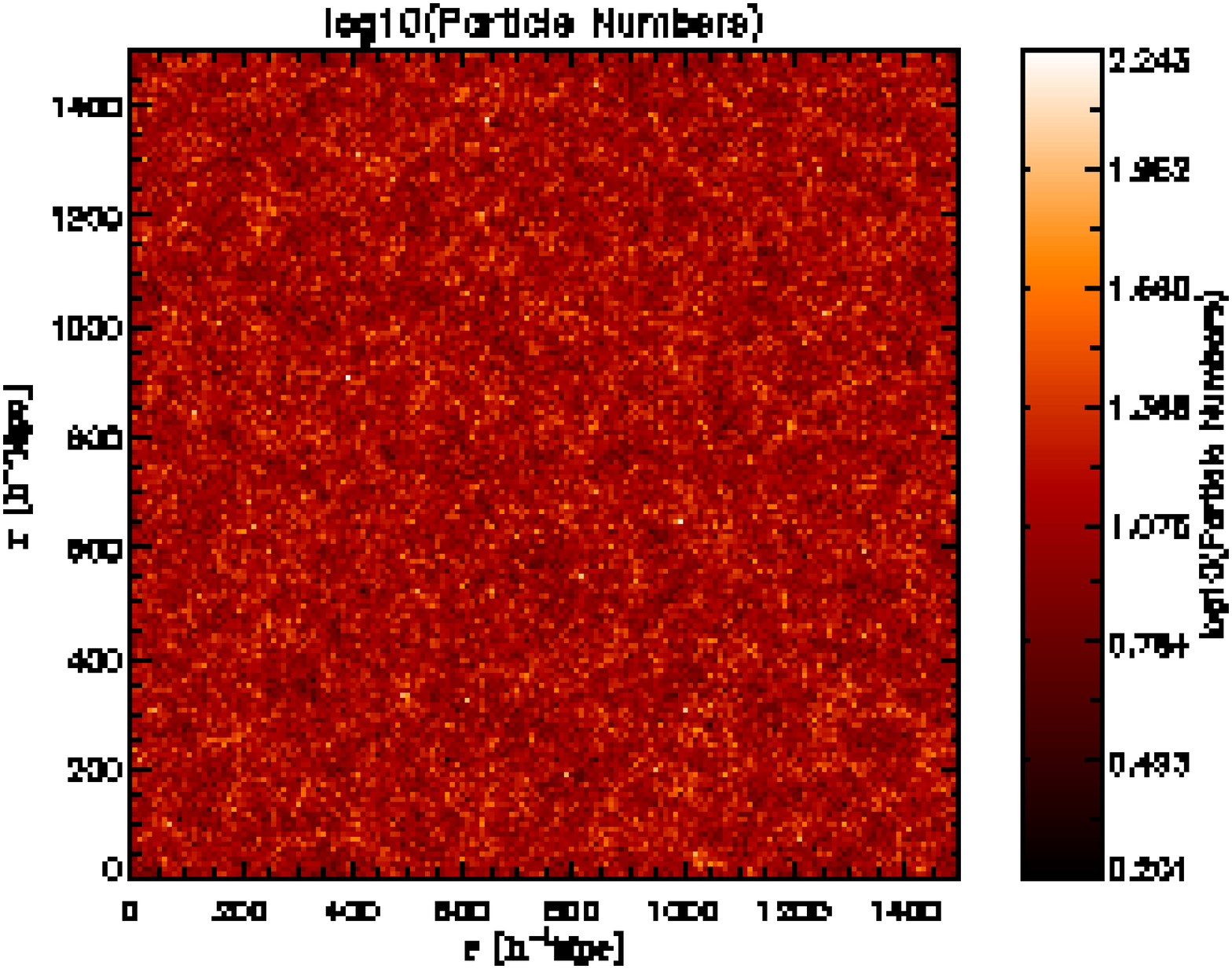}
  \includegraphics[width=8.8cm,clip=]{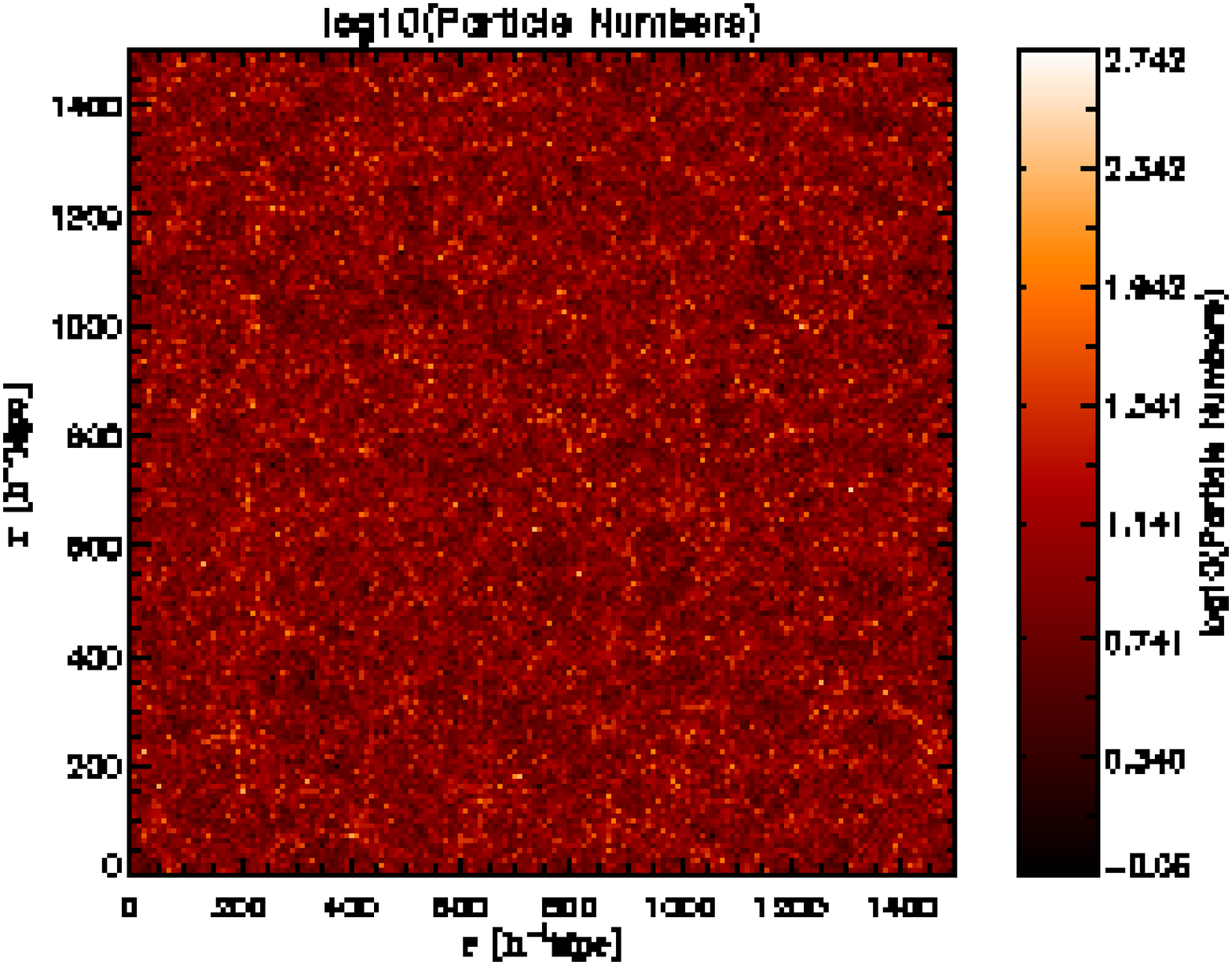}}
\caption{\small{Evolution of $\delta$ in a slab of thickness
 $\Delta x=100\Mpc$. The panels, going from left to right and top to bottom,
 represent redshifts: $z=\{15, 10, 5, 3, 1, 0\}$. \label{fig:MassMAP}}}
\end{figure*}


\begin{figure*}
\centering{
\includegraphics[width=8.8cm,clip=]{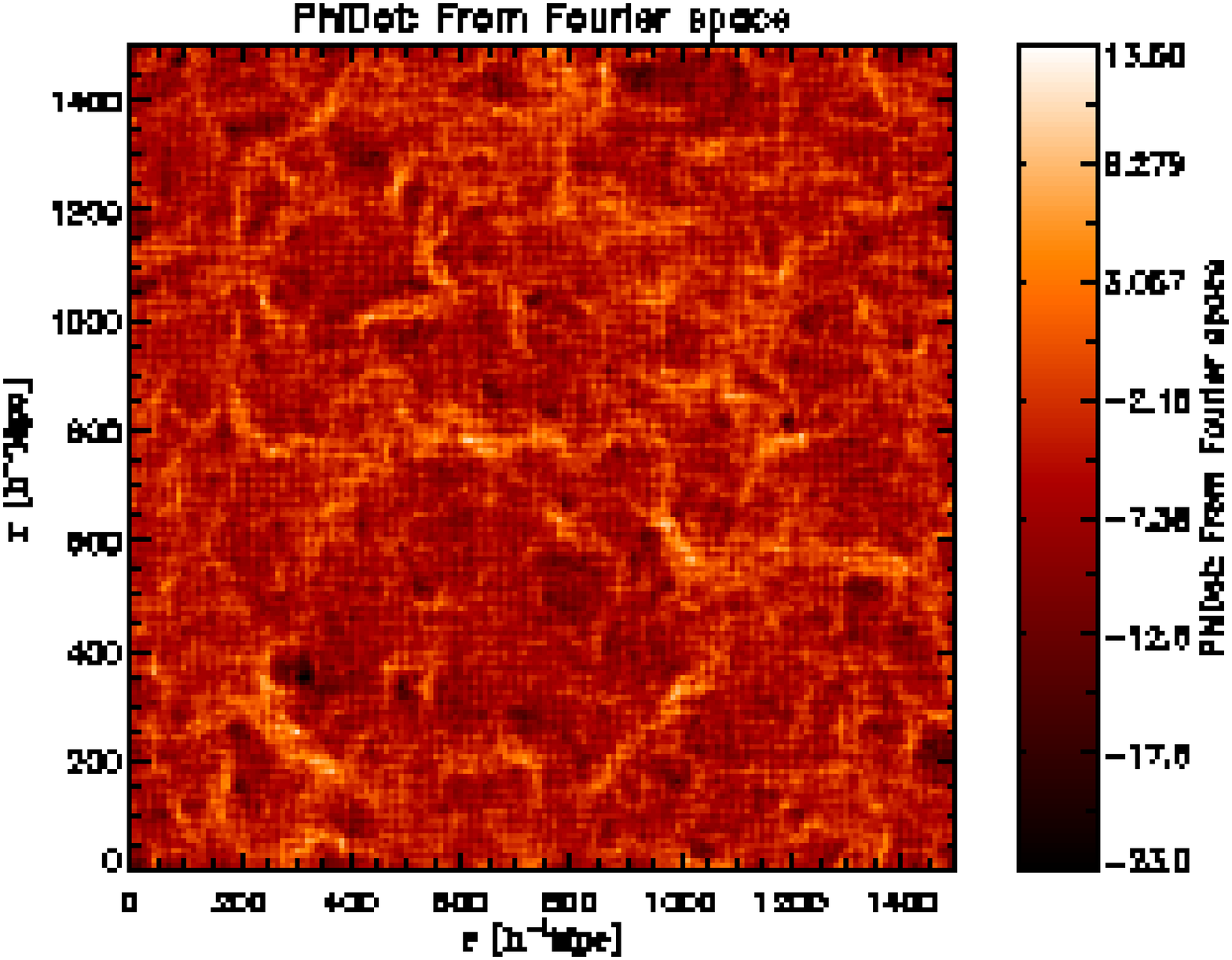}
\includegraphics[width=8.8cm,clip=]{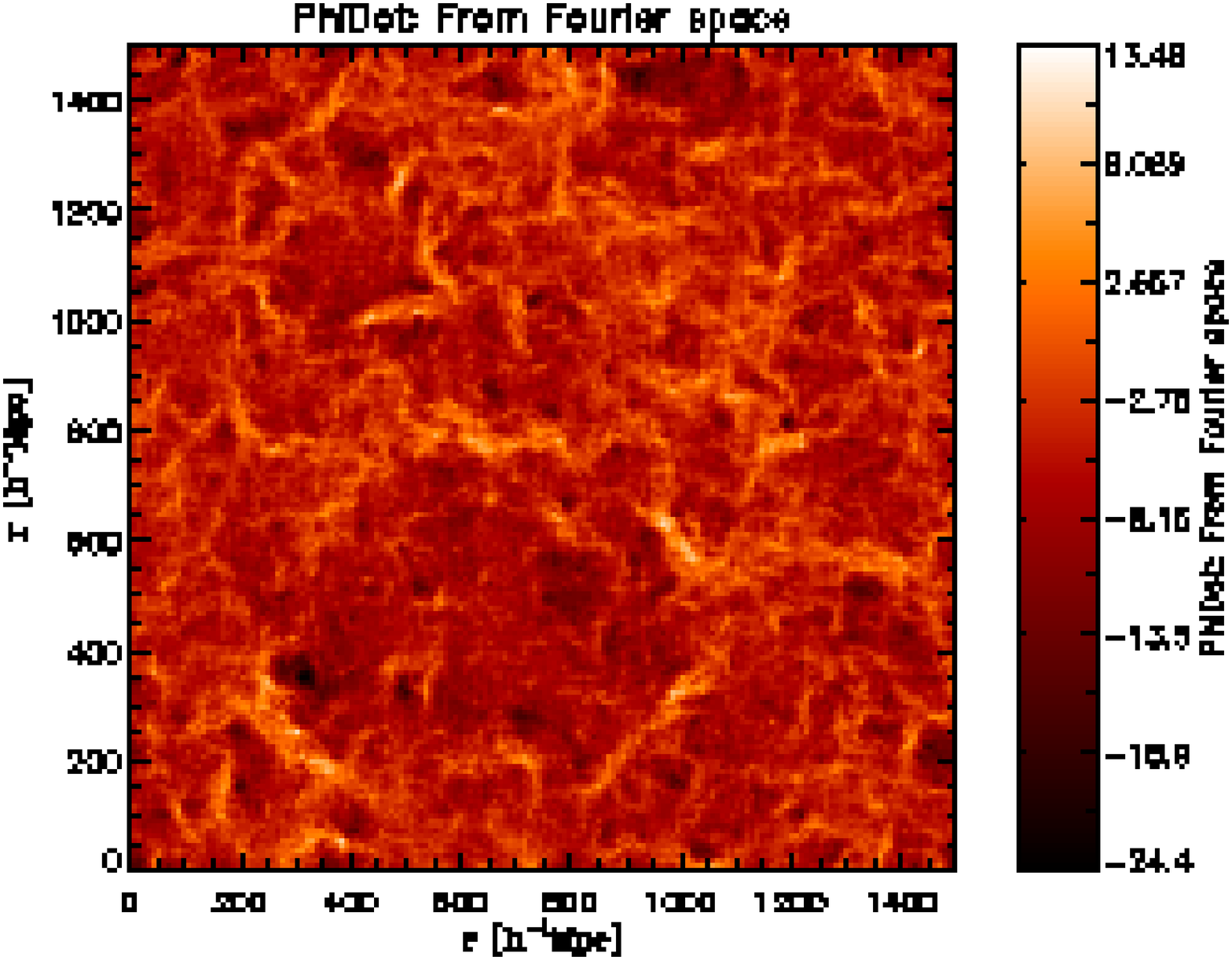}}
\centering{
\includegraphics[width=8.8cm,clip=]{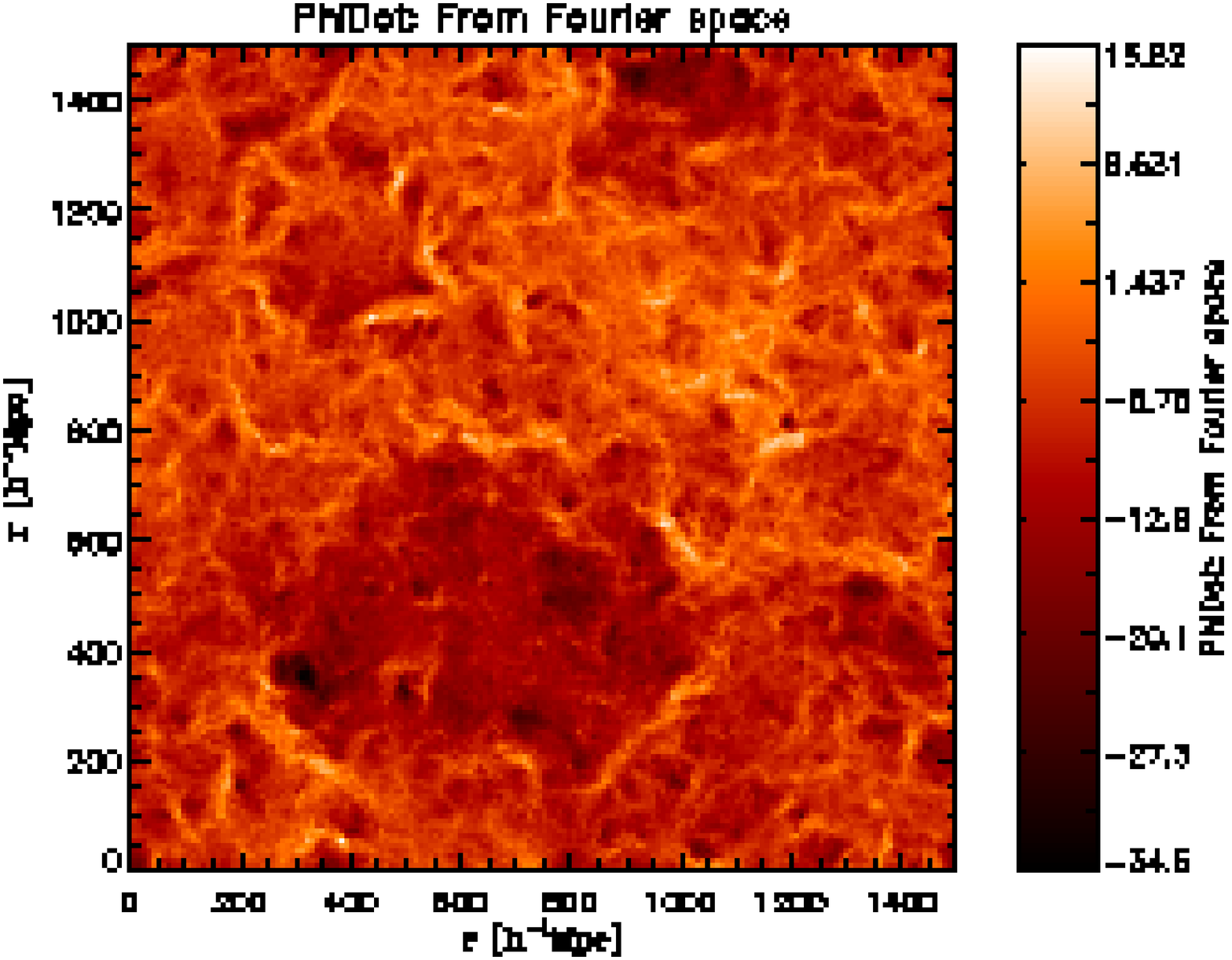}
\includegraphics[width=8.8cm,clip=]{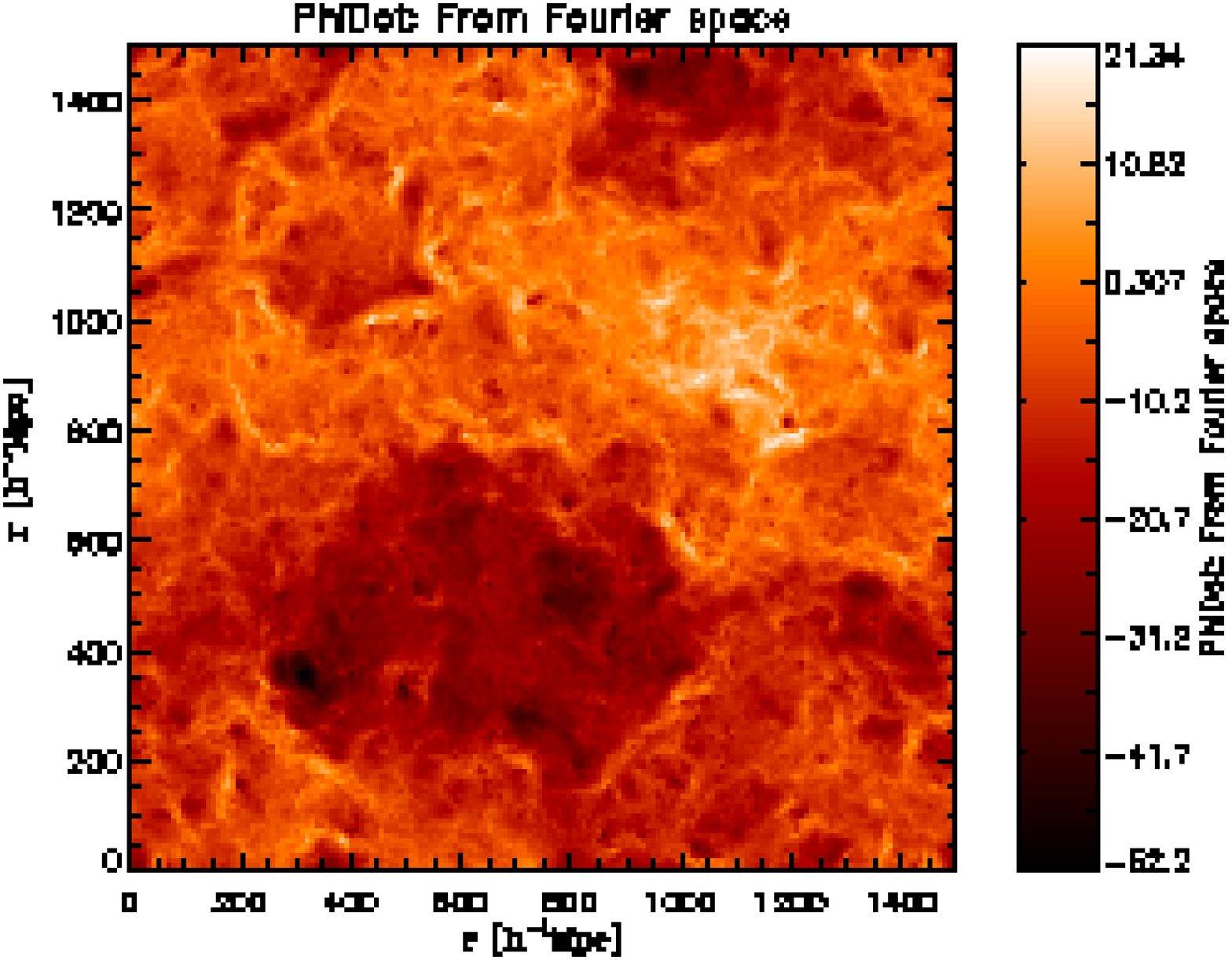}}
\centering{
\includegraphics[width=8.8cm,clip=]{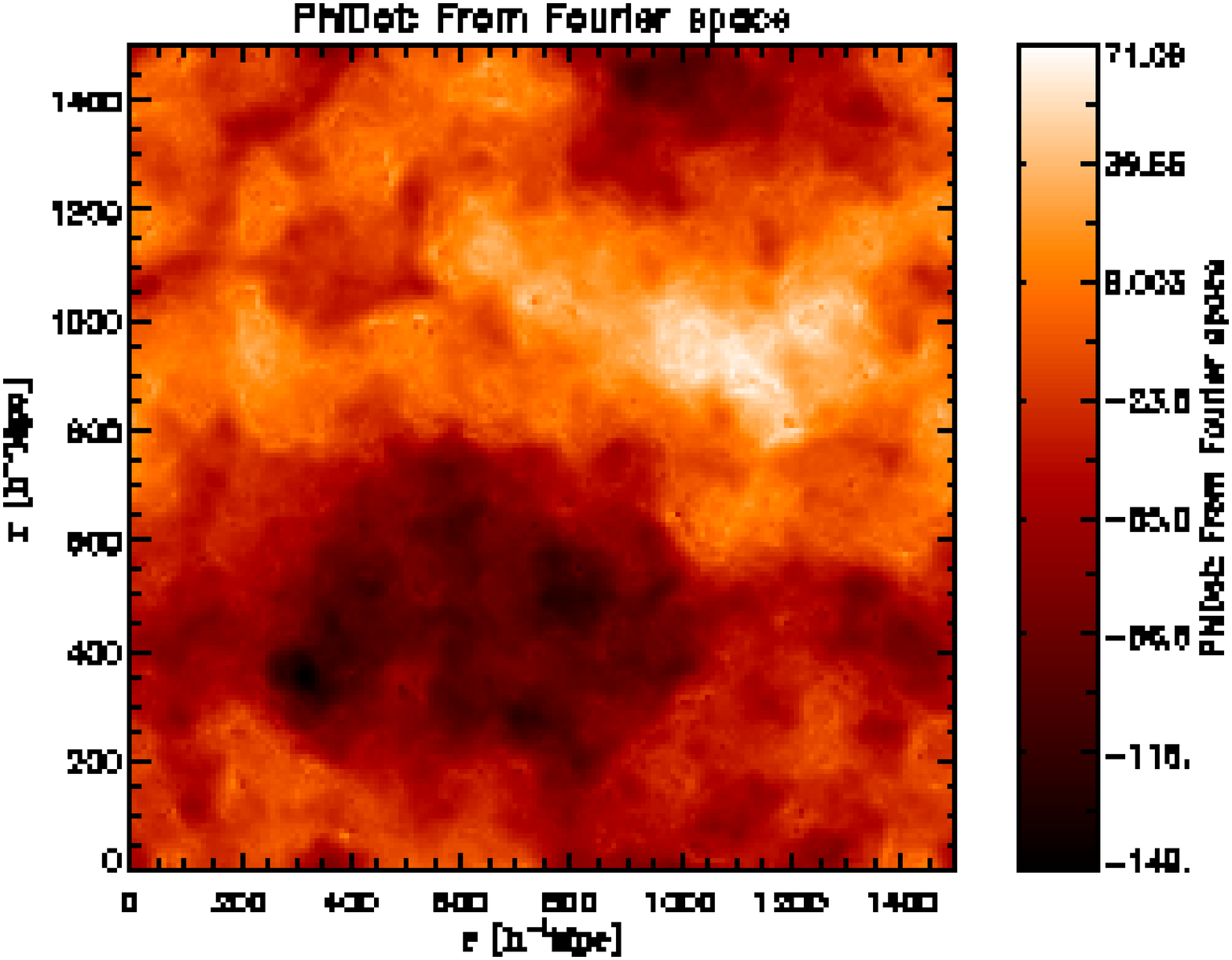}
\includegraphics[width=8.8cm,clip=]{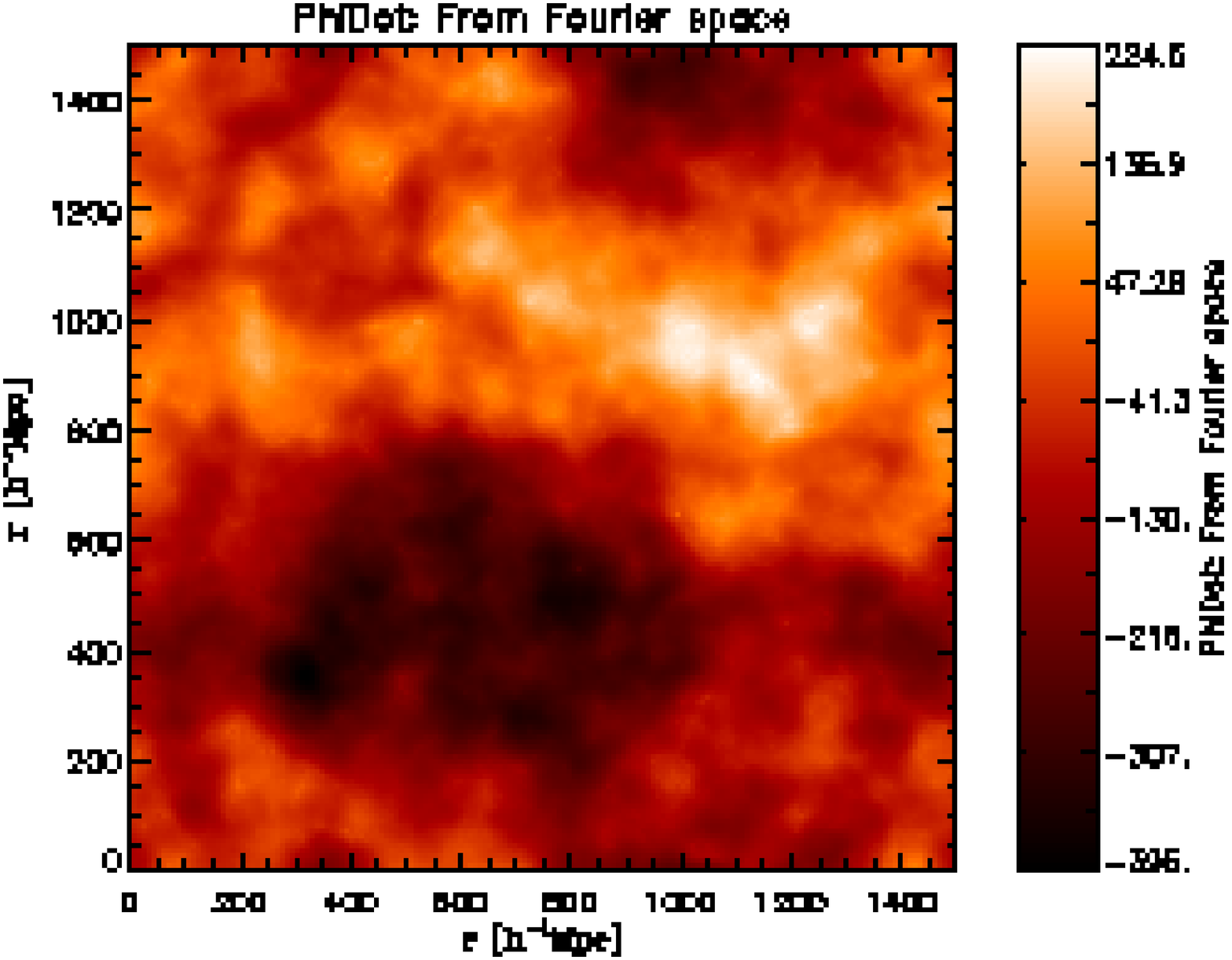}}
\caption{\small{Evolution of $\dot{\Phi}$ in a slab of thickness
 $\Delta x=100\Mpc$. The panels, going from left to right and top to bottom,
 represent redshifts: $z=\{15, 10, 5, 3, 1, 0\}$. \label{fig:PhiDotMAP}}}
\end{figure*}


\subsection{Visual representation of the evolution of $\pdot$}

Fig.~\ref{fig:MassMAP} shows how the dark matter particle number,
projected in a slab of thickness $\Delta x=100\Mpc$ and side length
$L=1500\Mpc$, evolves as a function of cosmic time from $z=15$ to the
present day. At early times, one can see that the Universe is regular
and homogeneous, and the imprint of the initial grid configuration is
still noticeable. At later times, gravitational instability of the
matter has led to the formation of a pattern of web like structures
with dense clumps at the vertices of the web -- the `Cosmic Web'.  The
point we wish to stress, is that it is difficult for the eye to pick out
features that are larger than $100\Mpc$.

Fig.~\ref{fig:PhiDotMAP} shows the evolution of $\pdot(\bx,t)$ as a
function of cosmic time.  At early times ($z\sim15$), when there is no
linear ISW, the maps are dominated by a small-scale foam-like
structure. At later times, $z\sim10$, the foam is sharpened and
transformed, with butterfly like features present at the high density
regions, as expected from the BG effect, i.e. a flow of mass moving
transversely across the sky. At later times the dominating structures
are on extremely large scales ($r>500\Mpc$), as expected by the linear
late time ISW effect.


\section{Density, momentum and potential power spectra in PT}\label{sec:TwoPoint}


\subsection{The 3D power spectra}

The perturbed fields of interest may be written as Fourier series,
\ba 
\psi_{\gamma}(\bx) & = & 
\sum_{j} \psi_{\gamma}(\bk_j) \exp\left[-i\bk_j\cdot\bx\right] \ ;\\
\psi_{\gamma}(\bk_j) & = & 
\frac{1}{V_{\mu}}\int \dx \psi_{\gamma}(\bx) \exp\left[i\bk_j\cdot\bx\right]\ ,
\ea
where $\psi_{\gamma}\equiv \{\delta(\bx),\nabla \cdot{\bf
  p}(\bx),\dot{\Phi}(\bx)\}$, and where $\Vu$ is some large region of
  the Universe over which we shall assume that the functions obey
  harmonic boundary conditions. Then, from translational invariance
  and isotropy, the correlation of different Fourier modes can be
  written
\be P_{\gamma_1\gamma_2}(k_i)\delta^{K}_{i,-j}\equiv V_{\mu}
\left<\psi_{\gamma_1}(\bk_i)\psi_{\gamma_2}(\bk_j)\right> \ ,\ee
where $P_{\gamma_1\gamma_2}$ is the power spectrum matrix of all of
the fields. Using Eq.~(\ref{eq:pdot}) we find, for example:
\ba P_{\pdot\pdot}(k,a) \!\! & = \!\! & \!\! [\Fka]^2
\left[
\frac{H^2(a)}{a^2(t)}P_{\delta\delta}(k) \right. \nonumber \\
& & 
\left. - 2\frac{H(a)}{a^3(t)}P_{\omega\delta}(k) + \frac{1}{a^4(t)} P_{\omega\omega}(k)
\right] \ ; \label{eq:PowPP} \\
P_{\delta\pdot}(k,a) \!\! & \!\!  = & \!\! \Fka
\left[\frac{H(a)}{a(t)}P_{\delta_b\delta}(k) -
\frac{1}{a^2(t)}P_{\delta_b\omega}(k) \right] \label{eq:PowPD} \ ; \ea
where we have defined $\omega(k;t)\equiv i \bk\cdot{\bf p}(\bk;t) =
\dot{\delta}(\bk;t)a(t)$.  


\subsection{Linear theory results}

The two-point statistics may be evaluated easily within the linear
theory: $\delta\ll1$ and $\nabla \cdot {\bf v}\ll1$. In this limit the
Fourier mode of the density and its time derivative evolve as:
\ba
\delta(\bk;t) 
& = & D(t) \delta(\bk;t_0)  \ ; \\
\dot{\delta}(\bk;t)
 & = & 
f(t)H(t)D(t) \delta(\bk;t_0)  \ ,  \label{eq:deltadot} 
\ea
where we have the usual definition of the logarithmic derivative of
the growth factor,
\be f(a)\equiv f(\Omega_m(a),\Omega_{\rm DE}(a)) \equiv 
\frac{\partial \log D(t)}{\partial \log a(t)}  \ .
\ee
Hence we have
\ba P^{\Lin}_{\omega\omega}(k,t) 
& = & \left[a(t)\dot{D}(t) \right]^2 
\left<\left|\delta(\bk;t_0)\right|^2\right> \Vu \ ;\nonumber \\
& = & \left[a(t) f(a)H(t)D(t)\right]^2 P^{\rm Lin}_{\delta\delta}(k;t_0) \ ; \nonumber \\
& = & \left[a(t) f(a)H(t)\right]^2 P^{\rm Lin}_{\delta\delta}(k;t) \label{eq:PwwLin}
\ea
and 
\ba P^{\Lin}_{\omega\delta}(k,t) 
& = & a(t)\dot{D}(t)D(t) \left<\left|\delta(\bk;t_0)\right|^2\right> \Vu \ ;\nonumber \\
& = & \left[a(t) f(a)H(t)\right] P^{\rm Lin}_{\delta\delta}(k;t) \label{eq:PwdLin} 
\ .\ea
Inserting these expressions into Eqs~(\ref{eq:PowPP}) and
(\ref{eq:PowPD}) gives:
\ba 
P^{\Lin}_{\pdot\pdot}(k,t) \!\!\! & = & \!\![\Fka]^2
\left[\frac{H(a)}{a}\left(1-f(a)\right)\right]^2P^{\rm Lin}_{\delta\delta}(k;t) \ ; 
\label{eq:LinISW1}  \\
P^{\Lin}_{\pdot\delta}(k,t) \!\!\!&  =  & \!\!\Fka
\left[\frac{H(a)}{a}\left(1-f(a)\right)\right]
P^{\rm Lin}_{\delta\delta}(k;t) \ .
\label{eq:LinISW2}
\ea
At this point we may note the well known result that, if the Universe
is in an Einstein--de Sitter (EdS) phase of expansion
(i.e. $\Omega_m(a)=1$, and $D(a) \propto a$), then $f(a)=1$ and the
bracketed terms in Eqs~(\ref{eq:LinISW1}) and (\ref{eq:LinISW2})
vanish, so the ISW effect vanishes. However, if the Universe is
under/overdense in gravitationally active matter, then we expect a
non-zero signal, which is positive for both spectra. In the currently
favored LCDM model, $\Omega_{m0}\approx 0.25$, and so
$1-f(a)<1$. However, at early times $\Omega_m\rightarrow1$ and the ISW
is shut off. In the next section we explore how this picture changes
as the fields are evolved into the mildly nonlinear regime.

Before moving on though, we point out that in the literature there are
a number of commonly used approximations for $f(a)$: for
example, $f(a)\approx\Omega^{0.6}$ \citep{Peebles1980}; and
somewhat better, $f(a)\approx
\Omega_m(a)^{0.6}+\frac{\Omega_{\Lambda}(a)}{70}
\left[1+\frac{1}{2}\Omega_m(a)\right]$ \citep{Lahavetal1991} for
models with a cosmological constant $\Lambda$; and better still the
previous formula, but with the power-index of the first term
$0.6\rightarrow 4/7$ \citep{Fry1985,Hamilton2001}. However, all these
approximations deviate at the few percent level when compared to the
exact result obtained from numerically solving the differential
equation for linear growth \citep[for further details see for
  example][]{Hamilton2001,LinderJenkins2003}. We therefore adopt the
exact numerical solutions for both $D(a)$ and $f(a)$ throughout
this study.


\subsection{Beyond linear theory: Nonlinear PT}\label{ssec:PT}

The collapse of cosmic structures can be followed into the nonlinear
regime using standard perturbation theory (PT) methods, applied to an
ideal fluid in a uniformly expanding spacetime \cite[for an
  excellent review see][]{Bernardeauetal2002}. The first application of
PT methods to estimate the impact of the nonlinear evolution of
$\pdot$ on the CMB, was given by \citet{Seljak1996a}. That work was
conducted within the context of the flat SCDM model, and hence only
provided an estimate for the Rees-Sciama contribution. Furthermore,
owing to the fact that $(1-f)=0$ at all times in the EdS model, it was
necessary only to calculate the PT up to 2nd order in $\delta$,
whereas in more general cosmologies, to be consistent at first order,
one requires the corrections up to 3rd order. We shall now calculate
the nonlinear ISW in the PT framework for the LCDM model.

To begin, we require from the PT theory the solutions for the fluid
overdensity, and in Fourier space these may be written as,
\be \delta(\bk,t) = \sum_{n=1}^{\infty}[D(t)]^n\delta_n(\bk,t_0)\ , \label{eq:PTsum}\ee
where the perturbative solutions at each order can be written 
\be \delta_n(\bk)=\int
\frac{\prod_{i=1}^{n}\left\{\dq_i\,\delta_1(\bq_i)\right\}}{(2\pi)^{3n-3}}
\left[\delta^D(\bk)\right]_n F^{(s)}_n(\bq_1,...,\bq_n) \label{eq:PTdelta} \ .
\ee
In the above expression $\delta_1(\bq_i)$ represents an initial field
at wavenumber $\bq_i$, and the $n$th order perturbed density depends
on $n$ initial fields. The quantities $F^{(s)}_n(\bq_1,...,\bq_n)$
represent the standard PT interaction kernels, symmetrized in all of
their arguments. Also we have adopted the short-hand notation
$\left[\delta^D(\bk)\right]_n=\delta^D(\bk-\bq_1-\dots-\bq_n)$. The
Dirac delta function ensures that the waves conserve momenta through
the interaction, i.e. $\bk=\bq_1+\dots+\bq_n$. For example, the second
order PT kernel can be written,
\ba 
F_2^{(s)}(\bq_1,\bq_2) & = & \frac{5}{7}+\frac{1}{2}\mu_{1,2}
\left[\frac{q_1}{q_2}+\frac{q_2}{q_1}\right]+\frac{2}{7} \mu_{1,2}^2\ , \ea
where $\mu_{1,2}\equiv\hat{\bq}_1\cdot\hat{\bq_2}$.

In the standard approach of nonlinear PT, the fluid equations are
solved for the flat EdS background model. In this case the spatial and
temporal parts of the evolution are fully separable and the
perturbative solutions at each order simply scale as powers of the
expansion factor $a(t)$ \citep{JainBertschinger1994}. However, this is
not the case for more general cosmological models, nevertheless a very
good approximation to the evolution can be obtained by replacing the
powers of $a(t)\rightarrow D(t)$. Strictly speaking, the PT
interaction kernels also inherit some time dependence, however this is
an extremely weak function of time and so to a very good approximation
we may use the kernels from the flat EdS case \citep[for deeper
  discussion of this see][]{Bernardeauetal2002}.

As \citet{Seljak1996a} showed, to calculate the ISW we simply require
the PT expansion for $\delta$ and its time derivative. Using
Eq.~(\ref{eq:PTsum}), this latter quantity may be written,
\be \dot{\delta}(\bk,t) = f(a)H(a) \sum_{n=1}^{\infty} n [D(t)]^n
\delta_n(\bk,t_0)\ \ .\ee
These quantities may now be used to compute the Next-to-Leading-Order
(NLO) corrections to the power spectra. Using the above expressions
plus the standard PT techniques we find, for pseudo-momentum: 
\ba 
P^{\rm NL}_{\omega\omega} & =  & P^{\rm Lin}_{\omega\omega} + P^{\rm 1Loop}_{\omega\omega} \ ; 
\label{eq:phiphiNLO}\\
P^{\rm NL}_{\delta\omega} & =  & P^{\rm Lin}_{\omega\delta} + P^{\rm 1Loop}_{\omega\delta} \ ;
\ea
where the one-loop corrections are,
\ba 
P_{\omega\omega}^{\rm 1Loop} & = & \left[af(a)H(a)\right]^2 
\left[4P^{22}_{\delta\delta}(k,a)+3P^{13}_{\delta\delta}(k,a)\right] \ , \\
P_{\delta\omega}^{\rm 1Loop} & = & \left[af(a)H(a)\right] 2 P^{\rm 1Loop}_{\delta\delta}(k,a) \ .
\ea
For the $\pdot$ we find:
\ba 
P^{\rm NL}_{\pdot\pdot} & =  & P^{\rm Lin}_{\pdot\pdot} + P^{\rm 1Loop}_{\pdot\pdot} ;
\label{eq:phiphiNLO}\\
P^{\rm NL}_{\delta\pdot} & =  & P^{\rm Lin}_{\delta\pdot} + P^{\rm 1Loop}_{\delta\pdot} ,
\ea
where the one-loop corrections are,
\ba 
P^{\rm 1Loop}_{\pdot\pdot}(k) & = & 
[\Fka]^2 \frac{H^2(a)}{a^2} \left\{\frac{}{}[1-2f(a)]^2P^{22}_{\d\d}(k) \right. \nonumber \\
& & \left. \frac{}{}+ [1-3f(a)][1-f(a)]P^{13}_{\d\d}(k)\right\}\ ; \label{eq:1LPP} \\
P^{\rm 1Loop}_{\delta\pdot}(k) & = & 
\Fka\frac{H(a)}{a} \left\{\frac{}{}1-2f(a)\right\}P_{\d\d}^{\rm 1Loop}(k,a)\ .
\label{eq:1LDP}\ea
In the above expression $P_{\delta\delta}^{22}$ and
$P_{\delta\delta}^{13}$ are the NLO corrections to the matter power
spectrum and we defined, $P^{\rm 1Loop}_{\delta\delta}(k,a)\equiv
P_{\delta\delta}^{13}(k,a)+P_{\delta\delta}^{22}(k,a)$ \citep[for
  explicit forms for the 1Loop expressions,
  see][]{JainBertschinger1994,ScoccimarroFrieman1996b,Smithetal2007}.

We may now learn how the NLO corrections entangle the pure linear ISW
decay of potentials with the nonlinear RS effects. The easiest way to
discern the changes is to consider the sign of the corrections in the
above equations. We notice that there are two ways the sign may
change: firstly, there is a sign flip with scale, since
$P_{\d\d}^{13}$ is negative and dominant on large scales and
$P_{\d\d}^{22}$ is positive and dominant on smaller scales; secondly,
the time dependent prefactors may change sign.


\begin{table}
\centering{
\caption{Sign of the NLO correction to the $P_{\delta\pdot}(k)$ power
  spectrum. Recall that a positive correction means an increase in the
  decay rate of the potentials. See text for further details.}
\label{tab:CrossSign}
\begin{tabular}{c|cc}
\hline Sign of correction & $|P_{13}(k)|>P_{22}(k)$ &
$|P_{13}(k)|<P_{22}(k)$ \\ \hline $\Omega_m>\Omega_m(a_{\rm RS})$ & $(+)$ &
$(-)$ \\ $\Omega_m < \Omega_m(a_{\rm RS})$ & $(-)$ & $(+)$ \\
\end{tabular}}
\end{table}


Considering the time dependent factors, we see that the cross-power
spectrum, Eq.(\ref{eq:1LDP}), will only change sign when, $1-2f(a)=0$,
which occurs when $\Omega\approx 0.3$. We shall label this time
$a_{\rm RS}$. Table~\ref{tab:CrossSign} summarizes the changes.  The
key point to notice is that at early times, there is an enhancement of
the ISW on very large scales (i.e. enhanced decay of gravitational
potentials) and on small scales there is a suppression of ISW effect
(growth of potentials). Then, at late times $z<z_{\rm RS}$ these
corrections invert themselves and ISW is suppressed on large scales
and small-scale potentials decay.

Turning now attention to the auto-power correction,
Eq.~(\ref{eq:1LPP}), we see that the prefactor multiplying
$P_{\delta\delta}^{22}$ is always positive, whereas the second term
only switches sign when $1-f<0$ or $1-3f>0$. However since
$\Omega_m<1$, the first bracket will never switch sign and will vanish
at early times. On the other hand the second bracket remains negative
until $\Omega_m\approx0.16$. Given current constraints on
$\Omega_{m0}\sim0.25$ the second bracketed term is always negative and
on multiplying by $P^{13}_{\d\d}$, we conclude that it too is always
positive.


\begin{figure}
\centering{
  \includegraphics[width=8cm,clip=]{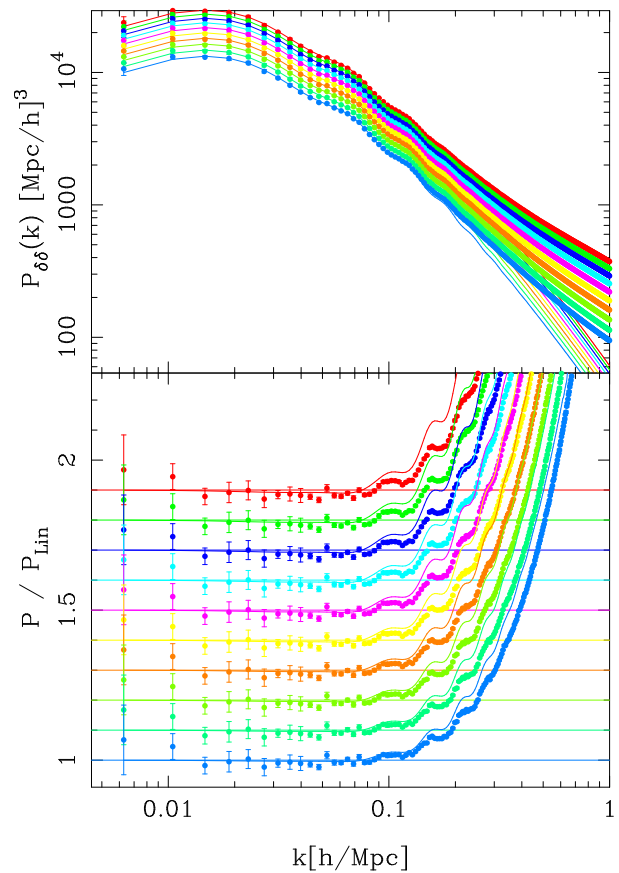}}
\caption{\small{The time evolution of the nonlinear CDM density power
    spectrum as a function of wavenumber. {\em Top panel}: colored
    points denote the absolute power and error bars are on the mean
    and are determined from the ensemble of simulations. The thin
    lines denote the linear theory and from top to bottom results are
    for expansion factors:
    $a=\{1.0,0.93,0.87,0.76,0.66,0.62,0.57,0.54\}$. {\em Bottom
      panel}: the ratio of the power spectra with respect to the
    linear theory prediction. The thick solid lines denote the
    predictions from the nonlinear Eulerian PT. Note that for clarity
    the measurements have been offset by 0.1 in the vertical
    direction.
    \label{fig:MM}}}
\end{figure}


\subsection{Evolution of density power spectrum}

Fig.~\ref{fig:MM} shows the evolution of the nonlinear matter power
spectrum measured from $z=1$ to the present day. The power is plotted
from the fundamental mode, $k=2\pi/L\approx 0.005\kMpc$ to half of the
Nyquist frequency of the mesh $k_{\rm NY}=\pi N_{\rm
  g}/L\approx2\kMpc$, where we use $N_{\rm g}=1024$ for all
transforms. Above this frequency the power in the Fourier modes is
affected by aliasing from smaller scales \citep{Pressetal1992}. In the
top panel we show the mean ensemble averaged absolute power from the
simulations at each epoch, colored points with errors. On the largest
scales $k<0.1$, the power grows by a factor of $\sim2$ from $z=1$ to
0, and there appears to be very good agreement with the linear theory
predictions on these scales (colored solid lines). On smaller scales
the power is significantly amplified.

In the bottom panel of the Fig.~\ref{fig:MM} we take the ratio of the
data with the linear theory, and to see clearly the effects for each
snapshot, we offset the curves by 0.1 in the vertical direction, with
the solid colored lines being the baseline for each corresponding
snapshot. We see that there is a small ($\approx2-3\%$) suppression of
power at late times for modes $0.05<k<0.1\,[\kMpc]$, this is termed
the `previrialization feature'
\citep[][]{Smithetal2007,CrocceScoccimarro2008,Anguloetal2008a}. On
smaller scales ($k>0.1\kMpc$) the power is strongly amplified,
compared to linear theory. In this panel we also present the
predictions from the standard PT (described in \S \ref{ssec:PT}) and
we see that it qualitatively captures the trends in the data. However,
in closer detail, we see that the PT over estimates the power on
smaller scales and that the predictions become progressively worse at
higher expansion factors and higher wavenumbers.


\begin{figure}
\centering{
  \includegraphics[width=8.cm,clip=]{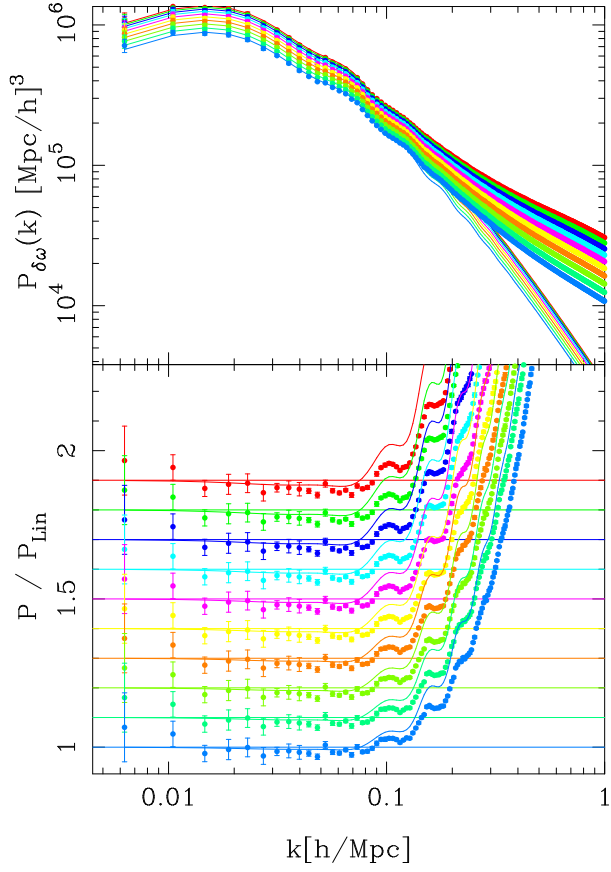}}
\caption{\small{The evolution of the pseudo-momentum--mass density
    cross-power spectra from $z=1$ to $0$, as a function of spatial
    wavenumber. Points and lines are as for Fig.~\ref{fig:MM}.
\label{fig:PsudoPM} }} 
\end{figure}


\begin{figure}
\centering{
  \includegraphics[width=8.cm,clip=]{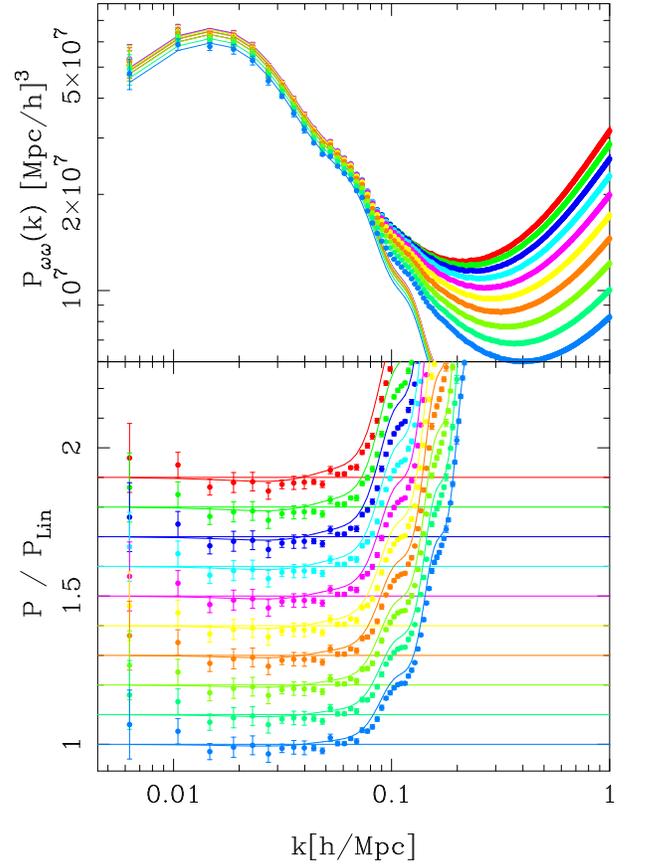}}
\caption{\small{The evolution of the pseudo-momentum auto-power
    spectra from $z=1$ to $0$, as a function of spatial
    wavenumber. Points and lines are as for Fig.~\ref{fig:MM}.
\label{fig:PsudoPsudoP} }} 
\end{figure}


\begin{figure}
\centering{
  \includegraphics[width=8cm,clip=]{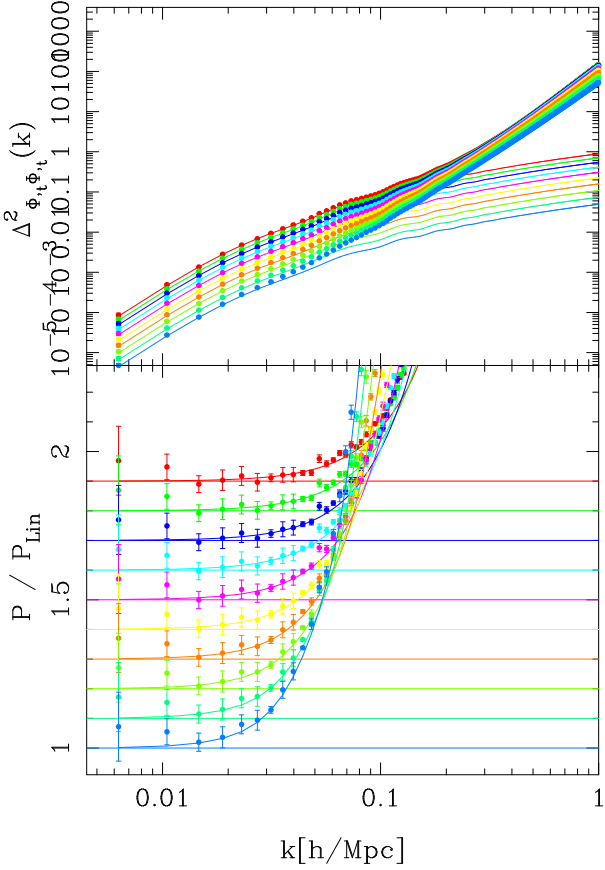}}
\caption{\small{Evolution of the $\pdot$ auto-power spectra from $z=1$
    to $0$ in dimensionless units, as a function of spatial
    wavenumber. Again points and lines are as presented in
    Fig.~\ref{fig:MM}.
\label{fig:PkPhiPhi}}} 
\end{figure}


\subsection{Evolution of the pseudo-momentum spectra}

In Figs~\ref{fig:PsudoPM} and \ref{fig:PsudoPsudoP} we present the
pseudo-momentum--density cross- and pseudo-momentum auto-spectra,
respectively.  Again the top part of each figure shows the absolute
power and the bottom the ratio with respect to the linear model
(c.f. Eqs~\ref{eq:PwwLin} and \ref{eq:PwdLin}). Note that the spectra
are amplified relative to the density spectrum, and that on large
scales this boost is well captured by multiplicative powers of
$af(a)H(a)$. In addition, we find that on very large scales, the
momentum spectra also display a previrialization feature and that the
suppression of power appears to be deeper in both cases. Furthermore,
on smaller scales the nonlinear amplification, which occurred at
around $k\sim0.1\kMpc$ for $P_{\delta\delta}$, appears at larger
scales in both cases, with the $P_{\omega\omega}$, strongly amplified
by $k>0.7\kMpc$. We compare these measurements with the predictions
from standard PT and find for $P_{\delta\omega}$ reasonably good
agreement on very large scales and an over prediction on smaller
scales. However, for $P_{\omega\omega}$ the agreement is much better.


\subsection{Evolution of the $\dot{\Phi}$ power spectra}

Having examined the individual components of the $P_{\pdot\pdot}$
spectrum we may now sum them together with weights as given by
Eq.~(\ref{eq:PowPP}).  Following \citet{Seljak1996a} and
\citet{Caietal2008}, we introduce the dimensionless and re-scaled form
of $P_{\pdot\pdot}$,
\ba 
\Delta^2_{\pdot\pdot}(k)\!\!\! & \equiv & \frac{4 \pi}{(2\pi)^3} \frac{k^3
  P_{\pdot\pdot}(k)}{[\Fka H(a)/a]^2}\ ;  \\
 & = & \frac{k^3}{2\pi^2}  \left[ P_{\d\d}(k)-\frac{2P_{\omega\delta}(k)}{H(a)a(t)} + 
\frac{P_{\omega\omega}(k)}{H^2(a)a^2(t)} \right] \label{eq:scaledP} \ .
\ea
Fig.~\ref{fig:PkPhiPhi} shows the evolution of the ensemble averaged
$\Delta^2_{\pdot\pdot}$, with errors on the mean. The top panel shows
the absolute spectra for the 10 snapshots from $z=1$ to the present
day. Also shown as the thin solid lines are the predictions from the
linear theory as given by Eq.~(\ref{eq:LinISW1}). Again, there appears
to be good agreement on large scales, and nonlinear amplification on
smaller scales. The bottom panel presents the ratio with respect to
the linear model, again we have offset different epochs by 0.1 in the
vertical direction for clarity. There is clear evidence for nonlinear
amplification of the spectrum on the very largest scales, and relative
to linear theory this becomes increasingly more important at higher
redshifts, as expected. Indeed by $k=0.03\kMpc$ and at $z\sim1.0$ the
power is more than 10\% in excess of the linear theory prediction,
whereas at $z=0$, a 10\% amplification is only achieved by
$k\sim0.07\kMpc$. Here we also show the predictions from the NLO PT
calculation from Eqs~(\ref{eq:phiphiNLO}) and (\ref{eq:1LPP}), and we
note a startlingly good agreement at all epochs.
 
That the nonlinear effects become increasingly important at higher
redshifts follows directly from the fact that $1-f(a)\rightarrow0$ as
$a\ll 1$. In this case, the only contribution to the spectrum comes
from the nonlinear Rees-Sciama effect, and in the limit $a\rightarrow
0$ it is given by
\be 
\Delta^{2}_{\pdot\pdot}(k ) \rightarrow
\frac{k^3}{2\pi^2}[1-2f(a)]^2P^{22}_{\d\d}(k)\ .
\ee

On comparing our results with Fig.~1 from \citet{Caietal2008}, we find
qualitatively good agreement. However, on the largest scales their
spectra do not appear to reproduce the linear theory at high
precision. The excess signal that they find compared to the linear
theory, we believe, is a result of using the approximation
$f\approx\Omega_m^{0.6}$. Some of the discrepancy may also be due to
cosmic variance, since they only show results for a single
simulation. In that work the authors also proposed a nonlinear
correction formula for $P_{\pdot\pdot}$, which has two free
parameters. Since the PT has no free parameters, and as it provides an
excellent match to the data for the scales $k<0.1$ we consider our
approach a sufficient description on these large scales. Such fitting
would most likely be necessary on smaller scales for good agreement,
but these scales are of diminishing importance for the calculation of
the CMB $C_l$ spectrum for $l<100$.


\section{Results: Impact on CMB spectrum}
\label{sec:results1}

The CMB temperature fluctuations arising from the ISW may be
decomposed using a spherical harmonic expansion, and the amplitude of
each harmonic can be written as
\citep{Cooray2002a,HernandezMonteagudo2008},
\be a_{lm}^{\rm T}=(-i)^{l} 4\pi \int \frac{\dk}{(2\pi)^3}
Y_{l,m}^{*}(\hat{\bk}) \Delta^{\rm T}_l(k,a_{ls})\ ,\ee
with, 
\be \Delta^{\rm T}_l(k)\equiv
\int_{0}^{\chi_{\rm max}} d\chi
\frac{2a}{c^3}j_{l}(k\chi)\pdot(k,\chi(t)) \ ,\ee
where we transformed Eq.~(\ref{eq:ISW}) to comoving geodesic distance
$\chi$ and $\chi_{\rm max}$ is the distance from the observer to the
surface at which the ISW first becomes significant.  The power in the
harmonic multipoles may be calculated using the standard methods, and
the ISW temperature spectrum may be written:
\ba C_{l}^{TT} & = & \frac{2}{\pi}\int dk k^2 \int_{0}^{\chi_{\rm max}}
d\chi_1d\chi_2 j_l(k\chi_1)j_l(k\chi_2) \nonumber \\
& & \times \ \frac{4a_1a_2}{c^6}P_{\pdot\pdot}(k;\chi_1,\chi_2) \ .
\label{eq:ClTT} 
\ea

In the limit $(l>10)$ we may use the Limber approximation to simplify
the above integrals \citep[see for
  example][]{Seljak1996a,BartelmannSchneider2001,Cooray2002b}. Assuming
that only modes transverse to the line of sight contribute to the
signal and also that the power spectra are slowly varying functions of
$k$, then the orthogonality of the spherical Bessel functions gives,
\ba 
& & \int dk k^2 j_{l}(k\chi_1)j_l(k\chi_2){\mathcal
  P}_{\alpha}(k,\chi_1,\chi_2) \nonumber \\
& & \hspace{0.5cm}\approx \hspace{0.5cm}
\frac{\pi}{2}\frac{\delta^D(\chi_1-\chi_2)}{\chi_1^2} {\mathcal
  P}_{\alpha}\!\left(\frac{l}{D_A(\chi_1)}\right)\ .\ea
where $D_A(a)$ is the comoving angular diameter distance
($D_A(a)=\chi(a)$ for flat space). On applying this approximation the
above expression reduces to the simple form:
\ba C_{l}^{TT}
& \approx & \int_{0}^{\chi_{\rm max}} d\chi\, \frac{4a^2}{c^6}
P_{\pdot\pdot}\!\left(k=\frac{l}{D_A(\chi)},\chi\right) \frac{1}{\chi^2} \ ; \\
& \approx & \frac{4}{c^5}\int_{a_{\chi(\rm max})}^{a_0} da\, 
P_{\pdot\pdot}\!\left(k=\frac{l}{D_A},a\right) \frac{1}{H(a)\chi^2}\ .
\ea
%


\begin{figure}
\centering{
  \includegraphics[width=7cm,clip=]{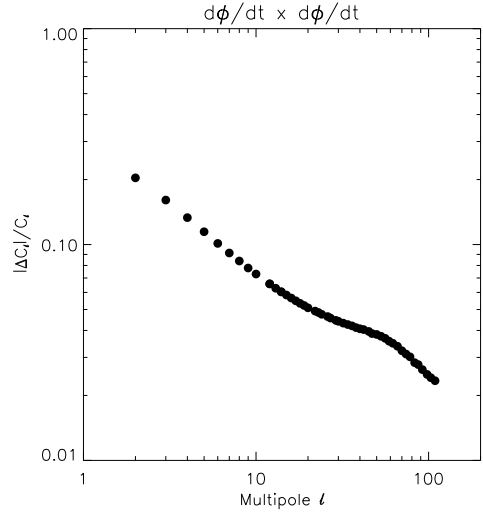}}
\caption{\small{\label{fig:LimberTestPhiPhi} Relative error $([C^{\rm
        Limber}_l-C^{\rm Exact}_l]/C^{\rm Exact}_l)$ between the
    Limber approximation and the exact $C_l$ computation of the CMB
    angular power spectrum. Results are shown for $k= (l+1/2)/D_A$
    replacement.}}
\end{figure}


\begin{figure}
\centering{
  \includegraphics[width=8cm,clip=]{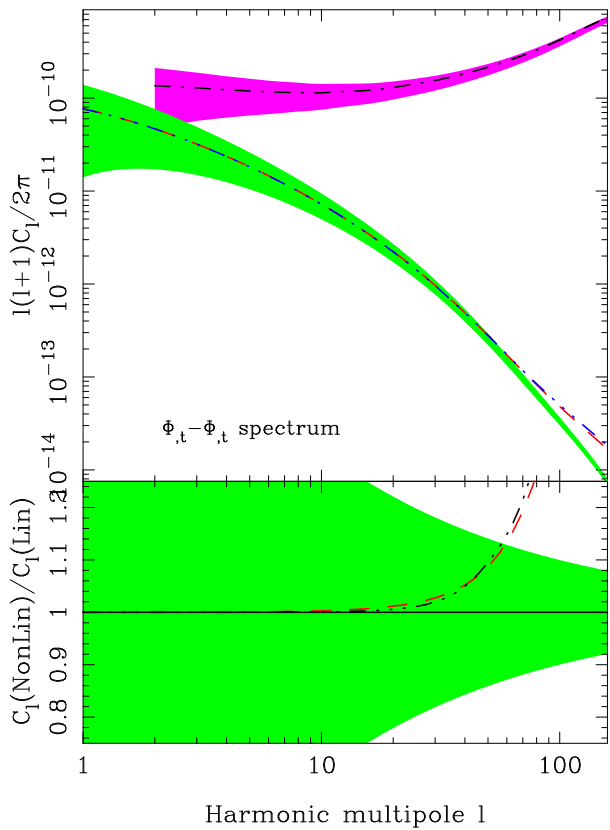}}
\caption{\small{{\em Top panel}: Angular power spectrum of CMB
    temperature fluctuations as a function of harmonic multipole. ISW
    contributions: solid green shaded region -- linear theory with
    encompassing 1-$\sigma$ Gaussian error domain; red dash line --
    nonlinear PT; ensemble average $N$-body measurement -- blue triple
    dot dash curve. The CMB primary anisotropy spectrum is given by
    the magenta dot-dash curve and its 1-$\sigma$ Gaussian error
    domain by the solid magenta shaded region. {\em Bottom panel}:
    ratio of the nonlinear ISW spectra to the linear theory
    spectrum. While this can exceed the 1-sigma error of ISW alone (as
    shown), it is always less than the 1-sigma error band of the
    overall CMB (not shown), hence the nonlinear effects are not
    detectable in the CMB power spectrum.
\label{fig:ClPhiPhi} }} 
\end{figure}


Fig.~\ref{fig:LimberTestPhiPhi} compares the Limber approximate
expressions for the angular power spectrum of temperature fluctuations
with the exact spherical harmonic line-of-sight integration. On scales
$l<10$, the Limber approximation is clearly poor with relative errors
being $>10\%$. The transformation $k= (l+1/2)/D_A$, as suggested by
\citet{Hoetal2008,LoverdeAfshordi2008}, improves the approximation,
but the errors still remain large. However for $l>10$, the error is
reduced and by $l=20$ it is of the order $\sim5\%$ (for a detailed
discussion on the validity of the Limber approximation for different
power spectra, see Appendix~\ref{sec:LimberApprox}).  We shall
nevertheless adopt the Limber approximation for our theoretical
analysis, but note that if significant effects are apparent on
multipoles $l<30$, then only a full spherical harmonic analysis will
give robust results. However, this would necessarily involve computing
the unequal time correlations of the Fourier modes of $\pdot(\bk,t)$.

Fig.~\ref{fig:ClPhiPhi} shows the results for the Limber approximated
ISW temperature angular auto-power spectrum. We scale the $C_l$
spectrum by $l(l+1)$ in the usual way and restrict our attention to
angular modes $l<100$. In the upper panel of the figure we compare
three predictions: the linear theory calculation with 1-$\sigma$
cosmic variance errors, denoted by the green shaded region; the
nonlinear PT, denoted by the red dash line; and the mean measurement
from the $N$-body simulations blue triple dot-dash curve.  The
1-$\sigma$ green shaded error region was computed using the Gaussian
variance formula:
\be \frac{\Delta C_l^{TT}}{C_l^{TT}} =\sqrt{\frac{1}{f_{\rm sky}}
  \frac{2}{2l+1}}\ ,\label{eq:CosmicVar} \ee
where $f_{\rm sky}$ is the fraction of sky covered, and we shall take
this to be of order unity. The estimate of the $C_l$ spectrum from the
$N$-body simulations was obtained by the following prescription: we
first made an array of the measured ${\mathcal P}_{\pdot\pdot}(k)$
spectra and divided this through by the linear theory ISW power
spectrum at that epoch.  On very large scales the ratios all
asymptotically approach unity and so the only evolution that remains
to be modeled is the higher $k$ domain.  To do this, we employ the
bi-cubic spline routine \citep{Pressetal1992} and interpolate the
spectra in $\log_{10}[a]$ and $\log_{10}[k]$. Note that on scales
greater than the fundamental mode of the simulation cube the bi-cubic
spline gives unity and we recover exactly linear theory. We emphasize
the importance of this step, since otherwise the $C^{TT}_l$ spectra
will be significantly reduced for $l<10$, owing to the finite volume
of the simulations. Note that in order to avoid extrapolating the
bicubic spline fits into regions where we have no measured data, the
upper redshift limit of the Limber integrals was set to $z=1$. We have
tested that this does not change our results in any significant way,
by computing the PT out to $z=5$.

In Fig.~\ref{fig:ClPhiPhi}, we see that all three theoretical
predictions converge for $l<30$, however for $l>30$ we find
enhancement of the signal for both the PT and $N$-body results and
that these agree to high precision, in agreement with expectations
from Fig.~\ref{fig:PkPhiPhi}. By $l=50$ they both show between
$\sim10-15\%$ increase in the power. We also show the CMB primary
anisotropy power spectrum as the black dot-dash line, with the magenta
shaded band giving the cosmic variance errors,
Eq.~(\ref{eq:CosmicVar}). The primary $C_l$ spectrum was obtained
using the {\tt cmbfast} routine with cosmological parameters to match
those of the {\tt zHORIZON} simulations. Note that, by default, this
spectrum already includes the linear ISW effect.

Comparing the primary with the ISW signal, we see that at $l=30$ the
primary signal is two orders of magnitude larger, and so the nonlinear
enhancement at these multipoles will induce changes to the CMB
spectrum that are $\ll\sim1\%$.  While the nonlinear effect exceeds 
cosmic variance in ISW for $l>50$, it never exceeds the 
cosmic variance from the total CMB, since ISW contribution to CMB decreases 
with $l$. 
Our findings are consistent with earlier results
\citep{Cooray2002b,Caietal2008}, but are established with improved
precision. We therefore do not expect large-scale nonlinear evolution
of the gravitational potentials to be responsible for any anomalies in
WMAP angular power spectrum. 


\section{ISW-dark matter cross-correlation spectrum}\label{sec:ISWdelta}


Having discussed the ISW auto-correlation spectrum we now move on to
discussing ISW correlation with the density field. We begin with the
dark matter density $C_l^{T\delta}$. This can be observed by
cross-correlating the CMB with the weak lensing signal of galaxies
\citep{SeljakZaldarriaga1999a,SeljakZaldarriaga1999b}, the weak
lensing of 21cm transitions \citep{ZahnZaldarriaga2006} or the weak
lensing of CMB itself (with information encoded in CMB bispectrum)
\citep{GoldbergSpergel1999,VerdeSpergel2002}.  In addition, there are
a number of advantages to be gained from studying this: firstly, there
exists an ``alternative'' method for estimating $P_{\pdot\delta}$, and
this provides us with an independent check on our ``standard method'',
described in \S\ref{sec:ISWtheory}; secondly, owing to the larger
number of dark matter particles, the effects of shot noise on the
spectra can be better assessed, and as we will show for the alternate
method, more easily corrected for.


\begin{figure*}
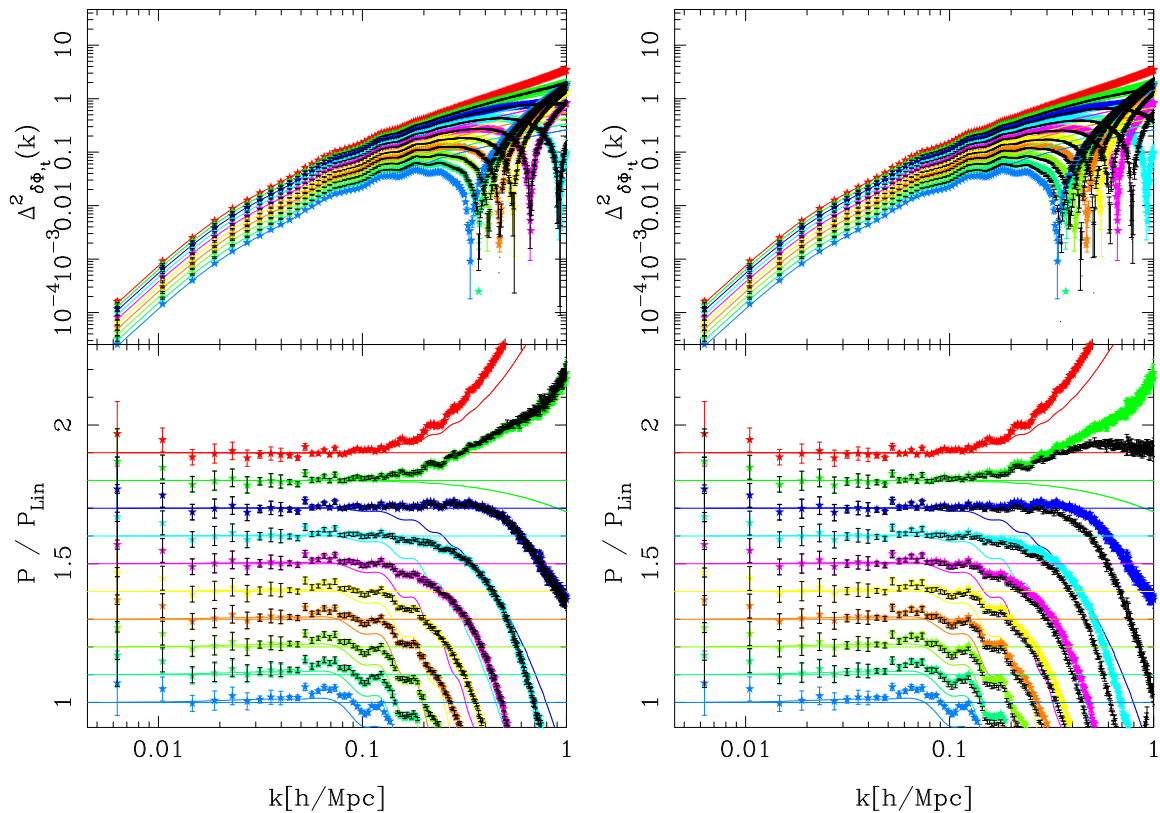

\centering{
\includegraphics[width=7.5cm,clip=]{FIGS/Pow.MASS-DPHI.UrosTest.LCDM750.ps}\hspace{0.2cm}
\includegraphics[width=7.5cm,clip=]{FIGS/Pow.MASS-DPHI.UrosTest.ShotCorr.LCDM750.ps}}
\caption{\small{Comparison of $P_{\pdot\delta}$ estimates obtained
    from standard method (continuity equation) and the alternate
    method (time derivative of $P_{\delta\delta}$,
    c.f. Eq.~\ref{eq:UrosMethod}). Colored points with errors denote
    the standard method, and black points with errors denote the
    alternate method.  {\em Left panel}: no shot noise
    correction. {\em Right panel}: shot noise correction applied to
    the alternative method. Top sections show absolute power, lower
    sections show the ratio with respect to the linear theory, and for
    clarity the results at each epoch have been offset from each other
    by 0.1 in the vertical direction. The lines in the bottom panels
    denote the predicitons form the standard PT and from top to bottom
    results are for expansion factors:
    $a=\{1.0,0.93,0.87,0.76,0.66,0.62,0.57,0.54\}$.}\label{fig:UrosTestMassPhi}}
\end{figure*}


\subsection{Alternative estimator for $P_{\pdot\delta}$}


Our alternative approach to estimating $P_{\pdot\delta}$ can be
understood as follows: Consider the ensemble average of the product of
$\delta(k)$ and $\pdot(k)$, using Poisson's equation we may rewrite
this as,
\ba 
P_{\delta\pdot}(\bk,a) & = & V_{\mu}\left<\delta(\bk,a)\pdot^*(\bk,a)\right> \nonumber\ , \\
& = & - a[{\mathcal F}(k)]^{-1} V_{\mu}\left<\Phi(\bk,t)\pdot^*(\bk,a)\right> \nonumber\ , \\
& = & - a[{\mathcal F}(k)]^{-1} P_{\pdot\Phi}(\bk,a) \ .
\ea
We now take advantage of the useful property that
\be P_{\Phi\pdot}(\bk,a) = 
\frac{1}{2}\frac{\partial}{\partial t} P_{\Phi\Phi}(\bk,a) =
\frac{1}{2}aH(a)\frac{\partial}{\partial a} P_{\Phi\Phi}(\bk,a)\ .\ee
Through further use of Poisson's equation, the last term in the above
equation may be rewritten in terms of the density power spectrum, i.e.
$P_{\Phi\Phi}(\bk,t)=\left[{\mathcal F}(k)/a\right]^2
P_{\delta\delta}(\bk,a)$. Putting this together, we arrive at the
result \citep{VerdeSpergel2002},
\ba 
P_{\pdot\delta}(\bk,a) & = & -\frac{1}{2}a^2H(a){\mathcal F}(k)
\frac{\partial}{\partial a} \left[\frac{P_{\delta\delta}(\bk,a)}{a^2}\right] \nonumber\ , \\
& = & 
-\frac{1}{2}aH(a)y(k,a){\mathcal F}(k)
\frac{\partial \ln y}{\partial \ln a} \ \label{eq:UrosMethod},
\ea
where 
$y\equiv\left[P_{\delta\delta}(\bk,a)/a^2\right]$.
This simple expression informs us that the ISW cross-correlation can
also be estimated from just two things: the matter power spectrum and
its evolution with time. We may check that the above expression is
consistent with our previous result (c.f. Eq.~\ref{eq:PowPD}).  On
assuming linear theory $P_{\delta\delta}(\bk,a)=D^2(a)P_{\rm Lin}(k)$,
then we find
\be \frac{\partial \ln y}{\partial \ln a}=2\left[f(a)-1\right] \ ,\ee
and on insertion of the above expression into
Eq.~(\ref{eq:UrosMethod}), we recover our earlier result.

The practical implementation of the above algorithm requires us to
estimate the time derivative of the power spectrum, and we do this
using the usual time-centered difference scheme:
\be \frac{\partial \ln y}{\partial \log a} \approx \frac{1}
{y_i}\frac{y_{i+1}+y_{i-1}}{\left[\log a_{i+1}-\log a_{i-1}\right]} \ ,\ee
where $y_i\equiv y(a_i)$ is the estimate at epoch $a_i$. Note that
since we employ a time-centered difference scheme, we do not show
results for for $z=0$ or $z=1$, the first and last epochs considered.


\subsection{Results: evolution of $P_{\pdot\delta}$}\label{ssec:UrosTest}

Fig.~\ref{fig:UrosTestMassPhi} compares the results for
$P_{\pdot\delta}$ obtained from our standard method
(c.f. Eq.\ref{eq:PowPD}) of solving the continuity equation (black
points with errors), with our alternative method (colored points with
errors). As was done for $P_{\pdot\pdot}$ we have introduced a
dimensionless and scaled form of the cross-power spectrum
(c.f. Eq.~\ref{eq:scaledP}):
\ba \Delta^2_{\delta\pdot}(k)\!\!\! & \equiv & \frac{4
\pi}{(2\pi)^3} \frac{k^3 P_{\delta\pdot}(k)}{[\Fka H(a)/a]}\ ; \\ 
& = & \frac{k^3}{2\pi^2} 
\left[P_{\delta\d}(k)-\frac{P_{\delta\omega}(k)}{H(a)a(t)}  \right] 
\ .  \ea
The left panel of Fig.~\ref{fig:UrosTestMassPhi} shows that the two
independent approaches produce results that agree to high
precision. We are therefore confident that both methods are consistent
and implemented correctly.

The top panel compares the spectra estimated from the simulations
(points with error bars) and the linear theory predictions (solid
lines). The lower panels show the ratio with respect to the linear
predictions. There are a number of important features that we draw
attention to: firstly, rather than nonlinear effects becoming
increasingly prominent with time, we see that they are stronger at 
earlier times and on larger scales. The explanation follows our
earlier discussion of Fig.~\ref{fig:PkPhiPhi}, and owes to the fact
that the linear ISW effect switches off as $a\rightarrow0$ and
$1-f(a)\rightarrow0$, leaving only the RS contributions
(c.f.~\S\ref{ssec:PT}).

Next, we note that there is a sign change in the spectra as one goes
from low to high $k$. Since we plot the absolute value of the power
the sign change is understood to be the point where the signal drops
to zero and bounces back up. The scale at which this sign change
occurs is a function of time, and it appears on larger scales at
higher redshifts \citep[see also][]{Caietal2008}.  The sign change is
due to the fact that the $\pdot$ signal becomes dominated by the RS
and BG effects. However, for the spectra with $z<0.3$ we see no sign
change over the $k$-range that we consider. Moreover, unlike the lower
redshift epochs we see an amplification relative to the linear
theory. This means that, at late times in LCDM model, nonlinear
evolution can actually {\em enhance} the decay of gravitational
potentials, consistent with our earlier discussion of the PT
(c.f.~\S\ref{ssec:PT}). Further support for the PT interpretation of
this phenomenon comes from the fact that if one considers the results
at high redshift, then around $k\sim0.1\kMpc$ there is a small
amplification of power with respect to the linear model.

In the discussion so far we have neglected the issue of the
discreteness correction due to finite number of dark matter particles.
It is unclear how to apply the shot noise correction to the momentum
density.  However, since we know the shot noise correction on the dark
matter power spectrum is $P_{\delta \delta}\rightarrow
P_{\delta\delta}-1/\nbar$, where $\nbar$ is the number density of dark
matter particles, the discreteness effects may be accounted for more
easily when using Eq.~(\ref{eq:UrosMethod}).
Fig.~\ref{fig:UrosTestMassPhi}, right panel, shows the results
obtained from this procedure.  Whilst we see that the correction
reduces the spectra by a small amount for all $k>0.2$, we nevertheless
see that both the small-scale late-time amplification and early-time
large-scale amplification of the $P_{\pdot\delta}$ remain
significant. We are therefore led to conclude that, nonlinear
evolution {\em may} lead to a small enhancement of the ISW in the LCDM
model.

Comparing our results with the measurements of $P_{\delta\pdot}$ from
\citet{Caietal2008}, we observe that these authors find no such late
time amplification. Owing to the fact that we have provided two
independent methods to obtain the estimates, and since we have a
significantly larger total simulation volume ($\sim12$ times larger)
furnishing smaller errors, we believe that our result is robust. In
the next section we shall investigate whether selecting highly biased
regions may influence these results further.


\section{ISW-halo cross-correlation spectrum}\label{sec:bias}

The cross-correlation between $\pdot$ and a density tracer field is
more easily observable than with the density field itself, which so
far is limited because of the small area or large errors in the weak
lensing reconstruction. One consequence of this is the added
complication of needing to understand the bias relation -- the mapping
from the tracer population to the underlying dark matter density.  In
this Section we shall explore whether the cross-correlation of $\pdot$
with cluster- and group-scale dark matter haloes, measured in the {\tt
  zHORIZON} simulations, between $z=[0.0,1.0]$, changes the ISW signal
in any significant way, beyond a linear bias. From the assumption that
all galaxies reside in dark matter haloes, it follows that the large
scale clustering properties of any galaxy sample are a weighted
average of the halo clustering statistics. Consequently, studying the
halo-ISW cross-correlations should provide representative results for
a number of plausible surveys. In particular, while we are limited by
the mass resolution of our simulations so that our analysis only
applies to biased halos with bias $b>2$, we note that most of the data
sets used for ISW detection so far are based on strongly biased
tracers \citet{Hoetal2008,Giannantonioetal2008}, so our results are
applicable to these.


\subsection{Linear Theory}

In nearly all ISW studies to date the bias has been assumed to be not
only constant in space, but also in time. As discussed in
\citet{Hoetal2008} and more recently \citet{Schaeferetal2009}, if one
wishes to go beyond detection and constrain cosmological models with
the ISW, then it is likely that this over simplification will
introduce a bias in the recovered parameters, especially when redshift
selection functions are broad. The next simplest scenario is a
time-dependent linear relationship:
\be \delta_{\alpha}(\bx,a)=b^{\rm \alpha}_1(a)\delta(\bx,a) \ , \ee
where $\delta_{\alpha}\rightarrow \left\{{\rm g},{\rm h},{\rm
c},\dots\right\}$ denotes the tracer type, e.g. galaxies, haloes,
clusters etc., $b^{\alpha}_1(a)$ is a linear bias parameter that
varies in time but is independent of scale. In this case the ISW
cross-spectra and biased tracer auto-spectra may be easily computed as
(cf. Eq.~\ref{eq:LinISW2}):
\ba 
P^{\rm Lin}_{\alpha\pdot}(k)  \!\!\!  & = & \!\! b_1^{\alpha}(a)\Fka 
\left[\frac{H(a)}{a}\left(1-f(a)\right)\right]
P^{\rm Lin}_{\delta\delta}(k;t)  \ ;\\ 
P^{\rm Lin}_{\alpha\alpha}(k) \!\!\!  & = & \!\! [b_1^{\alpha}(a)]^2 P^{\rm Lin}_{\delta\delta}(k;t) \ .
\ea
%


\subsection{Nonlinear theory for the bias}\label{ssec:bias}

Several recent theoretical and numerical studies of the bias of dark
matter haloes \citep{Coleetal2006,Smithetal2007,Anguloetal2008a}, have
revealed that the linear model is only likely correct on
asymptotically large scales. These predictions have been confirmed by
several observational studies of the relative bias of different galaxy
populations
\citep{Percivaletal2007a,SanchezCole2008,CresswellPercival2008}. In
\citet{Smithetal2007} it was shown that the scale dependence of halo
bias was a strong function of scale for $k>0.07 \kMpc$. In that work a
physically motivated analytic framework was developed to model these
scale changes. A similar approach was independently developed by
\cite{McDonald2006a,McDonald2006b}. The model utilizes a nonlinear
local bias model \citep{FryGaztanaga1993,Coles1993}:
\be \delta_{\alpha}(\bx,a)=\sum_{n=1}^{\infty} 
\frac{b_n^{\alpha}(a)}{n!}  \left[\delta^n(\bx,a)-\left<\delta^n(\bx)\right>\right] \
, \ee
where the constant term from the Taylor expansion was rewritten as
$b^{\alpha}_0=-\sum_{j=1}b_j^{\alpha}(a)\left<\delta^j\right>/j!$. The density
field may be expanded using the PT series expansions from
\S\ref{ssec:PT}. As was shown by \citet{Smithetal2007}, if one
transforms to Fourier space and collects terms to a fixed order, then
the density field of the biased tracers may be written as a
fluctuation series of the form:
\ba 
\delta^{\alpha}(\bk,a|R) & = &  \sum_{n=1}^{\infty} [D(a)]^{n}
\left[\delta^{\alpha}(\bk,a|R)\right]_n\ 
\label{eq:Halo-PT1a}\ ; \\
\left[\delta^{\alpha}(\bk|a,R)\right]_n   & = & \int
\frac{\prod_{i=1}^{n}\left\{d^3\!q_i\,\delta_1(\bq_i)\right\}}{(2\pi)^{3n-3}}
\left[\delta^D(\bk)\right]_n \nonumber \\
& & \times F^{\alpha}_n(\bq_1,...,\bq_n|a,R)\ ,
\label{eq:Halo-PT1b}\ea
where $[\delta^{\alpha}(\bk|a,R)]_n$ is the $n$th order perturbation
to the biased tracer density field.  The functions
$F^{\alpha}_n(\bq_1,...,\bq_n|a,R)$ are the bias tracer PT kernels,
symmetrized in all of their arguments. The kernels are described in
\citet{Smithetal2007}. Thus equations (\ref{eq:Halo-PT1a}) and
(\ref{eq:Halo-PT1b}) can be used to describe the mildly non-linear
evolution of the biased fields to arbitrary order in the dark matter
perturbation. There is a subtlety that we have skipped over: in order
to facilitate the Taylor expansion of the biased field it was
necessary to filter on a scale $R$, and hence all of the kernels
depend on the filter scale. To remove the filter dependence we adopt
the renormalization scheme suggested by
\citet{McDonald2006a,McDonald2006b}. The down side of this, is that
the parameters may not be derived {\em ab initio}, but must be
obtained through fitting to measured data and we shall do this in the
following section.

Using these relations, along with McDonald's renormalizations, we
find that the ISW--biased density tracer cross- and auto-power spectra
may be written:
\ba 
P^{\rm NL}_{\alpha\pdot}(k,a)  & = & P^{\rm Lin}_{\alpha\pdot}(k,a)+P^{\rm 1Loop}_{\alpha\pdot}(k,a) \ ; \\
P^{\rm NL}_{\alpha\alpha}(k,a) & = & P^{\rm
Lin}_{\alpha\alpha}(k,a)+P^{\rm 1Loop}_{\alpha\alpha}(k,a)\ , \label{eq:autospectra}
\ea
where the loop corrections are given by,
\ba 
P^{\rm 1Loop}_{\alpha\pdot} & = & {\mathcal
F}(k)\frac{H(a)}{a}\left[1-2f(a)\right]P^{\rm
1Loop}_{R,\alpha\delta}(k,a) \ ;\\
P^{\rm 1Loop}_{\alpha\delta} & = & b^{\alpha}_{R,1}
P^{\rm 1Loop}_{\d\d}+b_{R,2}^{\alpha} A(k,a) \ ; \\
P^{\rm 1Loop}_{\alpha\alpha} & = & 2 b^{\alpha}_{R,1}
 b^{\alpha}_{R,2} A(k,a) + \frac{[b^{\alpha}_{R,2}]^2}{2}B(k,a)
 +N^{\alpha}_{R}(a)\ \label{eq:HaloHaloRPT} ; \ea
and where we have introduced the auxiliary functions:
\ba A(k,a) & \equiv & \int \frac{\dq}{(2\pi)^3}\,P_{\Lin}(q)
P_{\Lin}(|\bk-\bq|)F_2(\bq,\bk-\bq) \
\label{eq:Ak} ; \\
B(k,a) & \equiv & \int \frac{\dq}{(2\pi)^3}  \,P_{\Lin}(q) \left[P_{\Lin}(|\bk-\bq|)-P_{\Lin}(q)\right] 
\label{eq:Bk} \ .
\ea
In the above equations we introduced the renormalized bias parameters
$b^{\alpha}_{R,i}(a)$ and the renormalized constant power term
$N^{\alpha}_{R}(a)$. This may be thought of as an arbitrary white
noise contribution. 

Before moving on, we notice that the sign reversal property of the
nonlinear cross-power spectrum of $\pdot$ with mass density, remains
unchanged. This owes to the fact that $b_{R,2}$ changes the scale at
which the loop corrections transit from large-scale power suppression
to small-scale enhancement (provided loop corrections are small
compared to linear theory).

%
\begin{table}
\centering{
\caption{Halo mass classes, and number densities.}
\label{tab:bins}
\begin{tabular}{l|ccc}
       &  Mass Range              &  $\nbar(z=0)$            &   $\nbar(z=1)$  \\
       &  $ [\times10^{13}\Msol]$  &  $[\Mpc]^3$ & $[\Mpc]^3$ \\
\hline
Bin 1 & $[1.50<M<10.0]$  & $3.5\times10^{-4}$ & $1.8\times10^{-4}$ \\
Bin 2 & $[10.0<M]$ & $2.5\times10^{-5}$ & $3.3\times10^{-6}$
\end{tabular}}
\end{table}
%

\subsection{Renormalized halo bias parameters}\label{ssec:rebias}

In order to use the nonlinear bias model we require the time evolution
of the bias parameters, $b_1$ and $b_2$.  These can be estimated
directly from the simulations in the following way. Firstly, we
divided the haloes at each expansion factor into two classes: (Bin 1)
group scale dark matter haloes and (Bin 2) cluster scale dark matter
haloes (see Table~\ref{tab:bins} for details).

These mass bins can be faithfully traced within our simulations out to
$z=1$.
We choose fixed mass bins at all epochs for simplicity,
but a reasonable association can be made between these
halo bins and tracer populations such as Luminous Red Galaxies (LRGs) or clusters. Then for each 
realization we compute the power spectra: $P_{\h\h}$, $P_{\h\d}$,
$P_{\h\omega}$, and $P_{\h\pdot}$, for all of the snapshots from
$z=[0,1]$.  The renormalized halo bias parameters were then directly
estimated from the $P_{\h\d}$ data in the following fashion. Firstly,
we fit for $b^{\h}_{R,1}$ on the largest scales using an inverse
variance estimator of the form:
\ba
\hat{b}^{\h}_{R,1}  & = & \frac{\sum_{i\in [k_1,k_2]} \hat{b}^{\h}_1(k_i)/\sigma^2_{b_1}(k_i)}
{\sum_{i\in [k_1,k_2]} 1/\sigma^2_{b_1}(k_i)}\  ; \label{eq:estimate}  \\ 
\sigma^2_{b_1} & = &   {\sum_{i\in [k_1,k_2]} 1/\sigma^2_{b_1}(k_i)}\ ,
\ea
where $\hat{b}^{\h}_1(k_i)=\hat{P}_{\h\d,i}/\hat{P}_{\d\d,i}$ and with
$[k_1,k_2]=[0.0,0.05] \kMpc$.  Note that we assume that there is
little covariances between $k$ bins on these large scales. Having
obtained $b^{\h}_{R,1}$, we next obtain our estimate for
$b^{\h}_{R,2}$.  Our estimator has exactly the same form as the above
equations except for the fact that $b^{\h}_{R,1}\rightarrow
b^{\h}_{R,2}$ and $\sigma_{b_1}\rightarrow\sigma_{b_2}$ and that
$[k_1,k_2]=[0.05,0.2] \kMpc$. The important quantity to specify is,
\be
\hat{b}^{\h}_2(k_i)=\frac{1}{A(k_i)}
\left[\hat{P}_{\h\d,i}-\hat{b}^{\h}_{R,1}P_{\d\d}(k_i)\right]\ .
\ee

It will also be useful later on for us to predict $P_{\h\h}$, and to
do this we are required to additionally estimate the renormalized shot
noise term: $N^{\h}_{R}(a)$. This may be obtained directly from our
estimates of $\hat{P}_{\h\h,i}$ along with Eq.~(\ref{eq:autospectra})
and by using Eq.~(\ref{eq:estimate}), but with
$b^{\h}_{R,1}\rightarrow N^{\h}_{R}$ and
$\sigma_{b_1}\rightarrow\sigma_{N^{\h}}$ and with $[k_1,k_2]=[0.0,0.2]
\kMpc$. Our estimate per mode is
\ba
\hat{N}^{\h}_{R} & = &
\hat{P}_{\h\h,i}- [\hat{b}^{\h}_{R,1}]^2 P_{\d\d}(k_i)-[\hat{b}^{\h}_{R,2}]^2B(k_i)\nonumber \\
& & \hspace{0.2cm}-2\hat{b}^{\h}_{R,1}\hat{b}^{\h}_{R,2}A(k_i)\ .
\ea

\begin{figure}
\centering{ \includegraphics[width=7cm,clip=]{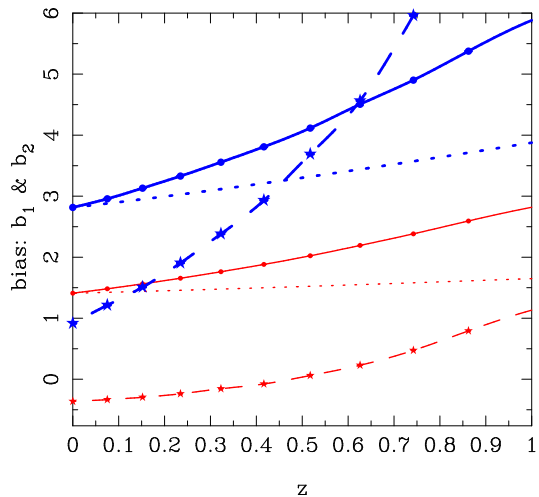}}
\caption{\small{Renormalized bias parameters $b_{R,1}$ and $b_{R,2}$
    as a function of redshift. Red (thin) and blue (thick) lines and
    points denote results from Bin 1 and Bin 2, respectively. The
    symbols are: $b_{R,1}$ estimates -- solid points; $b_{R,2}$
    estimates -- stars. The dotted lines show the bias evolution model
    of Eq.~(\ref{eq:biasevolve}). }\label{fig:bias}}
\end{figure}

Fig.~\ref{fig:bias} shows the time evolution of the best fit
renormalized halo bias parameters. As is evident from the figure, the
values of $b^{\h}_{R,1}$ for the two samples decrease with increasing
time. This is qualitatively consistent with the halo bias evolution
that emerges from Extended Press-Schechter formalism and the
Peak-background split argument (dotted lines), where linear halo bias
decays with time as
\citep{Fry1996,TegmarkPeebles1998,ShethTormen1999}:
\be
\left[b_1(a)-1\right]=D(a_0)/D(a)\left[b_1(a_0)-1\right]\ \label{eq:biasevolve}
. \ee
However as the figure clearly shows the actual measured halo bias
evolves much more strongly as a function of redshift. We also note
that the values of $b^{\h}_{R,2}$ are also similarly consistent with
this theory, which predicts that $b^{\h}_{R,2}<0$ for haloes around
$M_{*}(t)$ (the characteristic nonlinear halo mass at that epoch
$\sigma(M_{*},t)=1$), and that $b^{\h}_{R,2}>0$ for haloes with
$M>M_{*}(t)$ \citep{Scoccimarroetal2001}. 


\begin{figure*}
\centering{
  \includegraphics[width=7.cm,clip=]{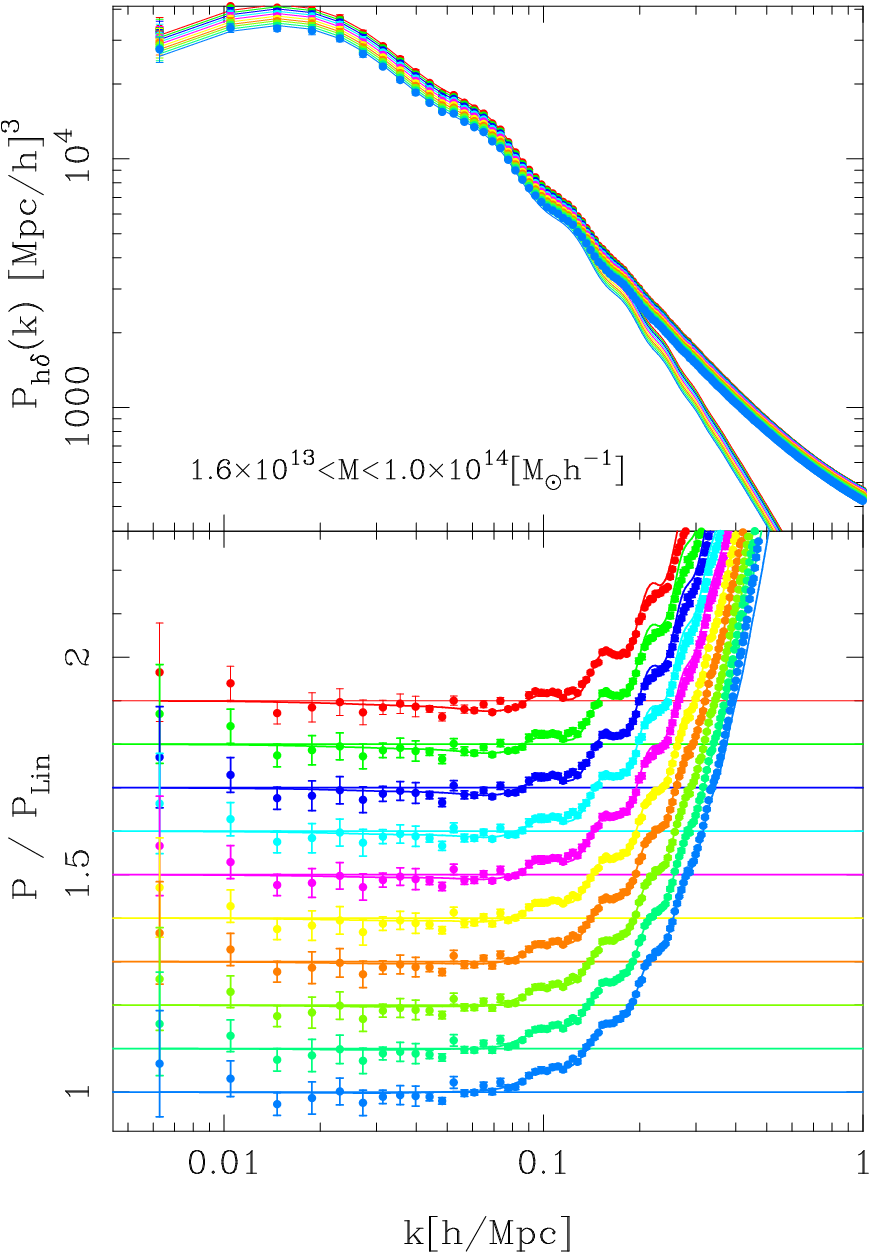}\hspace{0.2cm}
  \includegraphics[width=7.cm,clip=]{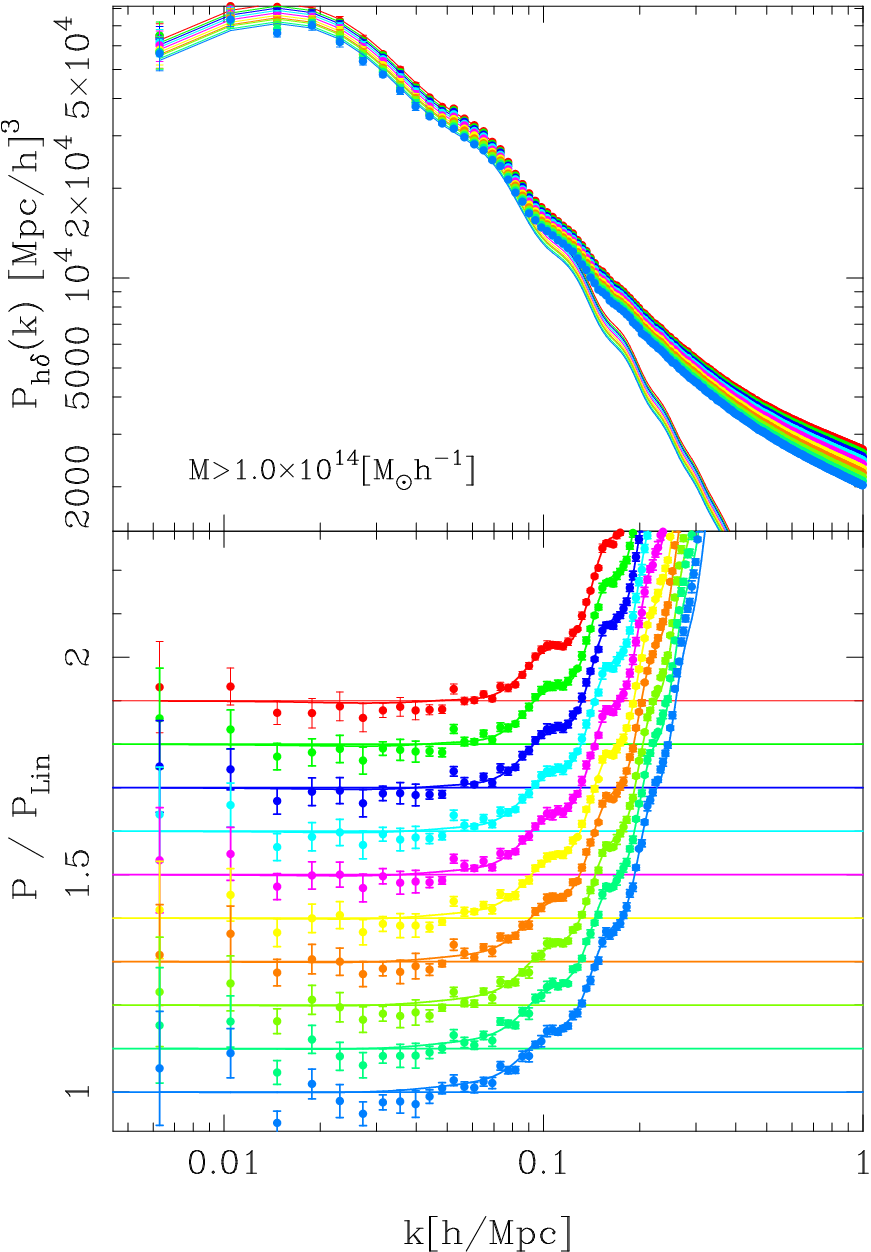}}
\caption{\small{Evolution of the ${\h}$--$\delta$ cross-power
spectra as a function of spatial wavenumber from $z=1$ to 0. Left
panel is for haloes in Bin 1 and the right panel shows results for
haloes in Bin 2. Points and lines are as presented in
Fig.~\ref{fig:MM}.\label{fig:HaloMass}}}
\centering{
\includegraphics[width=7.cm,clip=]{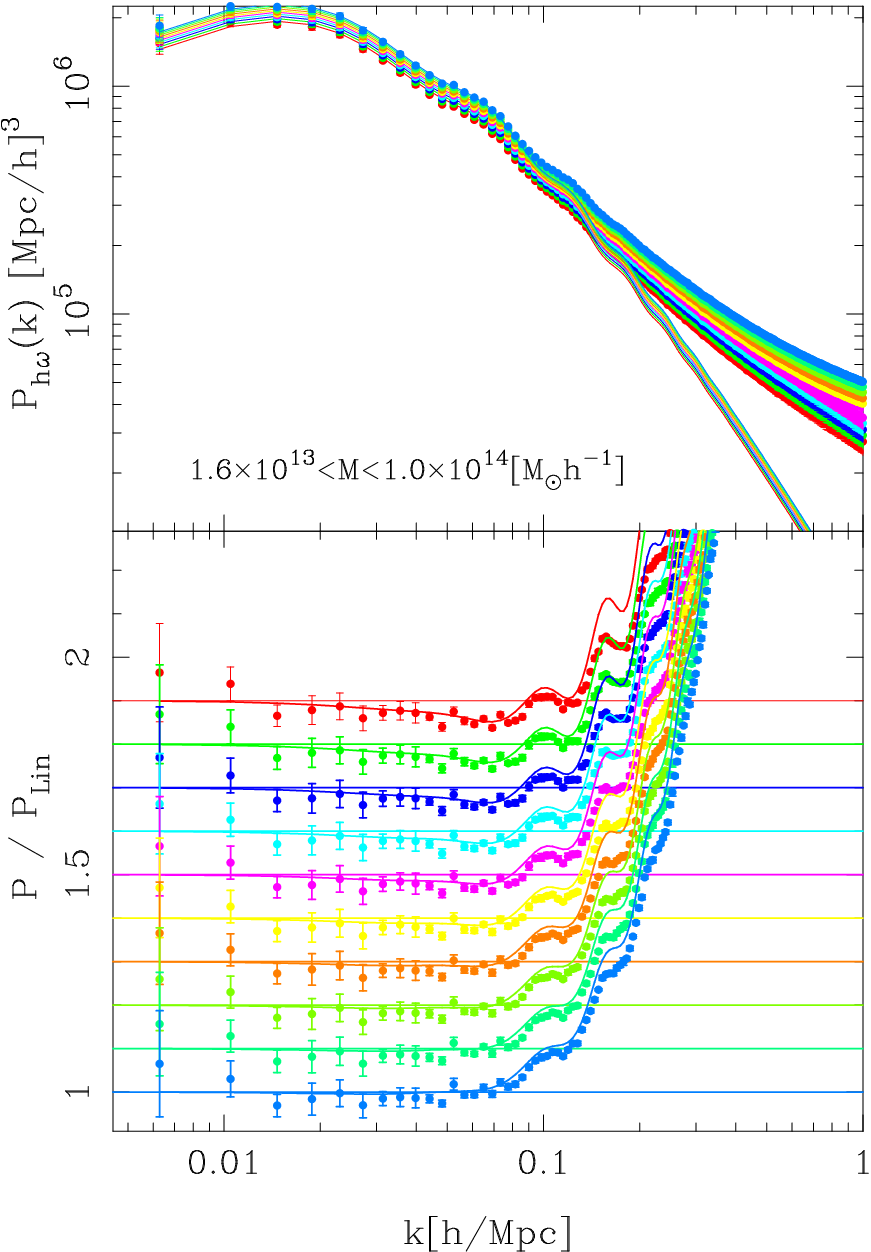}\hspace{0.2cm}
\includegraphics[width=7.cm,clip=]{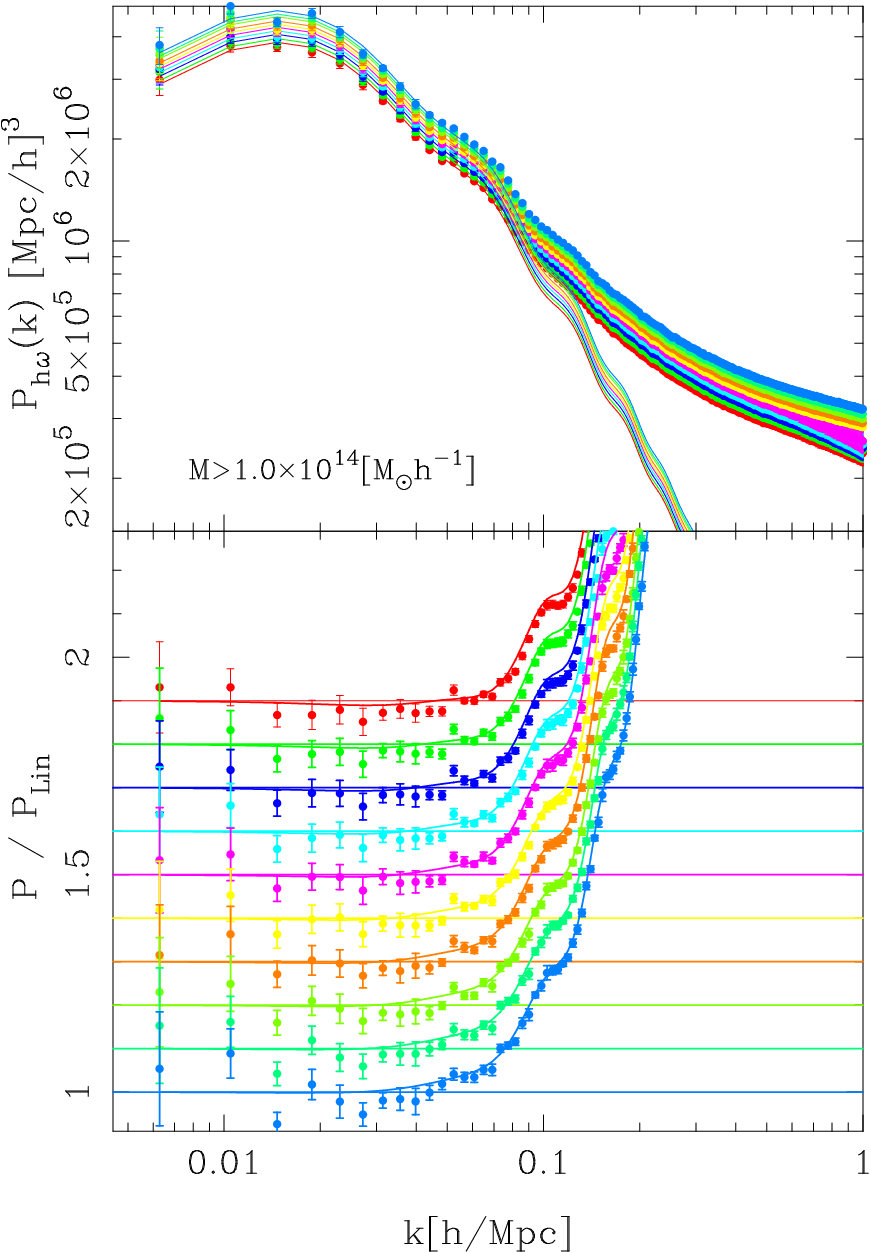}}
\caption{\small{Evolution of the ${\h}$-$\omega$ cross-power
spectra as a function of spatial wavenumber from $z=1$ to 0. Left
panel is for haloes in Bin 1 and the right panel shows results for
haloes in Bin 2. Points and lines are as presented in
Fig.~\ref{fig:MM}\label{fig:HaloSudoP}}}
\end{figure*}


\subsection{Results: Evolution of halo--density spectra}

Fig.~\ref{fig:HaloMass} shows the evolution of $P_{\h\delta}$ in the
simulations. The left panel presents the results for haloes in Bin 1
and the right Bin 2. The top sections show the absolute power and the
lower sections show the ratio with respect to the linear theory
predictions. We observe that all spectra exhibit a strong scale
dependence relative to the linear theory and that the departure is
characterized by a suppression of power on large scales
($k>0.05\kMpc$) followed by power amplification on smaller scales
($k>0.1\kMpc$), and this is exhibited in both mass bins and at all
times. The highest mass bin exhibits the strongest amplification with
scale, by $k\sim0.1 \kMpc$ the spectra are 10\% in excess of the
linear theory, whereas Bin 1 shows a slightly stronger large-scale
power suppression. In the lower sections of each panel in
Fig.~\ref{fig:HaloMass} we also show the predictions of the nonlinear
renormalized bias model from \S~\ref{ssec:bias} and we find
surprisingly good agreement over all of the scales of interest. We
note that for Bin 1 on the smallest scales $k>0.5\kMpc$, the
predictions appear to drop dramatically to zero. However, for the
computation of the $C_l$ we expect that this theoretical accuracy will
be sufficient. This owes to the fact that the spectra shown in
Fig.~\ref{fig:HaloMass} will all be premultiplied by ${\mathcal
  F}(k)\propto k^{-2}$ and so will be suppressed relative to larger
scales. We note that the scale dependence of the halo cross-power
spectra were investigated by \citet{Smithetal2007} and we confirm the
basic results presented in that study.
   

\subsection{Results: Evolution of halo--momentum spectra}

Fig.~\ref{fig:HaloSudoP} shows the evolution of the
halo--pseudo-momentum cross-power spectra as a function of
scale. Again left and right panels are for Bins 1 and 2,
respectively. As expected from our investigation of
$P_{\delta\omega}$, we again see nonlinear features in these spectra,
and that they are more enhanced relative to those in the
$P_{\h\delta}$ spectra. This can be inferred through considering the
ratios of the spectra with respect to linear theory (bottom section of
each panel). In particular, we note that for Bin 1 and the $z=0$
snapshot, the large-scale suppression feature is of the order
$\sim5\%$ at $k\sim0.7\kMpc$, in contrast to $\sim2\%$ supression in
$P_{\h\delta}$. Again, in the lower panels we over plot the
predictions from the renormalized bias model and the agreement is
again good, although for $k>0.2\kMpc$ small deviations of the model
from the data are more apparent. Also, the predictions for Bin 1 drop
to zero at higher $k$, and this occurs for the reasons previously
noted.


\begin{figure*}
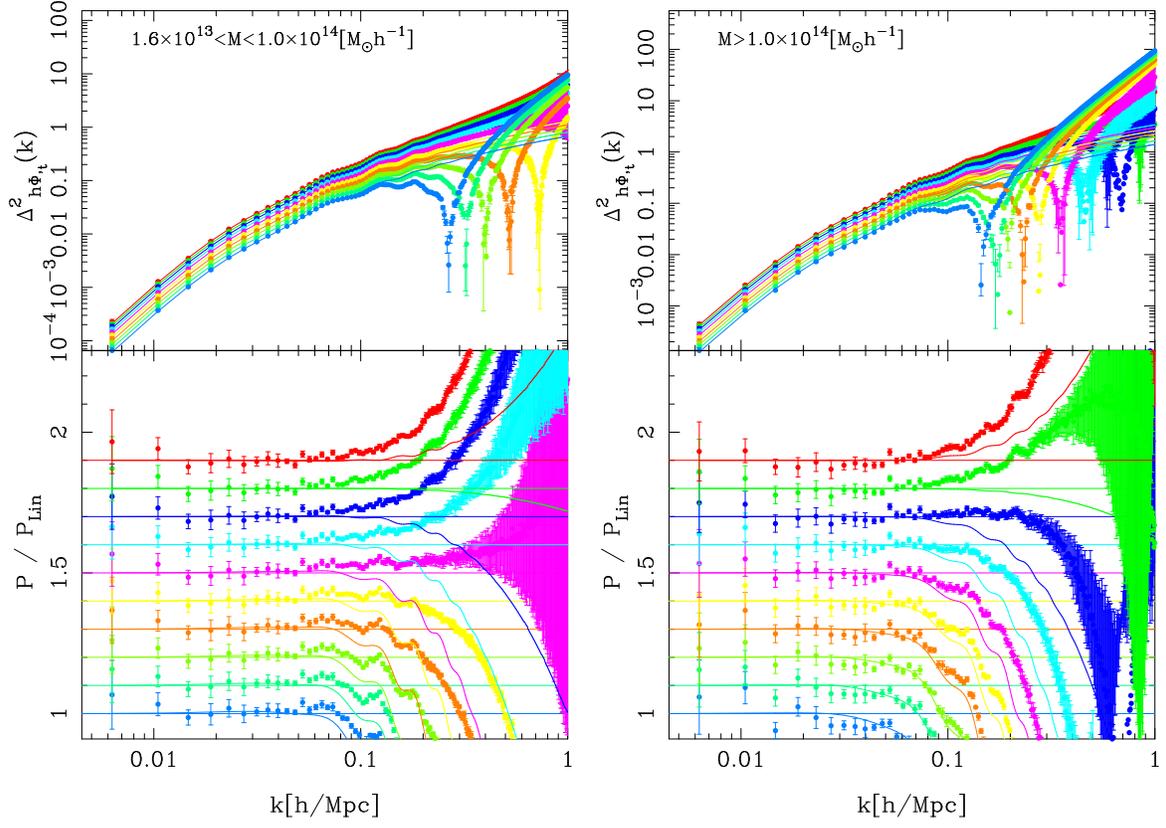

\centering{
\includegraphics[width=7.5cm,clip=]{FIGS/Pow.HALO-DPHI.Bar.LCDM750_HaloBin_2.ps}\hspace{0.2cm}
\includegraphics[width=7.5cm,clip=]{FIGS/Pow.HALO-DPHI.Bar.LCDM750_HaloBin_3.ps}}
\caption{\small{ Evolution of the dimensionless scaled ${\h}$-$\pdot$
    cross-power spectra ($\Delta^2_{\h\pdot}$) as a function of
    wavenumber from $z=1$ to 0. Left and right panels show results for
    haloes in Bin 1 and Bin 2, respectively. Points and lines are as
    presented in Fig.~\ref{fig:MM}.\label{fig:HaloDPhi}} }
\end{figure*}


\subsection{Results: Evolution of halo--$\pdot$ spectra}\label{ssec:halopdot}

In Fig.~\ref{fig:HaloDPhi} we combine the power spectra from the
previous two subsections to explore the evolution of $P_{\h\pdot}$. As
was done for the analysis of $P_{\pdot\pdot}$ and $P_{\delta\pdot}$ we
introduce a dimensionless and scaled form of the biased cross-power
spectrum (c.f. Eq.~\ref{eq:scaledP}):
\be \Delta^2_{\h\pdot}(k) =  \frac{k^3}{2\pi^2} 
\left[P_{\h\d}(k)-\frac{P_{\h\omega}(k)}{H(a)a(t)}  \right] 
\ .  \ee
The top panels compare the spectra estimated from the simulations
(points with error bars) and the linear theory predictions (solid
lines). The lower panels show the ratio with respect to the linear
predictions, and the lines show the predictions from the renormalized
nonlinear bias model. As was the case for our investigation of
$\Delta^2_{\pdot\delta}$ (c.f.~\ref{ssec:UrosTest}), departures from
linear theory are increasingly apparent as one considers higher
redshifts.  In addition, there is a sign change in the spectra as one
goes from low to high $k$. The explanation again follows our earlier
discussions surrounding Figs~\ref{fig:PkPhiPhi} and
\ref{fig:UrosTestMassPhi}. On comparing these results for the haloes
with those for the dark matter, Fig.~\ref{fig:UrosTestMassPhi}, we
find that the scale at which the spectra switch sign becomes larger
with increasing bias.

Considering the small-scale, late-time ISW boost relative to linear
theory, we see that for the haloes at $z=0$ the signal is stronger as
bias increases. However, we also note that the amplification is
present for the Bin 1 halo sample by $z\sim0.3$, compared to the Bin 2
sample where it is absent by $z>0.1$.  This result means that, at late
times in LCDM model, nonlinear evolution can {\em enhance} the decay
of gravitational potentials and that the rate of decay also depends on
the environment. Again, this result naturally emerges from the PT
(c.f.~\S\ref{ssec:PT}), although as is shown in the figure, the PT
struggles to capture the measured spectra precisely. In the next
section we shall investigate whether these nonlinear effects are
sufficiently large to impact the ISW-density tracer $C_l$'s.


\section{CMB-LSS angular power spectrum}
\label{sec:results2}


\subsection{Theory}

We now turn to the calculation of the ISW--biased density tracer
angular power spectrum. As described in \S\ref{sec:results1} for the
ISW auto-spectrum, we may also decompose the projected fluctuations in
our biased density tracer into spherical harmonics.  To do this, we
define the 2D biased density field as the weighted projection of the
3D density field along the line of sight and in a cone of solid angle
$d\Omega$. This we may write as,
\be \delta_{\alpha}^{\rm 2D}(\vec{\theta}) = \int_{\chi_i}^{\chi_j}
d\chi D_{A}^2(a) q(\chi) \delta^{\rm
3D}_{\alpha}(D_A(\chi)\vec{\theta},\chi)\ ,\ee
where $q(\chi)$ is a radial weight function, which is normalized such
that
\be \int_{\chi_i}^{\chi_j} d\chi 4\pi D^2_A(\chi) q(\chi)=1 \ .\ee
To proceed we must specify $q(\chi)$.  For a typical magnitude limited
survey the weight function would be $q(\chi)=n(>L_{\chi})/N_{\rm TOT}$
where $n(>L_{\chi},\chi)$ is the space density of galaxies above the
flux limit at a given redshift, and $N_{\rm TOT}$ is the total number
of galaxies, and so $D^2_A(\chi) q(\chi)\propto dN(z)/dz$ the number
redshift distribution. Therefore, in turn, one is required to specify
a model for the redshift distribution \citep[see for
example][]{Afshordietal2004,Padmanabhanetal2005c}. 

\begin{figure}
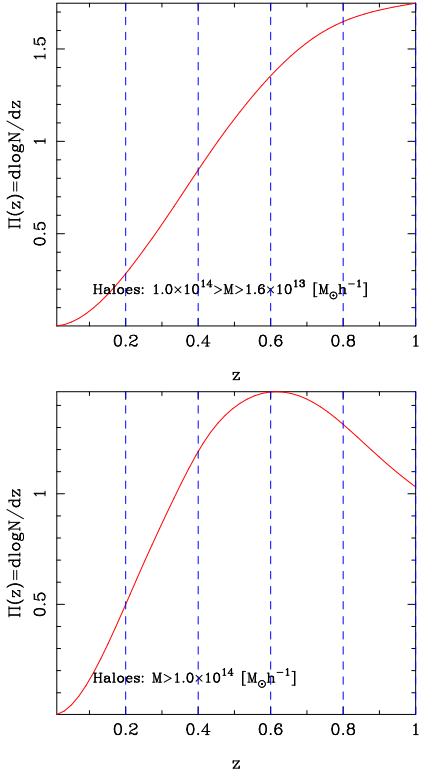

\centering{
  \includegraphics[width=5.5cm,clip=]{FIGS/SelectionFun.2.ps}}\vspace{0.1cm}
\centering{
  \includegraphics[width=5.5cm,clip=]{FIGS/SelectionFun.3.ps}}
\caption{\small{Mock LRG/cluster normalized number redshift
distributions as a function of redshift. Top panel shows results for
intermediate mass host haloes (Bin 1), and lower panel results for
cluster mass host haloes (Bin 2). Note that here we show the
normalized distributions over the entire sample range $z=0$--$1$. The
blue vertical dash lines show the 5 redshift bands for which we
compute the cross-correlations, and note that we renormalize the
distribution for each band.}\label{fig:selection}}
\end{figure}

Since we are more interested in {\em precisely} quantifying the
importance of nonlinear contributions to the cross-correlation signal
for biased tracers, which we can measure directly at all epochs in the
simulations, we shall therefore forgo attempting to fabricate certain
aspects of a real galaxy survey -- this level of detail may confuse
interpretation.  Instead we shall take a more simplified approach: we
assume that, above some fixed mass threshold, there is one and only
one galaxy (perhaps an LRG) per dark matter halo; that the mass
threshold is independent of redshift; and that we may construct a
volume limited sample of these objects from $z=0$ out to $z=1$. This
last condition implies that there is a tight relation between the mass
threshold of the host halo and the luminosity threshold for the
carefully selected target galaxy. Our model galaxy survey is therefore
equivalent to a target sample of haloes above some fixed mass from
redshift $z=0$ to $1$. Hence, we shall write the weight function,
$q(\chi)=n(>M,\chi)/N_{\rm TOT}(\chi_i,\chi_j)$, where $n(>M,\chi)$ is
the cumulative number density of dark matter haloes with masses
greater than $M$ at time $t(\chi)$; and where by our normalization
condition, for a redshift shell between $z_i$ and $z_j$ we have
\be N_{\rm TOT}(\chi_i,\chi_j)=\int_{\chi_i}^{\chi_j}
 d\chi 4\pi D^2_{A}(\chi) \int_{M}^{\infty} dM n(M;\chi) \ .\ee
In the above, $n(M,\chi)$, is the differential halo mass function at
time $t(\chi)$ and $\chi_i\equiv\chi(z_i)$. Figure~\ref{fig:selection}
shows the redshift distributions of our mock target samples, in the
two mass bins and as a function of redshift. In the figure we have
introduced the new weight function
\be \Pi_{ij}(\chi)\equiv 4\pi D^2_A(\chi)\Theta_{ij}(\chi)
\int_M^{\infty}dM \frac{n(M,\chi)}{N_{\rm TOT}(\chi_i,\chi_j)} \ ,
\ee
where
$\Theta_{ij}(\chi)\equiv\left[\Theta(\chi-\chi_i)-\Theta(\chi-\chi_j)\right]$,
is the top-hat function with $\Theta$ being the Heaviside step
function.


\begin{figure*}
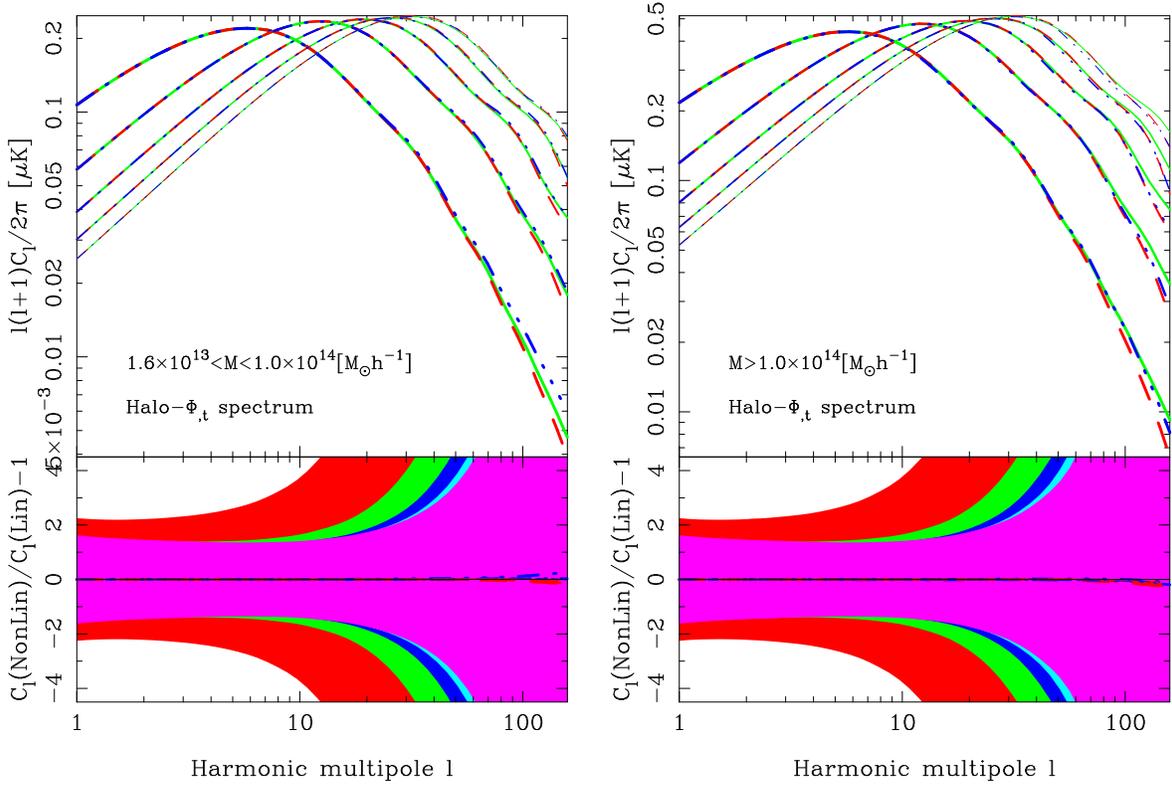

\centering{
\includegraphics[width=7.5cm,clip=]{FIGS/ClHPhi_Sim_HaloBin_2.ps_NEW}\hspace{0.4cm}
\includegraphics[width=7.5cm,clip=]{FIGS/ClHPhi_Sim_HaloBin_3.ps_NEW}}
\caption{\small{Angular cross-power spectrum of ISW effect and haloes
    as a function of spherical harmonic multipole $l$. {\em Left
      panel}: results for group scale dark matter haloes. {\em Right
      panel}: results for the most massive clusters. In each panel we
    show results for 5 equally spaced bins in redshift over the range:
    $z=[0.0,1.0]$. The predictions are differentiated by line
    thickness: thick lines -- low redshift; thin lines -- high. The
    line styles denote: linear theory -- solid green line; nonlinear
    PT -- red dash line; bi-cubic spline fit to the simulation data --
    blue triple-dot dash line. Top sections of each panel give the
    absolute power, and the lower sections the ratio with respect to
    linear theory. The shaded regions represents the 1-$\sigma$ error
    domains per multipole of the linear cross-spectra, where the
    central redshifts $z\in\{0.1, 0.3, 0.5, 0.7, 0.9\}$, correspond to
    the colours (red, green, blue, cyan, magenta).
\label{fig:ClHPhi}}}
\end{figure*}


The multipole amplitudes of the biased density tracers may therefore
be written,
\be a_{lm}^{\h}=(-i)^{l} 4\pi \int \frac{\dk}{(2\pi)^3}
Y_{l,m}^{*}(\hat{\bk}) \Delta^{\h}_l(k,\chi_{i},\chi_{j})\ ,\ee
with, 
\be \Delta^{\h}_l(k,\chi_i,\chi_j)\equiv \int_{\chi_i}^{\chi_j} d\chi
\Pi_{ij}(\chi) j_{l}(k\chi) \delta_{\h}^{\rm 3D}(k,\chi) \ .\ee
Following Eq.~(\ref{eq:ClTT}), the cross-angular-power of the ISW
temperature fluctuations and the projected density tracers may then be
written:
\ba C_{l}^{\h T} & = & \frac{2}{\pi}\int dk k^2 \int_{0}^{\chi_{\rm max}}
d\chi_1d\chi_2 j_l(k\chi_1)j_l(k\chi_2) \nonumber \\ 
& & \times \
\frac{2a_1}{c^3}\Pi_{ij}(\chi)P_{\h\pdot}(k;\chi_1,\chi_2) \ .  \ea
Under the Limber approximation (c.f. \S\ref{sec:results1}) this
expression reduces to:
\ba 
C_{l}^{\h T} & \approx & \int_{0}^{\chi_{\rm max}} d\chi\,
\frac{2a}{c^3}\Pi_{ij}(\chi)
P_{\h\pdot}\!\left(k=\frac{l}{D_A(\chi)},\chi\right) \frac{1}{\chi^2}\nonumber\ ; \\ 
& \approx & \frac{2}{c^2}\int_{a(\chi_{\rm max})}^{a_0} d\ln a\,
\Pi_{ij}(a) P_{\h\pdot}\!\left(k=\frac{l}{D_A(a)},a\right) \nonumber \\
& & \times \hspace{0.2cm} \frac{1}{H(a)\chi^2(a)}\ ; 
\ea
and for the halo auto-power spectrum we have
\ba 
C_{l}^{\h\h} & \approx & 
\int_{a(\chi_{\rm max})}^{a_0} d\ln a\,
\Pi^2_{ij}(a) P_{\h\h}\!\left(k=\frac{l}{D_A(a)},a\right) \nonumber \\
& & \times \hspace{0.2cm} \frac{1}{H(a)\chi^2(a)}\ . 
\ea

In Appendix~\ref{sec:LimberApprox} we present a short investigation of
the validity of the Limber approximation for predicting the ISW-LSS
cross-power spectrum. There we show that the relative error is $<10\%$
for $l\sim10$ and that for $l>10$ it is $<2\%$, and for a wide range
of survey window functions. These results are consistent with the
findings of \citet{Rassatetal2007} for the 2MASS survey. Since we are
interested in scales $l>10$, we shall therefore use the Limber
approximated expressions.


\begin{figure}
\centering{
\includegraphics[width=9cm,clip=]{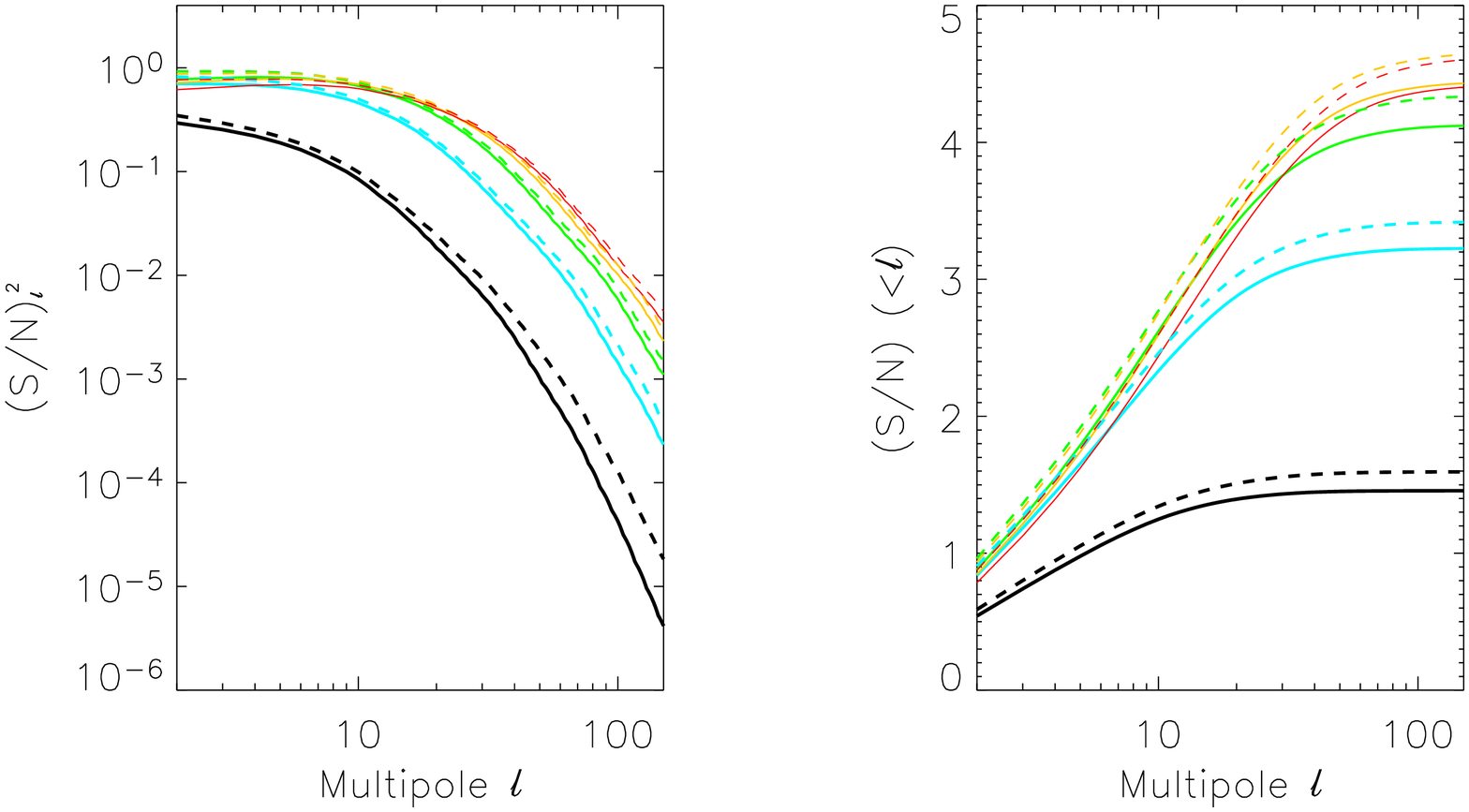}}
\centering{
\includegraphics[width=9cm,clip=]{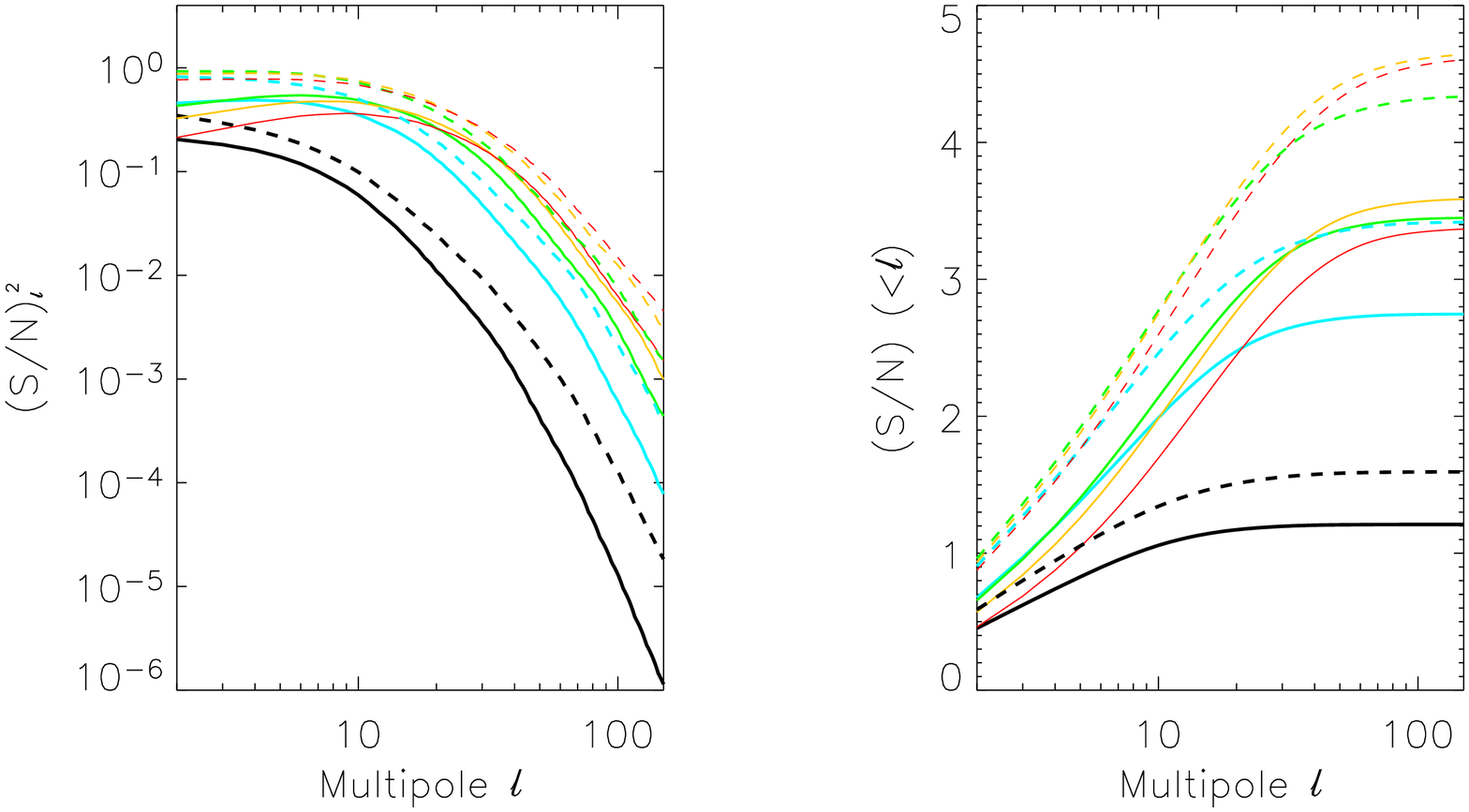}}
\caption{Top and bottom panels, $\SN$ results for Bins 1 and 2,
  respectively.  Left panels: Squared S/N for each multipole $l$ of
  the cross correlation of the CMB temperature with the most massive
  (Bin 2) halo population (solid lines) and the total matter density
  field (dashed lines). Results for different snapshots centered at
  $z=$ 0.1, 0.3, 0.5, 0.7 and 0.9 are displayed in black, blue, green,
  red and orange colors, respectively. Line thickness decreases with
  increasing redshift. Right panels: cumulative S/N. Lines are as in
  left panel.  \small{\label{fig:s2n}}}
\end{figure} 


\subsection{Results: ISW--biased tracer angular spectrum}

Figure~\ref{fig:ClHPhi} presents the results for the angular cross
power spectrum for the ISW and haloes in Bin 1 (left panel) and haloes
in Bin 2 (right panel). In each case we show the results for 5 narrow
bins in redshift, and where for each bin we weight by the redshift
distributions presented in Fig.~\ref{fig:selection}. The solid green
lines in the figure denote the linear bias predictions; the red dashed
lines correspond to our predictions from the nonlinear renormalized
PT, as described in \S\ref{ssec:bias} and \S\ref{ssec:rebias}; and the
blue triple-dot dash lines correspond to our bi-cubic spline fit to the
ensemble average measurements of $P_{\h\pdot}$ from the simulations,
and scaled by linear theory.

In the figure we see that for both Bins 1 and 2 the peak of the
angular power spectrum moves to the right and upwards as the mean
redshift of the sample increases. The rightward shift is due to the
fact that for a given physical scale the angular size decreases with
distance, in this case the scale is the peak of the $P_{\h\pdot}$
spectrum. The upward shift is more complex, if we were considering
unbiased tracers then we would expect that the signal would drop with
increasing redshift, owing to the fact that the ISW signal switches
off and also the amplitude of the power spectrum is decreasing with
$D^2(z)$. However, for a fixed mass range, the bias of the sample
increases with increasing redshift (c.f. Fig.~\ref{fig:bias}). For the
two host halo mass bins that we consider the bias evolves by a factor
of $\sim2$ from $z=0$--1.

Regarding the impact of nonlinearity on the predictions, we find that
for $l<100$ these are small, being at most $<10\%$. For Bin 1, the
deviations are characterized by a several percent boost around $l=50$,
followed by a several percent suppression by $l=100$. Whereas for Bin
2, the deviations are represented as a few percent suppression. For
$l>100$ the deviations are, in all but one case, characterized by a
much more significant suppression, and the signal rapidly drops to
zero. The case which does not conform to this picture is the lowest
redshift slice for Bin 1, here the signal estimated from the
simulations appears to be boosted by $\sim10\%$ at $l\sim 100$.
Unfortunately, this amplification is not mirrored in the predictions
from the PT, as also seen in Fig~(\ref{fig:HaloDPhi}) for the last
four spectra.

In Fig.~\ref{fig:ClHPhi} we also show the expected 1--$\sigma$ error
domains (shaded regions) of the cross-spectra, computed from using the
simple variance formula:
\be 
\Delta^2\!\!\left(C_l^{T\h}\right)=\frac{1}{f_{\rm sky}}
\frac{1}{(2l+1)}\left[C_{l}^{\h\h}C_{l}^{TT}+\left(C_{l}^{T\h}\right)^2\right] \ .
\ee
As in the case for $C_l^{TT}$, we again find that the cosmic and
sample variance errors dominate over the modeling errors on scales
$l<100$.


\begin{figure*}
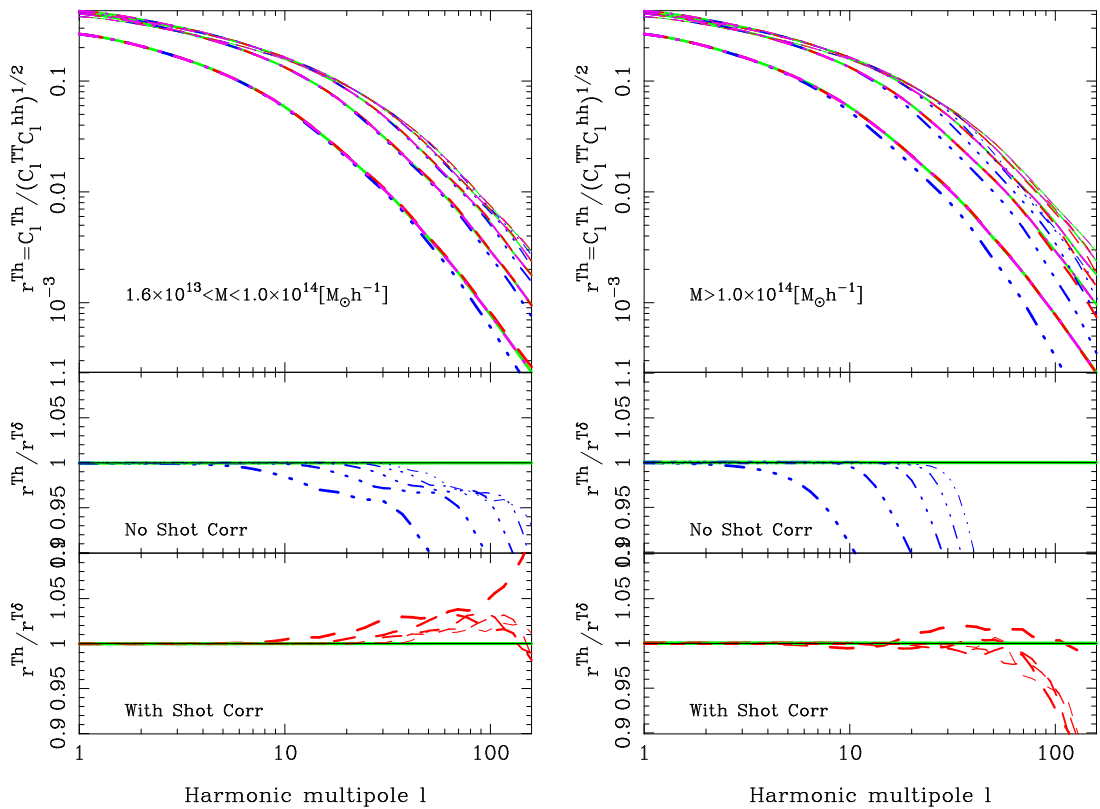

\centering{
\includegraphics[width=7.cm,clip=]{FIGS/Cl_Cross_Corr_NoShotCorr_Coef_HaloBin_2.ps_NEW}\hspace{0.4cm}
\includegraphics[width=7.cm,clip=]{FIGS/Cl_Cross_Corr_NoShotCorr_Coef_HaloBin_3.ps_NEW}}
\caption{\small{Cross-correlation coefficient between CMB and haloes
  ($r^{T\h}$) as a function of harmonic multipole $l$. Left panel
  shows results for group-scale haloes and the right for
  cluster-scale. {\em Upper panels}: results for 5 redshift bins
  centred on $z=\{0.1,0.3,0.5,0.7,0.9\}$ of thickness $\Delta
  z=0.2$. Thick to thin lines denote low to high redshift halo
  samples. Line styles are: green solid line -- linear theory; blue
  triple-dot dash line -- bi-cubic spline fit to simulation data
  without shot-noise subtraction; red dash line the same but with
  shot-noise subtraction. The magenta dot-dash curve shows the result
  for the dark matter $(r^{T\delta})$. Middle and lower panels show
  the ratio of halo to dark matter cross-correlation coefficients
  without and with shot noise subtraction, respectively.
  \label{fig:CrossCorr}}}
\end{figure*}


\subsection{Calculation of the $\SN$ for biased tracers}

The result from \citet{HernandezMonteagudo2008} is that for the
ISW--dark matter cross-correlation up to $90\%$ of the Signal-to-Noise
($\SN$) for the ISW comes from harmonic modes $l<50$. Here we shall
assess whether sampling biased density tracers can change these
conclusions. From the last equation we write the $\SN$ for the
ISW--dark matter cross-correlation, at a given multipole $l$, as
\be
\left( \SN\right)^2_l =  f_{\rm sky}(2l+1)\left[
\frac{ \left(C_l^{T\delta}\right)^2\; }{C_l^{TT}  C_l^{\delta\delta} 
+\left( C_l^{T\delta}\right)^2}\right] \ .
\label{eq:s2n1}
\ee
Similarly, this equation can be written for the halo distribution,
\be \left(\SN\right)^2_l = f_{\rm sky}(2l+1)
\left[ \frac{\left(C_l^{T\h}\right)^2} 
{C_l^{TT} C_l^{\h\h}  + \left(C_l^{T\h}\right)^2} \right]\ .
\label{eq:s2n2}
\ee
In the above, no shot noise subtraction on the halo
auto-power spectrum is assumed. 
We can define
the cumulative $\SN$ below a given multipole $l$ as
\be \left(\SN\right) [<l] = \sqrt{\sum_{l'= 2}^l (\SN)^2_{l'}}.
\label{eq:s2nc1}
\ee 
This addition is legitimate only under full sky coverage ($f_{\rm sky}
= 1$), since we assume that different multipoles are independent. In
the left panel of Fig.~\ref{fig:s2n} we show $(\SN)_l^2$, for the
cluster-mass halo population (Bin 2, solid lines) and the matter
density field (dashed lines) for the 5 different redshift shells. The
right panel of the figure shows the corresponding cumulative $\SN$
below each multipole $l$. We note that the $(\SN)^2_l$ is flat for low
multipoles, and declines rapidly with increasing $l$. The scale at
which the turn down occurs is a function of redshift. For $z=0.7$ the
turn down occurs at $l<20$, whereas for $z=0.1$ it has dropped by
$l\sim10$. From studying the cumulative $\SN$, we find that roughly
50\% of the total $\SN$ is achieved by $l\sim10$, and that $\sim 90$\%
is achieved by $l\approx40$
\citep[c.f.][]{HernandezMonteagudo2008}. On comparing these results
with the corresponding ones for the matter field (dashed lines), we
find slightly lower values for the haloes. This may be atributed to
the additional Poisson noise. Note that the redshift shell that gives
the highest $\SN$ is located at $z=0.7$, and that the total $\SN$ for
it is of order $\sim7$.

Based on these findings, we conclude that it is highly unlikely that
nonlinear evolution of the mass distribution or nonlinearities in the
scale dependence of bias can significantly affect the detectability of
the ISW.


\subsection{Results: Cross-correlation coefficient}

Finally, we investigate the cross-correlation coefficient of the CMB
temperature fluctuations and the halo samples. The cross-correlation
coefficient of two fields $A$ and $B$ is defined as,
\be r^{AB}_l \equiv \frac{C^{A B}_{l}}{\sqrt{C^{AA}_{l}C^{BB}_{l}}}
\ . \ee
Under the assumption of time independent linear bias we would have
$r^{T\h}\rightarrow r^{T\delta}$. Thus $r^{T\h}_l$ does not depend on
the bias of the tracer sample, nor the amplitude of the primordial
power spectrum. Instead it provides direct information on the Dark
Energy parameters and the curvature density:
$\{\Omega_{DE},w_0,\Omega_k,\}$. This approach was developed by
\citet{Giannantonioetal2008} to obtain cosmological parameter
constraints from current CMB and LSS data \citep[see also][for an
alternate method for removing bias, that uses CMB
lensing.]{Zhang2006}.

The validity of this analysis hinges on the fact that $b\sigma_8^2$
cancels out. However, since the bias is in fact time-dependent, we can
only have $r^{T\h}\approx r^{T\delta}$. Adding to this the fact that
the bias is scale-dependent it appears that such an approximation is
unlikely to be robust, and especially for LSS surveys with broad
selection functions. We may test their conjecture by estimating
$r^{T\h}$ for several samples of biased tracers, and if we do not find
that they match $r^{T\delta}$ within the same redshift shell, then the
modeling should be deemed to be insecure.  In that case one must
include the redshift evolution of the bias, as done by
\citet{Hoetal2008} in their analysis of the NVSS sample.

In Fig.~\ref{fig:CrossCorr} we present the measured cross-correlation
coefficients for the Bin 1 (left panel) and Bin 2 (right panel) halo
samples and for the 5 redshift bins previously considered.  The linear
theory predictions are represented by the solid green lines and note
that for these we use the time-dependent linear bias estimated
directly from the simulations. In the figure we also present two
different estimates for the full nonlinear cross-correlation
coefficient, estimated from bicubic spline fits to the measured
spectra: the blue triple-dot dash curves show the results for the case
where no shot noise subtraction was performed on the $C_l^{\h\h}$
data; the red dash curves show the same but with the shot noise
subtracted.  We also show the dark matter-CMB cross-correlation
coefficient, $r^{T\delta}$, measured in the same redshift bins as for
the haloes (magenta dot-dash curves). For the dark matter estimates,
we used the selection function
$\Pi_{ij}^{DM}(\chi)=D_A^2(\chi)/\int_{\chi_i}^{\chi_j}d\chi
D^2_{A}(\chi)\ $.

For these narrow redshift shells, $\Delta z=0.2$, we find that for
linear theory, neglecting the evolution of the bias does not lead to
significant errors. This can be seen from the middle panels of the
figures, where we plot $r^{T\h}/r^{T\delta}$ (solid green line for
linear theory). However, for the actual measured nonlinear $r^{T\h}$,
we find that the scale-dependence of the $C_l^{\h\h}$ spectra, leads
to a significant discrepancy between $r^{T\h}$ and $r^{T\delta}$ . The
discrepancy is $\approx10\%$ at $l=10$ for the lowest redshift
cluster-sized halo sample (Bin 2). For the group-scale haloes (Bin 1),
the deviation is smaller, being $\approx10\%$ at $l=50$, for the same
redshif shell. However, if one subtracts shot noise from the halo
auto-spectra (bottom panel of the figures), $P_{\rm shot}=1/\nbarh$,
then these effects can be mitigated, and the ratio
$r^{T\h}/r^{T\delta}$ is brought within $5<\%$ of unity. A note of
caution, is that we found that using the standard $P_{\rm
  shot}=1/\nbarh$ to correct for the shot-noise lead to negative power
spectra at high $k$. Since this is forbidden, we believe that such
simple corrections are in fact an {\em over-correction} and new more
accurate methods for accounting for the discreteness will be required
\citep[for a deeper discussion of this issue see][]{Smithetal2007}.

We thus conclude that the relation $r^{T\h}\approx r^{T\delta}$ holds
to within $5\%$ for $l<50$, for the halo samples considerd in this
study. This comes under the provision that the shot noise is accounted
for and the shells are narrow. 


\section{Conclusions}\label{sec:conc}

In this paper we have investigated the impact of the nonlinear
evolution of the time rate of change of the gravitational potentials on
the CMB temperature auto-power spectrum, and also on the
cross-correlation of biased density tracers and the CMB.  Linear
perturbation theory informs us that, for nearly the entire history of
the Universe, gravitational potentials are constant and there is no
net heating or cooling of the primordial CMB photons. However, at late
times in the LCDM model the symmetry between the growth rate of
density perturbations and the expansion rate is broken. The growth
slows, and potentials begin to decay. Using the {\tt zHORIZON}
simulations, a large ensemble of $N$-body simulations and analytic
perturbation theory methods, we explored how this picture changed.

In \S\ref{sec:simulations} we generated maps of the rate of change of
the gravitational potentials at different stages in the simulation. We
showed that, at redshifts $z\sim15-10$, whilst the ISW signal is
vanishingly small, the potentials are indeed evolving nonlinearly on
small scales giving rise to the Rees-Sciama and Birkinshaw-Gull
effects -- nonlinear infall and mass motion across the line of
sight. However, the amplitude of these effects, at these redshifts, is
too low to be detected directly in the CMB or through a
cross-correlation analysis. We then showed that at later times $z>3$
the potential evolution becomes dominated by the large-scale ISW
effect.

In \S\ref{sec:TwoPoint} we focused on investigating the impact on the
CMB temperature power spectrum. The late time ISW effect can be
quantified through a line-of-sight integral over three power spectra:
the auto-spectra of density and momentum, and their cross-spectrum. We
used the nonlinear PT to derive explicit expressions for each of these
quantities. Estimates were then measured from the ensemble of
simulations over the range $z=1$--$0$. In all cases there was evidence
for large-scale nonlinearity, the effects being strongest for the
momentum auto-spectra and at the lowest redshifts. However, when the
spectra were combined to produce the $\pdot$ spectrum, 
the nonlinear corrections to linear predictions
increased with increasing redshift. This was attributed to the fact
that the ISW vanishes at early times, so leaving only the RS and BG
effects.  The standard PT was able to reproduce the nonlinear behavior
at high precision over this redshift range. 

In \S\ref{sec:results1} we estimated the CMB spectrum using the Limber
approximation, we found that the nonlinear amplification of the ISW
effect was $<10\%$ of the linear theory on scales $l<50$, and was also
swamped by the cosmic variance of the linear ISW effect on these
scales. On smaller scales the effect was more significant, however the
primary CMB signal is more than $\sim10^3$ times larger at this
scale. We conclude that for the standard LCDM model, it is highly
unlikely that the nonlinear ISW effects could
contaminate the $l<1500$ multipoles of the CMB spectrum in any
traceable way. Our results support conclusions from earlier studies
\citep{Seljak1996a,Cooray2002a,Maturietal2007,Caietal2008}.

In \S\ref{sec:ISWdelta} we analyzed the cross-correlation of ISW with
the dark matter density field, showing that while the nonlinear
effects suppress this cross-correlation at early times, they may
enhance it at very late times.  This is further investigated in
\S\ref{sec:bias}, where we computed the ISW signal obtained from the
cross-correlation of the CMB with a set of biased tracers of the
density field. We modeled the bias using a time dependent linear model
and also a time- and scale-dependent nonlinear model
\citep{Smithetal2007,McDonald2006a,McDonald2006b}.  For the biased
samples we took the haloes measured in the simulations between $z=0$
and 1, with masses $M>10^{13}\Msol$. These were then sub-divided into
a high- and low-mass sample. The linear and nonlinear bias parameters
were then estimated from the halo-mass cross-power spectra.  The
angular power spectrum of the ISW depends on two spectra: the
cross-power spectrum of the biased tracer with the mass density and
the momentum. These spectra were estimated from the simulations. Again
there was evidence for large-scale nonlinearity, the effects being
strongest for the momentum cross-spectrum and at late times. The
predictions from the nonlinear analytic PT model were found to
qualitatively reproduce the power spectra.  On combining the two
spectra to produce the ISW-density tracer cross-spectrum, we again
found evidence of nonlinearity, and as for the case of the ISW
auto-spectrum, the effects were more noticeable at higher
redshifts. We also found that at late times there was an amplification
of the cross-power spectrum.  Thus at late times in the LCDM model,
nonlinear evolution can lead to a small {\em increase} in the decay
rate of the gravitational potentials.

In \S\ref{sec:results2} we computed the angular power spectra,
averaging over the halo spectra at various redshifts. We found that on
scales $l<100$ the departures from linear theory predictions were
$<10\%$, and these were characterized by a small amplification of the
signal, followed by a strong suppression. The departures are
sub-dominant to the cosmic variance. We then investigated the $\SN$
for the haloes and found good agreement with the linear theory
expectation: the presence of bias effectively cancels out in the $\SN$
expression and leads to negligible changes in the cross-correlation
detectability. We also showed that through the increased Poisson noise
of the biased sample, there was a reduction in the $\SN$, relative to
that for the mass.  Our analyses also demonstrated that the $\SN$ of
the ISW--large-scale structure cross correlation is localized to a
narrow angular range: more than $\sim 90$\% of the overall
significance arises from $l<50$, or angular scales larger than $\sim
4$ degrees.  We therefore conclude that the current power spectrum
analyses of \citet{Hoetal2008} and \citet{Giannantonioetal2008} are
not affected by nonlinear density evolution or scale-dependent bias to
influence the detectability of the ISW-LSS cross-correlation. Since we
do not repeat the exact analysis of \citet{Granettetal2008b} we cannot
directly address whether that result can be explained by nonlinear
effects or whether it requires an alternative explanation.

Finally, we compared the cross-correlation coefficient of the biased
density tracers and the CMB with that of the dark matter and the
CMB. We found that the relation $r^{T\h}\approx r^{T\delta}$ holds to
within $5\%$ for $l<50$, for the halo samples considerd in this
study. This comes under the provision that the shot noise is accounted
for and the shells are narrow. Otherwise the deviations can be large.

The power spectrum anslysis of ISW, therefore, appears to be a probe
relatively free from contamination by the pernicious effects of
late-time nonlinear evolution of the large-scale structures or scale
dependent bias, at least for $l<100$ where most of the signal is. It
therefore continues to be a useful probe for the presence of Dark
Energy or its alternatives \citep{Lombriseretal2009}.


\section*{Acknowledgments}
We acknowledge L. Marian for a careful reading of the draft. RES
kindly thanks the Argelander Institute, University of Bonn for
hospitality whilst some of this work was being performed. CHM
acknowledges the warm hospitality of the University of Zurich, where
this work was initiated. We kindly thank V. Springel for making public
{\tt GADGET-2} and for providing his {\tt B-FoF} halo finder; R.
Scoccimarro for making public his {\tt 2LPT} code. RES acknowledges
support from a Marie Curie Reintegration Grant.  This work is partly
supported by the Swiss National Foundation under contract
200021-116696/1 and WCU grant R32-2008-000-10130-0.



\bibliography{/Users/res/WORK/PAPERS/REFS/refs}


\vspace{0.2cm}

\appendix


\section{Validity of the Limber approximation}\label{sec:LimberApprox}


\begin{figure*}
\centering{ 
\includegraphics[width=13.cm,clip=]{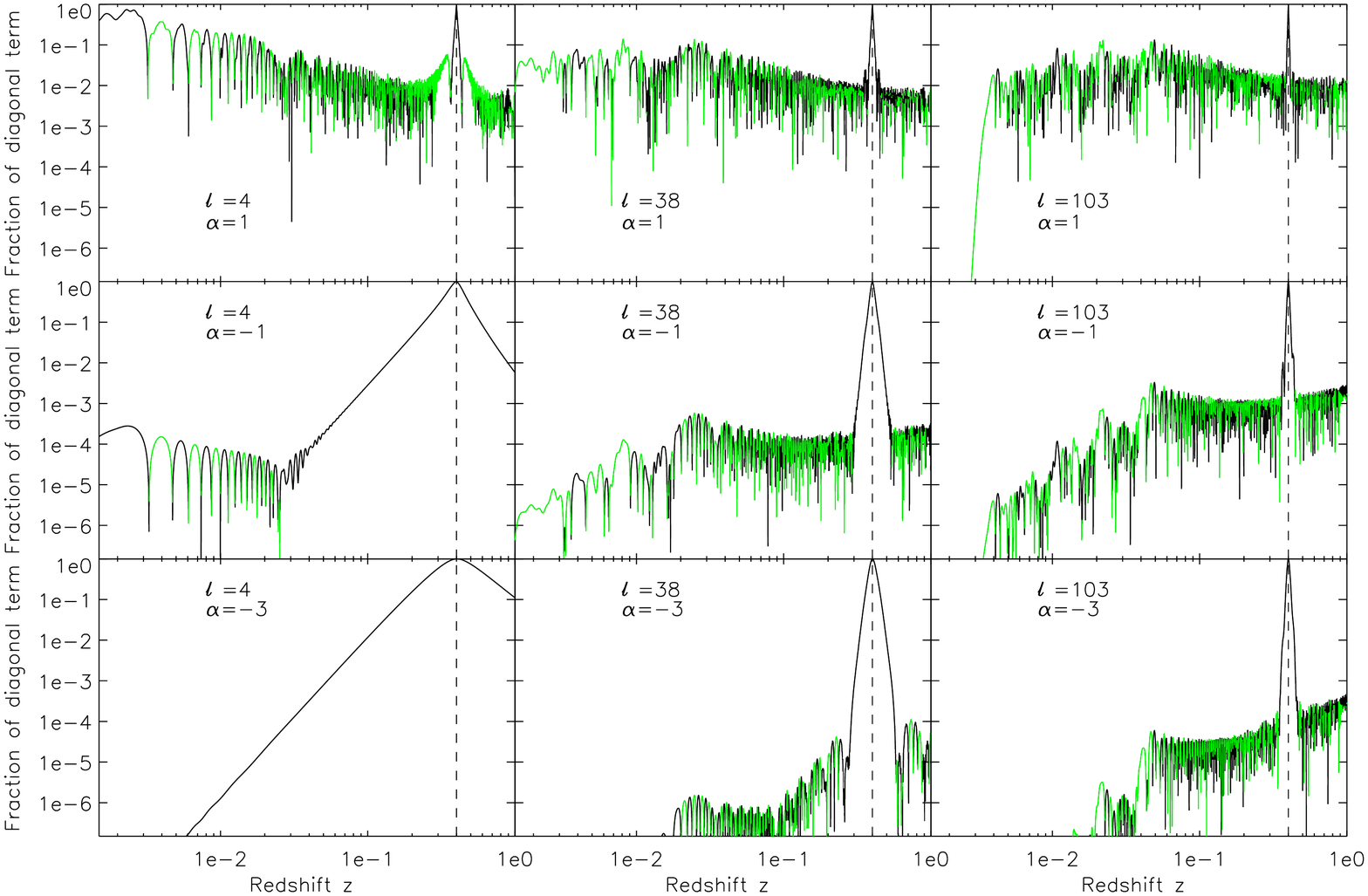}}
\caption{\small{Rows of the matrix ${\cal F}(l,\alpha,r_1,r_2)$
    (normalized by the diagonal term) corresponding to $r_1(z_0=0.4)$
    versus the redshift corresponding to $r_2$, under different
    choices of $l$ and $\alpha$. Given the logarithmic scale, green
    color displays negative values, black points positive ones.  }
\label{fig:limb1}}
\centering{ 
\includegraphics[width=10cm,clip=]{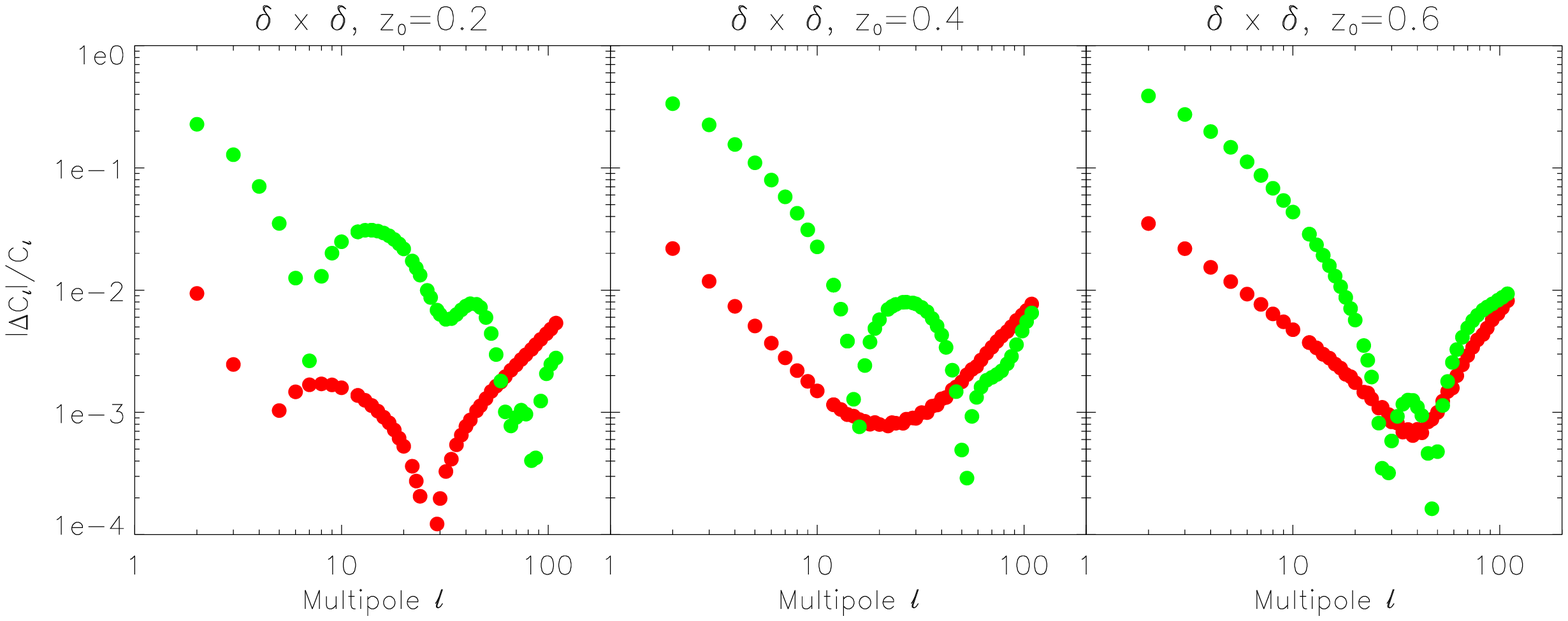}
\includegraphics[width=10cm,clip=]{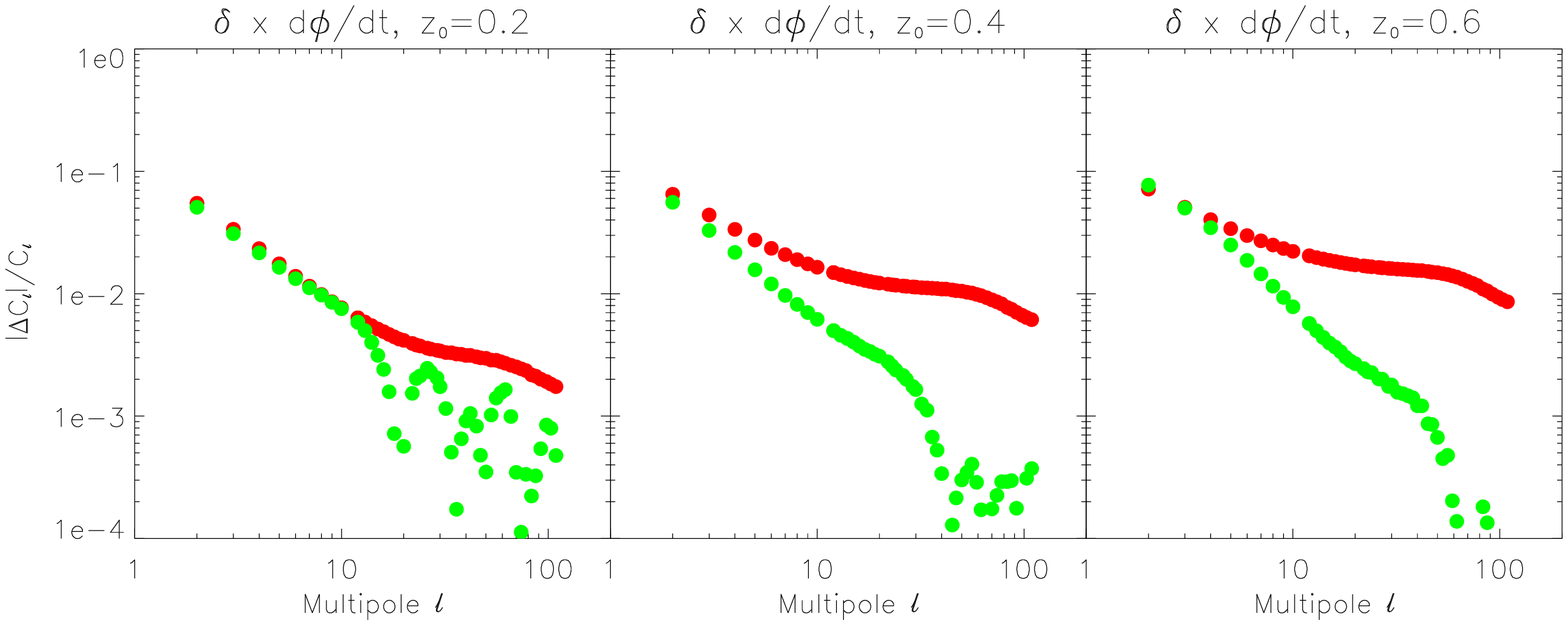}
\includegraphics[width=10cm,clip=]{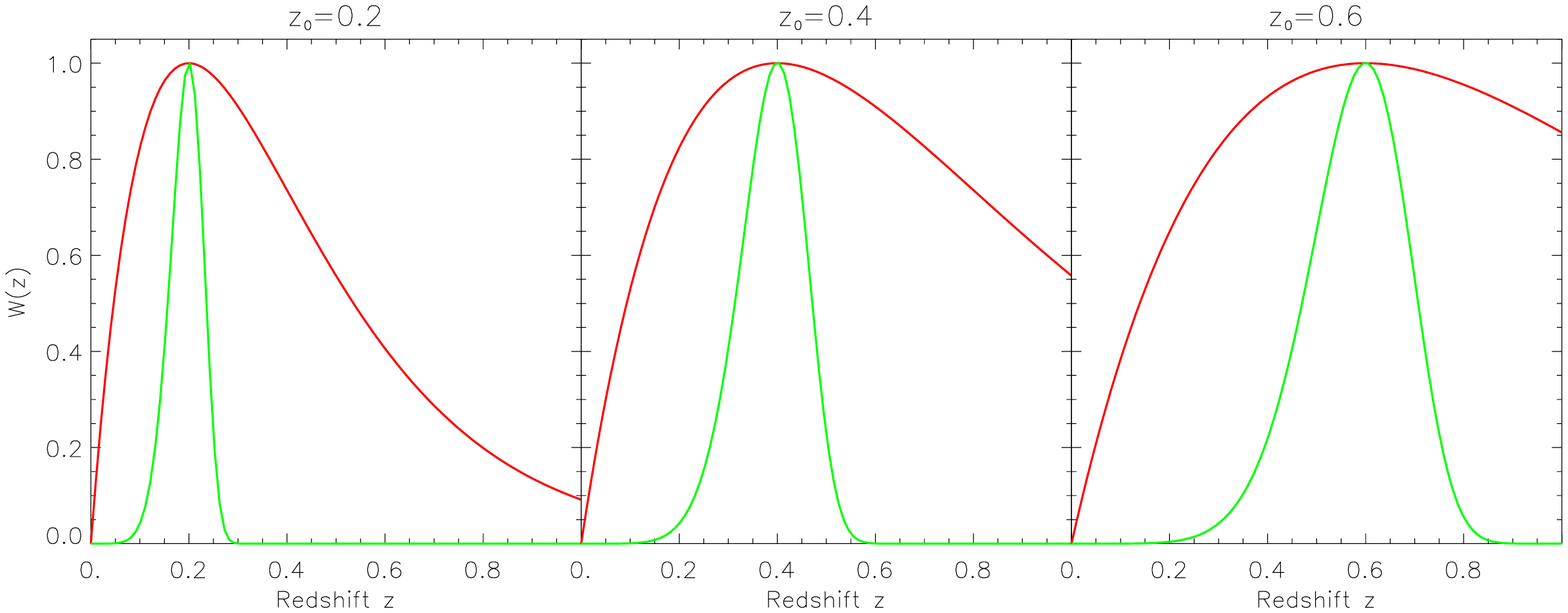} }
\caption{\small{Comparison of the exact $C_l$ evaluation for the
ISW--density tracer correlation with the Limber approximation
evaluation. The top three panels show the relative errors for a near,
intermediate and far density tracer survey. Solid lines denote
predictions for thick redshift shell and dash lines denote thin
redshift shells. The corresponding bottom panels show the redshift
distributions.}
\label{fig:LimberTestDPhi}}
\end{figure*}


The Limber approximation is motivated by:
\be
\int_0^{\infty}dk \; k^2 j_l(kr_1) j_l(kr_2) = \frac{\pi}{2}\frac{\delta^D(r_1-r_2)}{r_1^2},
\label{eq:limb1}
\ee
where the symbol $\delta^D$ denotes the Dirac delta function. Under
the assumption that the spherical Bessel functions $j_l(kr)$ are
rapidly oscillating for high enough $l$-s, then one can write an
integral over a generic power spectrum as
\ba
{\cal F}(l,\alpha,r_1,r_2) & \equiv & 
\int_{k_{min}}^{k_{max}}dk \; k^2 j_l(kr_1) j_l(kr_2) \; P(k)\nonumber \\
& \approx & \frac{\pi}{2}\frac{\delta^D(r_1-r_2)}{r_1^2}\; P\left(k=\frac{l+1/2}{r_1}\right) \ ,
\label{eq:limb2}
\ea
where the power spectrum $P(k)$ is assumed, in a cosmological context,
to be a power law times some transfer function $|T(k)|^2$, $P(k) =
k^{\alpha} |T(k)|^2$.  If seen as a four dimensional matrix with
indices running on $\{l,\alpha,r_1,r_2\}$, the deviation of ${\cal F}$
from a diagonal matrix in the last two indices, may be viewed as a
measure of the error introduced by the Limber approximation.

In Fig.~\ref{fig:limb1} we examine ${\cal F}$ for the case where we
have fixed $r_1$ to be the comoving distance to $z_0 = 0.4$ and where
$r_2$ varies on the X-axis. We consider three cases for the multipole
number: $l=\{4, 38, 103\}$; and three cases for the spectral index:
$\alpha = \{1,-1,-3\}$ which may be thought of as $P_{\delta\delta}$,
$P_{\pdot\delta}$, and $P_{\pdot\pdot}$. We take $k_{\rm min}=10^{-5}
\kMpc$ and $k_{\rm max}=1 \kMpc$. For the sake of clarity, the
elements of ${\cal F}$ have been normalized by the maximum value of
each row. Black points denote positive values and green ones negative
entries. The diagonal term (at $z_0 = 0.4$) has been marked by a
vertical dashed line. From the figure it is clear that the deviation
from a diagonal matrix is more apparent at low multipoles, and for
more negative values of $\alpha$.  At higher $l$, however, the width
of the ${\cal F}$ matrix shrinks around $z_0$, making the Limber
approximation more precise. The actual error on these multipoles is
related to how the off-diagonal terms are weighted by the time
dependent factors, and how their sum cancels within the integration
range.

Fig.~\ref{fig:LimberTestDPhi} presents the errors on
$C_l^{\delta\delta}$ and $C_l^{T\delta}$, at three different redshifts
$z_0 = \{0.2, 0.4, 0.6\}$ and for thin (green color) and thick (red
color) redshift shells (these are displayed in the bottom panels). In
all cases, for $l>20$, the errors are below 3\%. We find that for
$C_l^{\delta\delta}$, the net resulting error is larger for thin
redshift shells than thick ones. This is inverted for $C_l^{T\delta}$,
where the contribution to the off-diagonal terms are smaller for thin
shells. However errors remain always below the few-percent level. The
amplitude of the errors are defined by: the actual width of the peak
around $z=z_0$; the amplitude of the oscillating floor around the
wings of the peak at $z=z_0$; and the actual width of the redshift
integration range compared to the width of the peak at $z=z_0$.


\end{document}